\documentclass[a4paper,11pt]{article}
\usepackage{jcappub} 
\usepackage[usenames]{color}
\usepackage{changepage} 
\usepackage{amssymb}
\usepackage{amsmath}
\usepackage{rotating}
\usepackage{comment}
\usepackage{graphicx}
\usepackage{caption}
\usepackage{subcaption}
\usepackage[dvipsnames]{xcolor}

\title{\boldmath A Search for Dark Matter Annihilation in Stellar Streams with the \textit{Fermi}-LAT}

\author[a, b]{Cristina Fern\'andez--Su\'arez}
\author[a, b]{and Miguel \'A. S\'anchez--Conde}
\affiliation[a]{Instituto de F\'isica Te\'orica UAM-CSIC,
Universidad Aut\'onoma de Madrid, \\
C/ Nicol\'as Cabrera, 13-15, 28049 Madrid, Spain}
\affiliation[b]{Departamento de F\'isica Te\'orica, M-15,
Universidad Aut\'onoma de Madrid, \\
E-28049 Madrid, Spain}

\emailAdd{cristina.fernandezs01@estudiante.uam.es, miguel.sanchezconde@uam.es}

\abstract{Stellar streams are remnants of globular clusters or dwarf galaxies that have been tidally disrupted by the gravitational potential of their host galaxy. Streams originating from dwarfs can be particularly compelling targets for indirect dark matter (DM) searches, as dwarfs are believed to be highly DM-dominated systems. Although these streams are expected to lose most of their DM during the tidal stretching process, a significant amount may still remain in their core.
If DM is composed of \textit{Weakly Interacting Massive Particles} (WIMPs), their annihilation within the streams' cores could produce a detectable gamma-ray signal. In this study, we analyze nearly 15 years of data from the Large Area Telescope aboard NASA’s \textit{Fermi} Gamma-ray Observatory (\textit{Fermi} LAT) to search for potential WIMP annihilation signals from the direction of eleven stellar streams selected using DM-motivated criteria.
No gamma-ray emission is detected from any of the streams in our sample. In the absence of a signal, we place the first constraints on the WIMP parameter space based on these objects for two representative annihilation channels. Individual DM limits are first computed and then combined across all streams. The most reliable combined constraints, derived from a \textit{golden} sample and using a benchmark description of their DM content, are $\mathcal{O}(10)$ times above the thermal relic cross-section for the $b\bar{b}$ and $\tau^{+}\tau^{-}$ annihilation channels.
We also explore how these constraints vary under more conservative or optimistic yet realistic assumptions, finding variations of up to a few orders of magnitude. This large systematic uncertainty stems from key model limitations, which can and should be addressed in future observations and simulations to produce more robust DM limits from streams.
Despite uncertainties, this study demonstrates the feasibility of using stellar streams for indirect DM searches and highlights their significant potential to uncover the properties of DM particles.}

\begin{keywords}
{dark matter theory, gamma ray theory}
\end{keywords}

\begin{document}
\maketitle
\flushbottom

\newpage
\section{Introduction}
\label{sec:intro}
From galactic to cosmological scales there is multiple sources of evidence of the existence of a non-baryonic form of dark matter (DM) that accounts for 85$\%$ of the total matter in the Universe \cite{Bertone_2005, Garrett_2011, Bertone_2018}. Despite our efforts, we still do not know what it is made of, and revealing its nature is one of the most challenging goals of modern physics. 
Among the numerous candidates that have been proposed to form this unseen component of the Universe, \textit{Weakly Interacting Massive Particles} (WIMPs, \cite{Bertone_2010, Roszkowski_2018}) are one of the most promising and well-motivated. WIMPs would have been produced in the very early and hot Universe via a thermal freeze-out mechanism. It is assumed that, initially, Standard Model (SM) and DM particles were in thermal equilibrium. As the Universe evolved, it expanded and cooled, and WIMPs eventually froze out of equilibrium with the thermal plasma, the WIMP comoving relic number density remaining mostly constant since then. The amount of DM that we observe today requires that the thermal pair annihilation cross-section is $\langle \sigma v \rangle = 5.2 \times 10^{-26} \,  \mathrm{cm^{3} s^{-1}} $ at masses around 0.3 GeV, and $\langle \sigma v \rangle = 2.2 \times 10^{-26} \,  \mathrm{cm^{3} s^{-1}} $ for masses above 10 GeV \cite{Steigman_2012}. Remarkably, WIMPs are predicted to be produced with the right relic density to be DM. This coincidence is the so-called \textit{``WIMP miracle''}.
In addition, WIMPs arise in several beyond the SM (BSM) models in order to solve problems of particle physics. 

There are different strategies to look for WIMPs: direct production at colliders that seeks to detect DM particles produced in particle collisions \cite{Buchmueller_2017, Boveia_2018}, direct detection in Earth-based experiments aimed at detecting interactions between DM and SM particles \cite{Cerde_o_2010}, and {\it indirect} searches that look for the annihilation or decay products of DM particles, such as photons, antimatter and neutrinos \cite{conrad2014indirect}. 
Among them, photons are considered as the best probe for indirect DM searches, since they are not deflected by magnetic fields, as opposed to antimatter, and are easier to detect than neutrinos. WIMPs may be detectable in gamma rays since self-annihilation of WIMPs produce SM particles, which can eventually yield photons, among other possible by-products.
Many astrophysical targets have already been scrutinized at gamma-ray energies with the aim of detecting a hint of DM, with no unequivocal signal as of today (e. g., the Galactic center (e.g., \cite{Ajello_2016_GC, Ackermann_2017_GCE, Di_Mauro_2021_GCE, Zuriaga_Puig_2023}), dwarf spheroidal galaxies (dSphs; e.g., \cite{Ackermann_2015_dSphs, oakes2019combined, mcdaniel2023legacy}), dwarf irregular (dIrr) galaxies (e.g., \cite{Gammaldi_2021}), galaxy clusters (e.g., \cite{Ackermann_2015_galaxyClusters, Di_Mauro_2023}), or dark satellites (e.g., \cite{Ackermann_2012_dark_sat, Coronado_Bl_zquez_2019_1, Coronado_Bl_zquez_2019_2, Coronado_Bl_zquez_2022}). 

In this work, we will explore the potential that stellar streams may have as a new, complementary target for indirect DM searches with gamma rays. Stellar streams are tubular galactic structures with lengths from 1 kpc to more than 100 kpc. They are the remnants of ancient globular clusters (GCs) or dwarf galaxies (dGs) accreted during the process of formation of galaxies like our own, that have been heavily stripped by the tidal field of the host galaxy since then. Indeed, the streams' progenitor orbiting around the galaxy gets tidally stretched by the galactic potential, with the gravitational pull being stronger on the closer stars to the galactic centre. This way, the inner stars become a leading arm of the stellar stream while the outer stars form a trailing arm. Given both their low surface brightness and typical large angular extensions in the sky, stellar streams are generally difficult to find and to study. However, since the first stellar stream was discovered almost three decades ago \cite{1994Natur.370..194I}, the search for streams has become an active field of research. 
With the advent of wide and deep sky surveys, such as the Sloan Digital Sky Survey (SDSS, \cite{York_2000}), 
Pan-STARRS \cite{Hodapp2014},
Gaia \cite{Gaia2016},  
and the Dark Energy Spectroscopic Instrument (DESI, \cite{desicollaboration2016desi, desicollaboration2016desi_b}), 
the number of discovered streams has grown rapidly and today around a hundred are known to orbit around the Milky Way (MW) \cite{2023MNRAS.520.5225M}. Streams have already been studied in a large variety of contexts: as the majority of these objects reside in our galactic halo, they can be used to reconstruct the assembly history of the MW (e.g., \cite{Malhan_2022_MW}), as well as to infer the shape and the mass of the galaxy (e.g., \cite{Malhan_2019_MW}). Furthermore, those streams exhibiting low star velocity dispersions, the ``cold'' stellar streams, have been proposed as a powerful laboratory to search for DM subhalos (e.g., \cite{Bonaca_2019, Banik_2021, mcgee2022dark, Carlberg_2012, Erkal_2016, Mirabal_2021}). Indeed, in some cases, significant inhomogeneities in the density of stream stars (like gaps, blobs and spurs) have been observed and characterized in detail, that could be due to the impact of low-mass, dark subhalos with the streams, that would gravitationally disturb them. The effects that such encounters would cause can be used not only to infer the mass and orbit of the perturber (e.g., \cite{Bonaca_2020_pert}), but also to test the underlying cosmological model. For example, Cold DM (CDM) and Warm DM (WDM) models predict a completely different abundance of dark subhalos in the MW, which would result in different morphologies of galactic stellar streams (e.g., \cite{Yoon_2011}). However, these studies are currently challenging due to lack of precision in kinematics data and the existence of multiple statistical and systematic uncertainties \cite{Banik_2018}.

In this work, we will focus only on those streams whose progenitor is a dG, since they are DM-dominated objects, while GCs -- the other possible progenitor for streams -- are expected to be devoid of DM according to the standard picture. Although we expect streams to have lost most of their DM during the stretching process, they should still contain a non-negligible amount of DM in their core \cite{Aguirre_Santaella_2022}, which becomes the region of interest in our search. Indeed, 
our aim is to detect a potential WIMP annihilation signal in the form of gamma rays from the cores of an optimized sample of streams with the Large Area Telescope (LAT) onboard the NASA \textit{Fermi} Gamma-Ray Observatory (\textit{Fermi}-LAT) \cite{Gehrels:1999ri, Atwood_2009}. The LAT, launched in 2008, is a high-energy gamma-ray telescope covering the energy range from 20 MeV to more than 300 GeV, that surveys the whole sky every 3 hours with unprecedented sensitivity. Not only this operating energy window makes the LAT the ideal instrument for this type of search, since WIMPs are expected to have GeV-TeV masses \cite{Bertone_2005, Bertone_2010}, but also its full-sky coverage is of particular importance, since it  allows us to perform a data analysis for any stream in our sample. 
 
The paper is organized as follows: in Section \ref{sec:sample} we discuss the sample of stellar streams that we chose for our study, explaining the adopted selection criteria and providing a brief description of each stream in the sample. Different samples are built according to our level of confidence in the properties of the streams. In Section \ref{sec:modeling} we detail the DM modelling that has been performed for our selection of streams and, under different scenarios of their DM content, derive the astrophysical factors that enter in the computation of expected DM annihilation fluxes. Section \ref{sec:analysis} is devoted to the LAT data analysis: events selection, methodology, and results. We then show in Section \ref{sec:constraints} the obtained results for a joint likelihood analysis of all the streams in our sample, and, in the absence of a signal, the derived constraints on the WIMP parameter space, both for each considered stream and for the whole sample. Finally, in Section \ref{sec:concl}, we discuss our main results, place them into the more general DM landscape, and address the most significant caveats behind our work. 

\section{Sample selection}
\label{sec:sample}

The first step is to build our best sample of stellar streams for gamma-ray DM searches, among all those discovered, according to their most relevant known properties. Indeed, we impose several criteria as to include a given stream in our sample or not, namely:
\begin{itemize}
    \item \textbf{Progenitor}: We are interested in those streams whose progenitor is a dwarf galaxy (dG).  
    These objects have been widely used in the context of DM searches as they are expected to be highly DM-dominated systems, with typical mass-to-light (M/L) ratios of $10-1000$ \cite{Mateo_1998, Sanchez_Conde_2011, Ackermann_2015}. They are also expected to be free from bright astrophysical gamma sources \cite{Winter_2016}. We discard those stellar streams whose progenitor is a GC, since they are not expected to harbour a significant amount of DM within them and thus are not interesting targets in this context.
    
    \item \textbf{Distance}: Since the annihilation flux is inversely proportional to the source distance, the closest streams are the most promising ones for gamma-ray DM searches. 
    Given the typical masses of the streams in our sample (see below), and taking into account tidal stripping arguments~\cite{aguirresantaella2023viability}, we decide not to include in our sample streams whose core is further than $\sim$ 100 kpc from us. The flux decreases by two orders of magnitude from the closest streams in our sample (located at $\sim$10 kpc) and those at 100 kpc, so the latter are not expected to be competitive anymore.
    
    \item \textbf{Stellar mass}: Knowing the stellar mass of each stream is key to achieve our purposes,  since the annihilation flux is proportional to the DM mass and, as we will see in Section \ref{sec:modeling}, we will make a correspondence between the stellar and the DM mass. In order to be conservative, we discard those streams whose stellar mass is currently (too) uncertain. 
\end{itemize}

\begin{figure}[h!]
    \centering
    \includegraphics[width=1.0\textwidth]{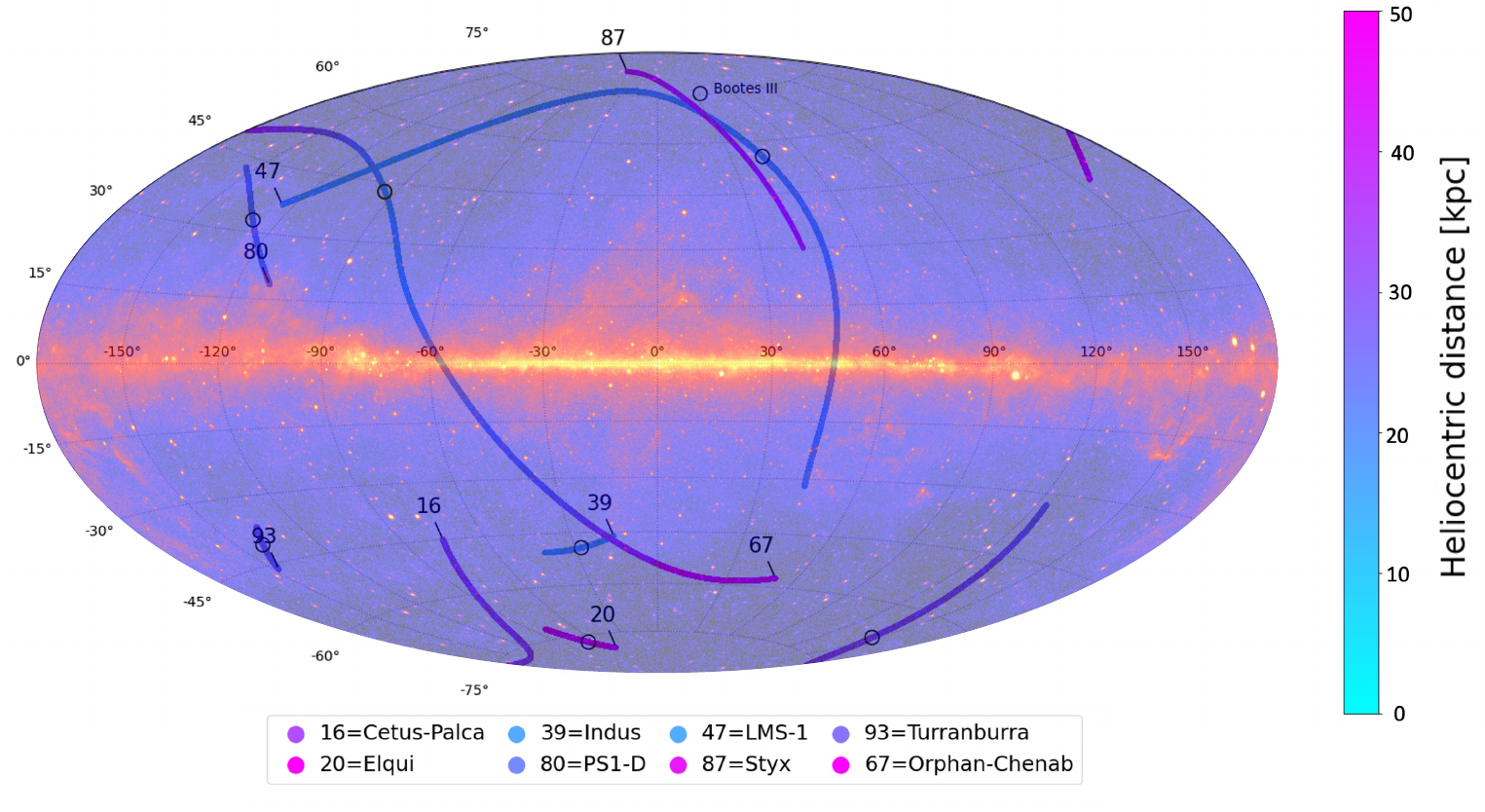}
    \caption{Sky map, in Galactic coordinates and Hammer projection, of the celestial tracks of stellar streams in our \textit{golden} sample, superimposed on the 12-years \textit{Fermi}-LAT gamma-ray skymap (NASA/DOE/Fermi-LAT Collaboration). For each stream track, we represent its core position with a black empty circle. The color bar depicts the heliocentric distance of each stream along its track in kiloparsecs. Plot made using the \textit{galstreams} library \cite{2023MNRAS.520.5225M}.    }
    \label{fig:sample}
\end{figure}

In order to coherently build our sample, we have made use of the \textit{galstreams} library \cite{2023MNRAS.520.5225M}, which provides the most up-to-date compilation of stellar streams in the Milky Way according to the information available in the literature. We summarize in Table \ref{table:table_sample2} the list of known streams whose most probable progenitor is a dG, distinguishing them into three groups:
\begin{itemize}
    \item First, we present our ``\textbf{\textit{golden}}'' sample of streams, i.e., those that meet all the criteria set out above and that will therefore be the ones providing the most reliable results in this work. Figure \ref{fig:sample} shows a skymap of the \textit{golden} sample of streams superimposed on the gamma-ray sky map as seen by the \textit{Fermi}-LAT. 
    
    \item Second, the ``\textbf{\textit{silver}}'' sample contains all the streams in the \textit{golden} sample plus three additional streams: Monoceros, AntiCenter and Sagittarius. While these extra streams meet all our criteria, they have associated certain caveats that lead us to include them separately in a \textit{silver} sample. In the case of Monoceros and AntiCenter, there is an ongoing debate in the literature about whether they are stellar streams or structures formed due to perturbations in the Galactic disk (see discussion in Section~\ref{sec:silver}). 
    Regarding Sagittarius, despite also meeting all the selection criteria, we have decided to include it in the \textit{silver} sample since it is known to be in a very advanced and complicated dynamic state of stretching (further information is also given below). Later in our work we analyze and provide results for both the streams in the \textit{golden} and the \textit{silver} samples. Yet, we consider the results obtained for the \textit{golden} sample the most reliable ones, and this is the way they will be presented and discussed throughout the manuscript.

    \item Finally, the streams listed in Table \ref{table:table_sample2} under the category ``\textbf{\textit{Other streams whose progenitor is a dG}}'', despite being originated from a dG, do not meet the other two imposed selection criteria (see Section \ref{sec:others} for details). Therefore, they will not be considered for our analyses, but remain potentially interesting objects.
\end{itemize}

There are other coherent structures reported in the literature as stellar streams originating from dGs (the Helmi streams \cite{Helmi_1999}, S1-S4 \cite{Myeong_2017}, Nyx \cite{necib2022evidencevastprogradestellar}, and Icarus \cite{Fiorentin_2021}), although they are not included here because they are so close to the Sun that their tracks can not be well-defined. For the same reason, these streams are also not included in \textit{galstreams} either. Next, we briefly describe the most relevant information pertaining each stream listed in Table \ref{table:table_sample2}.

\begin{table}[h!]
\begin{adjustwidth}{-1.3cm}{}
\begin{center}
\begin{tabular}{ c  c  c  c  c  c c}
\hline\hline
Stream & (l, b) ($^{\circ}$) & $d_{Sun}$ (kpc) & Length ($^{\circ}$) & $\frac{M_{*}}{10^4} \, (M_{\odot})$ & Selected Refs. \\ \hline\hline

\multicolumn{6}{c}{\rule{6.5cm}{0.4pt} \textit{Golden sample} \rule{6.5cm}{0.4pt}}\\
Indus & (332.26, -49.19) & 16.6 & 18.2 & 3.40  & \cite{Shipp_2018, Shipp_2019, Ji_2020, Malhan_2021}  \\
LMS-1 & (43.27, 55.46)  & 18.1 & 179.2 & 10.00  & \cite{Malhan_2021, Yuan_2020} \\
Orphan-Chenab & (264.90, 43.60)  & 20.0 & 230.6 & 16.00 & \cite{Shipp_2018, Shipp_2019} \cite{Grillmair_2006orphan, Belokurov_2007, koposov2019, Mendelsohn_2022} \\
PS1-D & (230.95, 32.67) & 22.9 & 44.9 & 0.75  & \cite{Bernard_2016}  \\
Turranburra & (219.72, -40.79)  & 27.5 & 13.7 & 0.76 &\cite{Shipp_2018, Shipp_2019}  \\
Cetus-Palca & (147.90, -67.80)  & 33.4 & 100.9 & 150.00 & \cite{Shipp_2018} \cite{Newberg_2009, Yuan2019, chang2020, Thomas_2022}  \\
Styx & (35.40, 75.40)  & 46.5 & 60.4 & 1.80 & \cite{Grillmair_2009, carlin2009kinematics, Carlin_2018}  \\
Elqui & (293.88, -77.20) & 50.1 & 10.9 & 1.04 & \cite{Shipp_2018, Shipp_2019}  \\
\multicolumn{6}{c}{\rule{6.5cm}{0.4pt} \textit{Silver sample} \rule{6.5cm}{0.4pt}}\\
Monoceros & (180.0, 25.0)  & 10.6 & 46.9 & 600.00 & \cite{Morganson_2016} \cite{Newberg_2002, refId0} \\ 
AntiCenter & (140.0, 35.0)  & 11.7 & 57.7 & 0.93  & \cite{refId0, Grillmair2006}  \\
Sagittarius & (6.01, -14.89)  & 25.0 & 280.0 & 13000.00 & \cite{1994Natur.370..194I} \cite{Malhan_2022_MW} \cite{Mateo_1996, Antoja_2020, Majewski_2003, ramos2020, venville2023prospective}  \\
\multicolumn{6}{c}{\rule{4cm}{0.4pt} \textit{Other streams whose progenitor is a dG} \rule{4cm}{0.4pt}}\\
Jhelum-a & -  & 13.0 & 30.0 & - &\cite{Shipp_2018, Shipp_2019} \cite{Bonaca_2019jhelum}  \\
Jhelum-b & -  & 13.0 & 30.0 & - & \cite{Shipp_2018, Shipp_2019}  \cite{Bonaca_2019jhelum}  \\
Parallel & (263.88, 61.26)  & 14.3 & 37.7 & - & \cite{Sohn_2017, Weiss_2018} \\
C-19 & (102.29, -38.54)  & 18.0 & 29.7 & 0.32 & \cite{ibata2023chartinggalacticaccelerationfield} \cite{Martin_2022, Errani_2022} \\

\end{tabular}
\end{center}

\caption{Known stellar streams whose progenitor is most likely a dwarf galaxy. For each stream, we provide the galactic longitude ($l$) and galactic latitude ($b$) of the most likely location for its core;
the distance from this core to the Sun ($d_{Sun}$), the angular length of the whole stream, and the stellar mass ($M_{\star}$) of each stream. In the last column, we provide a set of selected references for each stream that provide the most relevant information (including the discovery work). Different subsamples are presented in the table according to our level of confidence in each stream; see text for details. Only streams in both the \textit{golden} and \textit{silver} samples will be considered for our gamma-ray analysis and DM results. We have used the \textit{galstreams} library \cite{2023MNRAS.520.5225M} as a starting point for building this table.}
\label{table:table_sample2}
\end{adjustwidth}
\end{table}

\subsection{Golden Sample}
We present below detailed information on the 8 stellar streams that meet all our selection criteria and thus form our \textit{golden} sample (see Table \ref{table:table_sample2} for details).  These streams make up the primary sample of this work, yielding the main results of the analysis.

\begin{itemize} 
    \item \textbf{Indus}: It was proposed as a stellar stream by Ref.~\cite{Shipp_2018} and interpreted as a low-mass dG stream by Ref.~\cite{Ji_2020}. In addition, in Ref. \cite{Malhan_2021} they suggest that the progenitor of Indus was a satellite dG of the progenitor of LMS-1 due to the similarity in their dynamics, metallicity and stellar population. The exact location of the progenitor of Indus has not been clearly identified yet. Then, motivated by other works in which they take the mid-point of the stream as the position of its core in the absence of a clear progenitor ~\cite{Malhan_2021}, we will assume that the core of Indus is located at the mid-point of the stream, i.e., (l, b) = (332.26$^{\circ}$, -49.19$^{\circ}$), whose position is 16.6 kpc from the Sun.
    
    \item \textbf{LMS-1}: This stream was first reported in Ref.~\cite{Yuan_2020} and recently studied in detail by Ref.~\cite{Malhan_2021}. In the latter work, the authors conclude that the progenitor of LMS-1 is a dG and assume that its core  is currently at the mid-point of the stream (although it has not been detected so far, presumably because it is completely disrupted). We will follow this same assumption and place the core in the middle of the stream, i.e., (l, b) = (43.27$^{\circ}$, 55.46$^{\circ}$), which is located 18.1 kpc from the Sun.

    \item \textbf{Orphan-Chenab (OC)}: The Orphan stream was discovered by the authors of Ref.~\cite{Grillmair_2006orphan} and Ref.~\cite{Belokurov_2007} independently. Besides, Ref.~\cite{Shipp_2018} discovered the Chenab stream  using data from the Dark Energy Survey (DES, \cite{Abbott_2018}). Strong evidence was found later that both streams come from the same dG, which is now named as the Orphan-Chenab stream \cite{Shipp_2019,koposov2019}. Although some potential progenitors of the OC stream have been proposed, its exact location is still unknown. Yet, according to Ref.~\cite{Mendelsohn_2022}, there is an overdensity of stars located at (l, b) = (264.90$^{\circ}$, 43.60$^{\circ}$), coinciding with the position of its more likely progenitor, so we will take this position as the location of the OC's core. This position is 20.0 kpc away from the Sun.
    
    \item \textbf{PS1-D}: In Ref. \cite{Bernard_2016}, authors report the discovery of PS1-D suggesting that its progenitor is a low-luminosity dG. In the absence of a known progenitor as of today, we will assume once again that it is located at the mid-point of the stream, i.e., (l, b) = (230.95$^{\circ}$, 32.67$^{\circ}$), at a distance of 22.9 kpc to the Sun.
    
    \item \textbf{Turranburra}: This stream was discovered in data from DES and first reported in Ref.~\cite{Shipp_2018}. Its properties suggest that its most likely progenitor is a dG, however no potential progenitor has been detected as of today. For our work, we will assume that the progenitor's core lies at the mid-point of the stream, i.e., (l, b) = (219.72$^{\circ}$, -40.79$^{\circ}$), also motivated by an overdensity of stars in that location in Figure 1 of Ref.~\cite{Li_2022}, which corresponds with a distance of 27.5 kpc from the Sun.
    
    \item \textbf{Cetus-Palca}: The Cetus stream was discovered by Ref.~\cite{Newberg_2009} using SDSS data. Later, Ref.~\cite{Yuan2019} showed that Cetus is comprised of two parts on opposite sides of the Galactic disk, while Ref.~\cite{chang2020} suggested that half of its members are located in the southern sky. This southern region overlaps with the location of the Palca stream discovered in the DES data by Ref.~\cite{Shipp_2018}. Moreover, Palca members are kinematically consistent with Cetus members, strongly suggesting the association of these two systems as being two parts of the same stellar structure: the so-called Cetus-Palca stream \cite{Thomas_2022}. Its properties suggest that its progenitor is a small dG and, according to Ref.~\cite{Thomas_2022}, if the progenitor were completely disrupted today, it would likely be located in the position where the leading and trailing arms of the stream meet, i.e., (l, b) = (147.90$^{\circ}$, -67.80$^{\circ}$), whose position is 33.4 kpc from the Sun.
    
    \item \textbf{Styx}: Ref.~\cite{Grillmair_2009} discovered the Styx stream and proposed the Bootes III dG as its progenitor. Later works \cite{carlin2009kinematics, Carlin_2018} show that Bootes III is a dG satellite in a transitional state between a bound entity and an unbound stellar stream, and prove its relation with Styx. Therefore, we will consider the current position of Bootes III, (l, b) = (35.40$^{\circ}$, 75.40$^{\circ}$), as the location of the Styx's progenitor (although due to its current state of transition to stream, its coordinates do not exactly match those given for the Styx track, as can be seen in Figure \ref{fig:sample}), which is 46.5 kpc from the Sun.
    
    \item \textbf{Elqui}: This stream was discovered by Ref.~\cite{Shipp_2018} using DES data and classified as the remnant of a dG. Its progenitor is still unknown, thus as in other cases we will assume the core of the stream to be at its mid-point, i.e., (l, b) = (293.88$^{\circ}$, -77.20$^{\circ}$). This position corresponds with a distance of 50.1 kpc to the Sun.

\end{itemize}

\subsection{Silver Sample}
\label{sec:silver}

We dedicate this subsection to provide information corresponding to the streams belonging to the \textit{silver} sample. We remember that this sample includes the streams belonging to the \textit{golden} sample plus three more streams (see Table \ref{table:table_sample2}), that we will also analyze. However, it is important to note that the results obtained will not be as reliable as those given by the streams of the \textit{golden} sample.

\begin{itemize}

    \item \textbf{Monoceros Ring (MR)}: A stellar substructure located in the outer disk of the MW that extends more than one hundred degrees in longitude (100$^{\circ}$ $<$ l $<$ 270$^{\circ}$), both in the northern and southern hemispheres \cite{Morganson_2016}, with an overdensity of stars at (l, b) = (180.0$^{\circ}$, 25$^{\circ}$) \cite{ramos2020}. We take this position as its core, which is located at 10.6 kpc from the Sun, thus being this stream the closest to us of all those in the sample. The origin of MR is controversial, with two main hypotheses being debated. Originally, MR is believed to be a tidal stream formed from the disruption of a satellite galaxy \cite{Yanny_2003}. Under this hypothesis, several works tried to find its potential progenitor (e.g., \cite{refId0}) and even estimate upper limits for its total mass ($\sim 10^{10} \, M_{\odot}$ \cite{10.1093/mnras/stx3048}). Later works suggest that the interaction of a Sagittarius-like dG with the Galactic disk can lead to the formation of a stream composed of disk stars consistent with the observations \cite{Kazantzidis_2008, laporte}. As its origin remains unclear, in order to be conservative we have decided not to include it in our \textit{golden} sample, but in the \textit{silver} one instead.\footnote{Indeed, as it will be shown later in Section~\ref{sec:constraints}, the inclusion or not of MR impacts our combined DM limits significantly.}
   
    \item \textbf{AntiCenter (ACS)}: As in the case of the Monoceros Ring, ACS is a structure in the outer disk of the MW, whose origin is still under debate. Although initially thought to be a tidal stream \cite{Grillmair2006} showing an overdensity of stars at (l, b) = (140.0$^{\circ}$, 35.0$^{\circ}$) \cite{ramos2020}, some works suggest that the whole stream could actually be a feature produced by perturbations in the Galactic disk \cite{laporte,refId0}. Once again we consider the location of this overdensity  of stars as the core of the stream, which is at a distance of 11.7 kpc to the Sun. As in the previous case, and in order to be conservative, we have decided to also discard ACS from our \textit{golden} sample and include it in the \textit{silver} one.

    \item \textbf{Sagittarius (Sgr)}: The Sagittarius stream, whose progenitor is the Sagittarius dG, was the first stellar stream discovered \cite{1994Natur.370..194I,Mateo_1996}. It is the most prominent stream of the MW, spreading all over the sky, and has been the subject of several studies aimed at reconstructing the assembly history of our galaxy and its gravitational potential (e.g., \cite{Law2016}). This object is a complicated case to study since it is in a very advanced state of tidal disruption, very far from dynamical equilibrium. This means that some of its properties (e.g., the Sgr core mass and its very own dynamical state) are still very uncertain. Remarkably, the model presented by Ref.~\cite{Wang_2022} reproduces most of the Sgr core properties, even showing how the stars and the DM are being tidally removed from the core. These authors conclude that most of the DM is completely stripped after two pericentric passages, and predict that the Sgr core will disappear within the next 2 Gyr. In addition, this stream also constitutes a special case since a population of GCs associated with its core was discovered \cite{Minniti2021a, Minniti2021b}. In particular, M54 is the GC primarily associated with the Sgr core (located at (l, b) = (6.01$^{\circ}$, -14.89$^{\circ}$), 25.0 kpc away from the Sun), as it lies just outside the 95\% radius of M54. Some gamma-ray studies of the Sgr/M54 region report a signal, e.g., \cite{Viana2012, Crocker_2022}, and, since no HI gas associated with its core has been detected \cite{Grcevich_2010}, the only sources capable of producing gamma rays with energies $\gtrsim$ 100 MeV are thought to be either millisecond pulsars (MSPs) or DM self-annihilation \cite{Evans_2023}. 
    Regarding this last possibility, the study recently presented in Ref.~\cite{venville2023prospective} shows that, should the observed gamma-ray emission in Sgr be originated from DM annihilation, this would imply inconsistencies with current DM constraints. Overall, with all these considerations in mind and in order to remain conservative, we decided to include Sgr in the \textit{silver} sample.\footnote{As it will be seen in Section~\ref{sec:constraints}, our DM modelling would make Sgr the dominant stream in terms of expected DM annihilation flux, thus its inclusion in our combined analysis would have a critical impact in our DM limits.}
    
\end{itemize}

\subsection{Other streams whose progenitors are dGs}
\label{sec:others}
We list below the streams that, despite their most likely progenitor being a dG, do not meet the other criteria we have established and therefore will not be included in our subsequent analysis.

\begin{itemize}

    \item \textbf{Jhelum}: Discovered using DES data by Ref.~\cite{Shipp_2018} and proposed as the remnant of a dG, this stream has a very complex structure with two different branches (Jhelum-a and Jhelum-b) \cite{Shipp_2019}, located at 13.0 kpc from the Sun,  and no consensus about its tracks. Authors in Ref.~\cite{Bonaca_2019jhelum} suggest that the two components have a common progenitor, yet there is still room for two different progenitors. Since this is still under debate and there is not information about the mass of each branch separately either, we have decided not to include Jhelum in our two samples of streams for analysis. The ongoing discussion regarding the number of progenitors have also prevented us from estimating their location.

    \item \textbf{Parallel}: Ref.~\cite{Sohn_2017} reports the discovery of the Parallel stream, claiming that its progenitor is likely a dG instead of a GC. In the absence of a known progenitor, it may be reasonable to assume that its location should coincide with the stream's mid-point, as in previous cases, i.e. (l, b) = (263.88$^{\circ}$, 61.26$^{\circ}$), which is at a distance of 14.3 kpc from the Sun. We discard the stream for the analysis, however, because there is not yet information about its stellar mass. 

    \item \textbf{C-19}: C-19 was initially reported as a stream whose progenitor would be the most metal-poor GC ever discovered \cite{Martin_2022}. However, Ref.~\cite{Errani_2022} proposed that C-19 is a DM-dominated stellar system similar in structure to dwarf spheroidal galaxies but with GC-like abundance patterns. Indeed, the origin of C-19 is remarkably intriguing. On the one hand, its kinematic properties suggest a dG origin more likely than a GC one, but its element abundances coincide with those found in GCs. Although a dG-like progenitor seems to be favored, a GC origin cannot be ruled out at the moment. As we do not know its progenitor, we could assume that it may be located at the mid-point of the stream, i.e., (l, b) = (102.29$^{\circ}$, -38.54$^{\circ}$), at a distance of 18.0 kpc to the Sun.
    In the end, we have decided to include it in Table \ref{table:table_sample2} as a potentially interesting object, but we exclude it from our final sample of streams in the analysis.

\end{itemize}

\section{Dark matter annihilation flux from stellar streams}
\label{sec:modeling}
In this work, we assume the DM to be composed of WIMPs. Their self-annihilation can give rise, among other products, to photons at gamma-ray energies that can be potentially detected with the \textit{Fermi}-LAT. 
The expected photon flux produced by WIMP annihilations can be computed as \cite{Bergstr_m_1998}: 
\begin{equation}
    \mathcal{F} (E>E_{th}, \Delta\Omega, l.o.s) = f_{pp} (E>E_{th}) \times J(\Delta\Omega, l.o.s) \hspace{0.5cm} \mathrm{(ph\, cm^{-2}\,s^{-1})},
    \label{eq:flux}
\end{equation}
where $E_{th}$ is the threshold energy (set by the instrument),  $f_{pp} (E)$ is the ``particle physics term'', and $J(\Delta\Omega, l.o.s)$ is the astrophysical J-factor computed within a given solid angle, $\Delta\Omega$, and along the line of sight (l.o.s.).

The particle physics term is energy dependent and encodes all the information about the underlying DM particle theoretical model. It is computed as:
\begin{equation}
    f_{pp} (E) = \frac{1}{4\pi}\frac{\langle \sigma v \rangle}{\delta \, m_{\chi}^{2}} \sum_{f} B_{f} \int_{E_{th}}^{E} \frac{dN_{f}}{dE} dE,
    \label{eq:particlefactor}
\end{equation}
where $\langle \sigma v \rangle$ is the thermally-averaged annihilation cross-section; $\delta = 2$ if we assume DM Majorana particles (DM particles are their own antiparticles) and $\delta = 4$ if we assume DM Dirac particles (DM particles are not their own antiparticles); $m_{\chi}$ is the DM particle mass; the subscript $f$ refers to the annihilation channel; $B_{f}$ is the branching ratio to the channel; $E_{th}$ is the threshold energy, while the upper limit is $E = m_{\chi}$;
and $\frac{dN_{f}}{dE}$ is the differential spectrum from the annihilation of WIMPs via the channel $f$. In this work, we consider DM Majorana particles, adopt $B_{f} =1$ for each channel, and the differential annihilation spectrum is computed using PPPC4DMID \cite{Marco_Cirelli_2011}. 

The astrophysical J-factor is independent of energy and has only a spatial dependence. It provides information about the spatial morphology of the DM signal and it is computed as follows:
\begin{equation}
    J(\Delta\Omega, l.o.s) = \int_{0}^{\Delta\Omega} d\Omega \int_{l.o.s.} \rho^{2}(r) dl,
    \label{eq:jfactor}
\end{equation}
being the DM density profile of the considered astrophysical object. Hence, in order to compute the J-factor we need to model the DM distribution, as we will do in the following subsection in detail.

\subsection{Stellar streams' DM modelling: general procedure} 
We explain the DM modelling performed for all the stellar streams in our sample. 
Despite the stretching process driven by tidal stripping, we assume that the streams maintain the same DM density distribution as their progenitors at least within the core ($r \leq r_s$, being $r_s$ a certain scale radius in the DM profile; see below), i.e., we expect the DM content to remain nearly unaltered within the progenitor core from infall time to present \cite{Coronado_Bl_zquez_2019_1,Coronado_Bl_zquez_2022,aguirresantaella2023viability}. The rest of the DM outside the core is assumed to be lost due to tidal stripping. Although reasonable, this DM modelling should be considered only as a first-order approximation for each stream. A more rigorous treatment of the evolution of the bound mass fraction, considering specific subhalo orbital parameters and infall masses and times, may lead to significant modifications of the internal structure even within the original scale radius, e.g., ~\cite{Pe_arrubia_2010, Errani_2019, Aguirre_Santaella_2022, aas_masc_o}. Such careful modelling of the DM density profile in streams is left for future work. 
With this in mind, and following results from DM-only cosmological simulations, which predict an inner cusp-like DM density profile at the center of dGs, we model the streams' core with a truncated form of the Navarro-Frenk-White (hereafter NFWt) DM density profile \cite{Navarro_1996, Navarro_1997}, i.e.: 

\begin{equation}
    \rho_{NFWt} (r) =
    \begin{cases}
         \frac{\rho_{0}}{\left(\frac{r}{r_s}\right)\left(1+\frac{r}{r_s}\right)^{2}},& \text{if } r \leq r_s\\
        0,  & \text{if } r > r_s
    \end{cases}
    ,
    \label{eq:NFW}
\end{equation}
where $r_s$ is the scale radius and $\rho_{0}$ is a characteristic DM density. For each stream, we compute both parameters and build its DM density profile following the procedure detailed below.

First, the NFWt scale radius of the subhalo can be computed as:
\begin{equation}
    r_{s} \equiv R_{200}/c_{200},
\label{eq:rs}
\end{equation}
being $R_{200}$ the virial radius of the subhalo at time of accretion, i.e. the radius within which the enclosed density is 200 times the critical density of the Universe:
\begin{equation}
    R_{200} = \left(\frac{3 M_{200}}{4\pi\Delta_{200}\rho_{crit}} \right)^{1/3},
\label{eq:R200}
\end{equation}
where $\Delta_{200}=200$ represents the overdensity with respect to the critical density, $\rho_{crit}= 137 \, M_{\odot} kpc^{-3}$ (for $H_{0} = 70 \, km s^{-1} Mpc^{-1}$). $M_{200}$ is the initial mass of the DM subhalo that hosts the stream and is defined as:
\begin{equation}
    M_{200} = \int_{0}^{R_{200}} \rho_{NFW}(r) r^{2}drd\Omega,
\label{eq:M200}
\end{equation}
where we use the NFW profile instead of NFWt, since the subhalos have not been yet tidally stripped at time of accretion and, thus, their DM profile is not truncated.

The other parameter involved in Eq. \ref{eq:rs}, $c_{200}$, is the subhalo concentration. In particular, we will use the median subhalo concentration-mass relation derived in Ref.~\cite{moline2017}:
\begin{equation}
    c_{200} (M_{200}, x_{sub}) = c_{0} \left[ 1+\sum_{i=1}^{3} \left[a_{i}\log\left(\frac{M_{200}}{10^8 \, h^{-1} \, M_{\odot}} \right) \right]^{i} \right] \times [1+b\log(x_{sub})],
\label{eq:c200}
\end{equation}
with $c_{0}=19.9$, $a_{i}= (-0.195, 0.089, 0.089)$ and $b = -0.54$. $x_{sub}$ is defined as $x_{sub}=R_{sub}/R_{MW}$, being $R_{sub}$ the distance of the subhalo to the host halo center and $R_{MW}$ the virial radius of the host, which we assume is $R_{MW}=220$ kpc for the Milky Way. 

As for $\rho_{0}$, we make use of Eq. \ref{eq:M200} and perform the integral to obtain:
\begin{equation}
    \rho_{0} = \frac{2\Delta_{200}\rho_{crit}c_{200}}{3 f(c_{200})},
\label{eq:ro0}
\end{equation}
with:
\begin{equation}
    f(c_{200}) = \frac{2}{c_{200}^{2}}\left(\ln(1+c_{200}) - \frac{c_{200}}{1+c_{200}} \right) .
\label{eq:fc200}
\end{equation}

Finally, we can compute an angular extension in the sky for each DM subhalo hosting a stream, which will be particularly useful for subsequent sections. We adopt as such extension the one given by the angle subtended by $r_{s}$: 
\begin{equation}
    \theta_{s} = \arctan\left(\frac{r_{s}}{d_{Earth}}\right) ,
\label{eq:theta}
\end{equation}
where $d_{Earth}$ refers to the distance from the streams' core to the Earth. Note that $\theta_s$ will depend on the considered mass, as the scale radius itself varies with mass (Eq. \ref{eq:rs}).
This definition of DM subhalo extension is motivated by the fact that, for DM halos well represented by an NFW profile, 90\% of their annihilation flux is expected to come from the region within the scale radius $r_{s}$~\cite{Sanchez_Conde_2011}. This still holds in our case, since our subhalos are modeled by a truncated NFW profile and thus they are NFW-like exactly up to $r_{s}$.

\subsection{Considered mass-to-light ratio scenarios}
In order to build the DM density profile of the DM subhalos hosting the streams, we need to know the DM mass.  Since the value of $M_{200}$ is uncertain in most of these objects, we explore three scenarios.
We start from existing estimates of the stream's baryonic mass, and adopt three different mass-to-light ratios (M/L) to estimate the DM mass of the system at present time. We recall that typical M/L ratios for dGs -- progenitors of our selected streams -- are $10-1000$ \cite{Mateo_1998,Sanchez_Conde_2011,Ackermann_2015}, yet we may expect lower M/L ratios for streams because of additional loss of DM suffered during the extreme stretching process driven by ongoing tidal interactions with the host. With this in mind, and also with the intention to stay conservative given the large involved uncertainties, in the following we propose three different M/L scenarios from which we will draw $M_{200}$ estimates:

\begin{itemize}
    \item \textbf{Low}: In our most conservative approach, we assume that the DM mass is the same as the baryonic mass. This scenario corresponds to $M/L = 2$. 

    \item \textbf{Benchmark}: We consider $M/L = 5$ as an intermediate case. This scenario will represent our benchmark one, since we believe it represents a very reasonable choice for these objects (probably still on the conservative side).

    \item \textbf{High}: We also explore an upper value of $M/L = 50$. Although being the most optimistic case considered in our work, this M/L value could still be conservative in many cases taking into account the high M/L values reached by some dGs. Notwithstanding the above, high-resolution cosmological simulations particularly focused on studying subhalo tidal stripping (e.g. \cite{Aguirre_Santaella_2022}) indicate that some subhalos may have lost up to 99$\%$ of their initial masses at present time, making this scenario potentially overly optimistic in certain cases.
    
\end{itemize}

For each of the above M/L scenarios and for all the stellar streams in our sample, we apply the DM modelling described in Sec.~\ref{sec:modeling}, this way ending up with three different $M_{200}$ estimates for each stream at present time. We show the results for our \textit{golden} sample (Tab.~\ref{table:table_sample2}) and the Benchmark M/L case in Table \ref{table:table_modeling}. As it can be seen, the $M_{200}$ values found are $\mathcal{O}(10^4) - \mathcal{O}(10^6) \, M_{\odot}$ , whereas the $r_s$ values are $\mathcal{O}(10) \, pc$. 
However, $M_{200}$ is the ``original'' mass of each stream's core at accretion time, but it was removed long time ago due to tidal stripping. Therefore, it is only used to construct the DM profile, but the actual remaining mass will be the one enclosed within $r_s$, i.e., $M_s$. We also include the values of $M_s$ in Table \ref{table:table_modeling}.
The results obtained for the \textit{golden} sample in the Low and High M/L cases are left for Appendix \ref{sec:appendix_golden}. We also present the results for the \textit{silver} sample and the three considered M/L scenarios in Appendix \ref{sec:appendix_silver}.

Note that, in the literature available to date, we do not find the stellar mass for all the streams in our sample. For example, in the case of LMS-1, Ref.~\cite{Malhan_2021} provides both the stellar and the DM mass of its progenitor instead of those of the current stream. In these cases, we will consider the stellar mass of the progenitor to be the stellar mass of the stream, this way assuming that during the stretching process the stream loses a significant fraction of its DM mass, while its baryon matter content remains the same\footnote{According to the standard $\Lambda$CDM cosmological model, baryons are concentrated in their innermost regions of halos and subhalos. As a result, tidal forces first strip away the outer, DM-dominated regions, whereas the inner regions, where baryons are more tightly bound, remain intact for longer before being significantly affected.}.
For the PS1-D stream, Ref.~\cite{Bernard_2016} reported its luminosity instead of its stellar mass, thus in this case we will assume that the luminosity corresponds to the stellar mass. For Styx, Ref.~\cite{carlin2009kinematics} provides its total mass as well as its luminosity. Again, we take its luminosity as its stellar mass.
Finally, in the case of Orphan-Chenab, authors in Ref.~\cite{Mendelsohn_2022} estimate the baryonic and the DM mass of its progenitor within 300 pc of its center. As in the case of LMS-1, we decided to take the progenitor's baryonic mass as the stream's stellar mass.

\begin{table}[h!]
\begin{adjustwidth}{-0.9cm}{}
\begin{center}    
    \begin{tabular}{|c|c|c|c|c|c|c|}

         \hline
         Stream & $\frac{M_{200}}{10^4} \, (M_{\odot})$ & $R_{200}$ (kpc) & $c_{200}$ & $\frac{\rho_{0}}{10^8}\, (\frac{M_{\odot}}{kpc^3})$ & $\frac{M_{s}}{10^4} \, (M_{\odot})$ & $r_{s}$ (pc) \\
         \hline
         Indus & 17.00 & 1.14 & 58.2 & 5.78 & 1.12 & 20.0 \\
         LMS-1 & 50.00 & 1.63 & 58.2 & 5.82 & 3.10 &28.0 \\
         Orphan-Chenab & 80.00 & 1.91 & 58.4 & 5.86 & 4.97 & 32.7 \\
         PS1-D & 3.75 & 0.69 & 53.8 & 4.71 & 0.24 & 12.8 \\
         Turranburra & 3.80 & 0.69 & 50.6 & 4.00 & 0.27 & 14.0 \\
         Cetus-Palca & 750.00 & 4.03 & 53.8 & 4.70 & 47.93 & 74.9 \\
         Styx & 9.00 & 0.92 & 47.8 & 3.43 & 0.60 & 19.3 \\
         Elqui & 5.20 & 0.77 & 47.8 & 3.40 & 0.66 & 20.0 \\
         \hline
    \end{tabular}
    \end{center}
   \caption{DM modelling parameters for our \textit{golden} sample of stellar streams (Tab.~\ref{table:table_sample2}) and our Benckmark M/L scenario, computed following the procedure described in Section~\ref{sec:modeling} and Eqs. (\ref{eq:rs}-\ref{eq:fc200}). The values of $M_s$ are also given. 
   }
    \label{table:table_modeling}
    \end{adjustwidth}
\end{table}

\subsection{Astrophysical J-factors and angular sizes}
Once we have implemented the DM modelling for the stellar streams, we can compute the astrophysical J-factor associated with each stream in our sample. Integrating the general expression of the J-factor (Eq. \ref{eq:jfactor}) and adopting an NFW DM density profile we can obtain the J-factor for any given radius:
\begin{equation}
    J (r) = \frac{4\pi \rho_{0}^{2}r_{s}^{3}}{3D_{Earth}^{2}} \left(1-\frac{1}{1+\left(\frac{r}{r_s} \right)^3} \right),
    \label{eq:jrfactor}
\end{equation}
where $D_{Earth}$ is the distance to the target. 

\begin{figure}[h]
\begin{center}
		\begin{subfigure}{0.49\textwidth}
            \includegraphics[width=\textwidth]{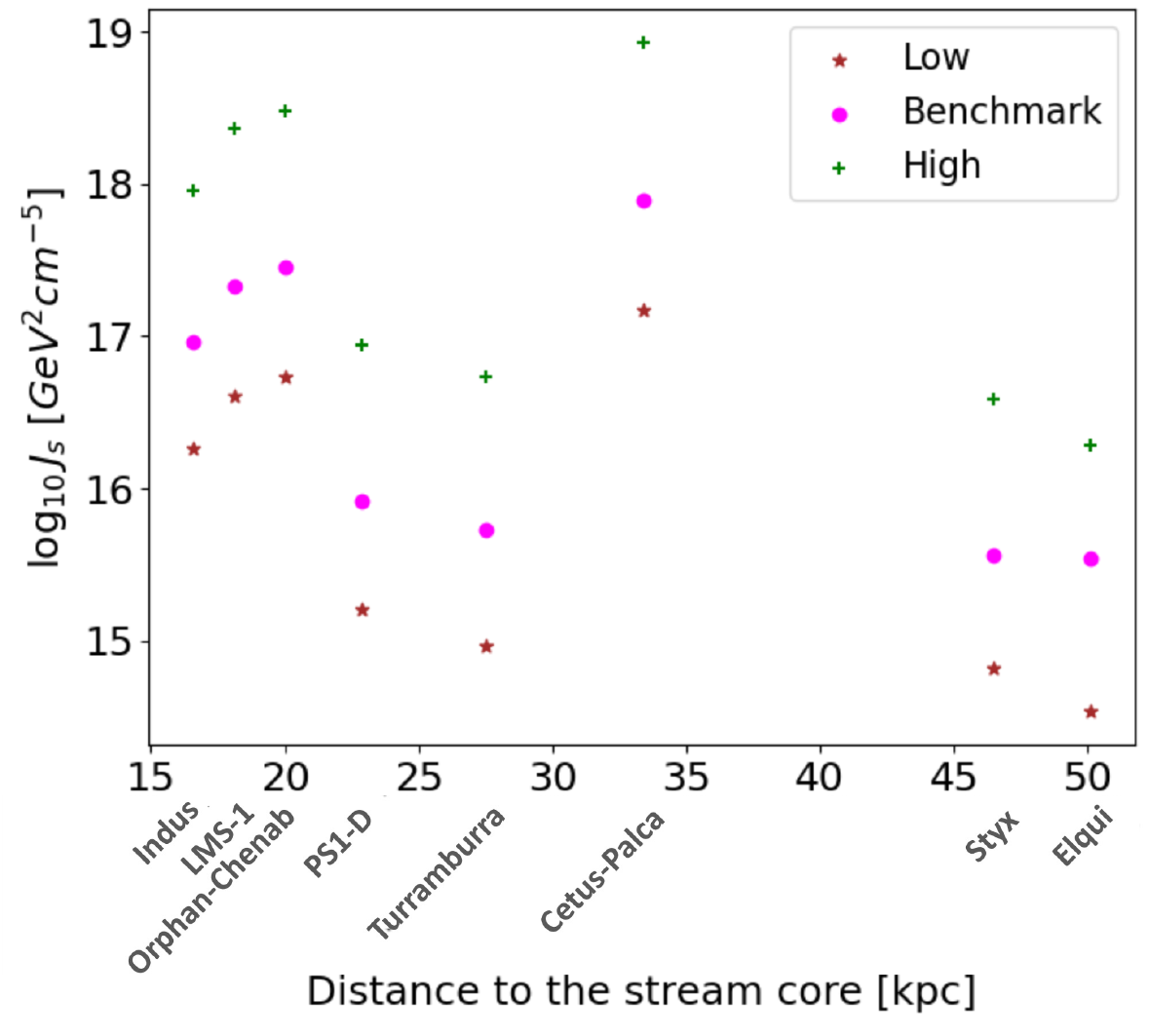}
			
		\end{subfigure}
		\hspace{0.02cm}
		  \begin{subfigure}{0.49\textwidth}
		      	\includegraphics[width=\textwidth]{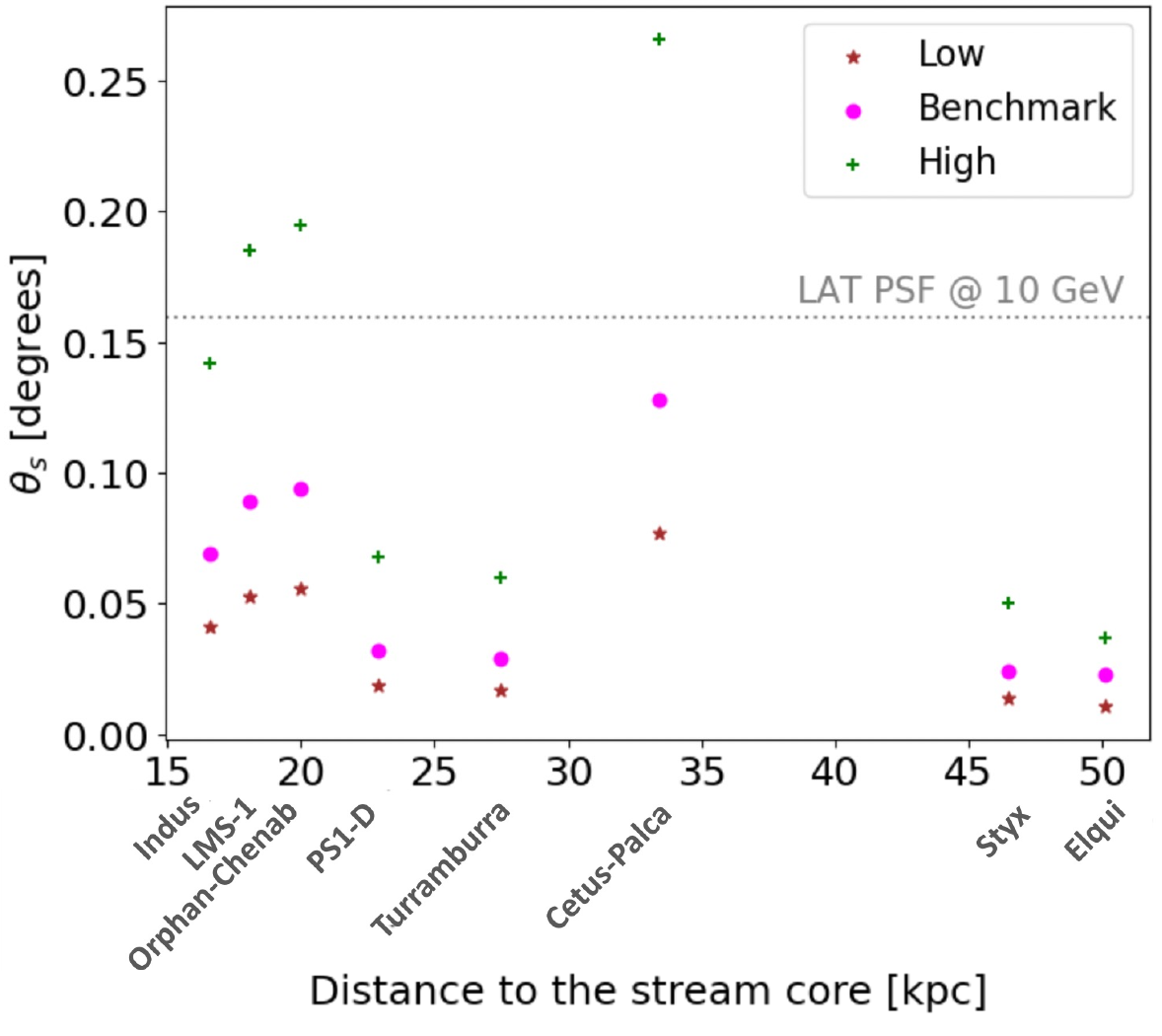}
			
		  \end{subfigure}
        \caption{$J_s$ values (left panel) and $\theta_{s}$ values (right panel) versus the distance from the streams' core to Earth, for all streams in our \textit{golden} sample and the three M/L considered scenarios. For comparison, the PSF of the \textit{Fermi} LAT at 10 GeV is also depicted as a horizontal dotted line.}
        
		\label{fig:js_theta}
		
\end{center}
\end{figure}

We summarize in Table \ref{table:table_jfactor} the astrophysical J-factor up to the scale radius $r_s$ ($J_s$) computed for each stream in our \textit{golden} sample using the DM density profile parameters obtained in Section~\ref{sec:modeling} and summarized in Table \ref{table:table_modeling}. The obtained results for the \textit{silver} sample are presented in Appendix \ref{sec:appendix_silver}. Three J-factor values are provided, each of them corresponding to each of the three $M_{200}$ values calculated for each stream (Low, Benchmark, High). 
$J_s$ values are also depicted in the left panel of Figure \ref{fig:js_theta} for all the streams in the \textit{golden} sample, for the three different M/L scenarios considered in this work. For our main case of interest, the Benchmark M/L scenario, we find values for $J_s$ between $\log_{10}J_s = 15.54$ for the Elqui stream, and $\log_{10}J_s=17.89$ for the Cetus-Palca stream. The latter presents the highest $J_s$ of the \textit{golden} sample, which in turn will dominate the limits on the DM annihilation cross-section when doing the combined analysis of all the streams of this sample, as we will see in Section \ref{sec:constraints}. We could compare these values with the ones obtained for dwarf spheroidal galaxies, which provide the most robust DM constraints so far and are found to be between $\log_{10}J = 16.40 - 19.70$ \cite{mcdaniel2023legacy}. While the J-factors for stellar streams are very competitive when compared to the dSphs ones, the latter are still the most stringent when considering the whole samples. Yet, when we refer to other targets widely used for DM searches (such as galaxy clusters \cite{Di_Mauro_2023}, dwarf irregular galaxies \cite{Gammaldi_2021}, or dark satellites \cite{Coronado_Bl_zquez_2022}), we find comparable or higher individual J-factor values. We discuss this in more detail in Section \ref{sec:concl}, where we compare our DM constraints with those obtained for other targets.

\begin{table}[h!]
    \begin{center}
        
    \begin{tabular}{| c | c | c | c | c | c | c |}
        \hline
        Stream & \multicolumn{3}{c|}{$\theta_{s} \, (^{\circ})$} & 
        \multicolumn{3}{c|}{$\log_{10}J_s \, (GeV^{2} cm^{-5})$} 
        \\
        \cline{2-7}
        & \textit{Low} & \textit{Bench.} & \textit{High} & 
        \textit{Low} & \textit{Bench.} &\textit{High}
        \\
        \hline
        Indus & 0.04 & 0.07 & 0.14 & 16.26 & 16.96 & 17.96\\
        LMS-1 & 0.05 & 0.09 & 0.19 & 16.61 & 17.33 & 18.36 \\
        Orphan-Chenab & 0.06 & 0.09 & 0.20 & 16.73 & 17.45 & 18.48\\
        PS1-D & 0.02 & 0.03 & 0.07 & 15.21 & 15.92 & 16.94 \\
        Turranburra & 0.02 & 0.03 & 0.06 & 14.97 & 15.73 & 16.73\\
        Cetus-Palca & 0.08 & 0.13 & 0.27 & 17.17 & 17.89 & 18.93\\
        Styx & 0.01 & 0.02 & 0.05 & 14.82 & 15.56 & 16.59\\
        Elqui & 0.01 & 0.02 & 0.04 & 14.54 & 15.54 & 16.28\\
        \hline
        
    \end{tabular} 
    \end{center} 
    \caption{Values of the angular extension of the streams' core ($\theta_{s}$), and the astrophysical J-factors integrated up to $r_{s}$ ($J_s$) for each stellar stream in the \textit{golden} sample and under the three considered M/L scenarios (Low, Benchmark, High).}
    \label{table:table_jfactor}
\end{table}

Additionally, we include in Table \ref{table:table_jfactor} the results obtained via equation \ref{eq:theta} for $\theta_{s}$, i.e., the angular extension of the streams' core, this one defined as the region within $r_{s}$. We also show these results in the right panel of Figure \ref{fig:js_theta}.  
In the same figure, we compare such $\theta_{s}$ values to the point spread function (PSF) of the \textit{Fermi} LAT at 10 GeV. From this comparison we can conclude that, if detected, the core of the streams in our sample would appear as point-like sources for \textit{Fermi} LAT in almost all considered scenarios, indeed typical $\theta_{s}$ values being well below 0.15 degrees in the Benchmark M/L case. As we will see later on in the next section, this has important implications in order to define the best DM search analysis strategy, as it justifies a point source analysis of each stream's core against an extended, more complicated one.

\section{Data analysis}
\label{sec:analysis}
In this section, we present the analysis of LAT data from the direction of our sample of streams. We explain our data selection, the analysis pipeline, as well as the main results.
We perform the data analysis with the \texttt{Fermipy} Python package (version 1.2) \cite{wood2017fermipyopensourcepythonpackage}, which uses the underlying \texttt{Fermitools} (version 2.2.0).

\subsection{Technical setup}
\label{subsec:setup}
We use almost 15 years (Mission Elapsed Time (MET): from 239557417 to 702032312; Gregorian: from 2008-08-04 to 2023-04-01) of data collected by the \textit{Fermi}-LAT, with an energy range of 500 MeV - 500 GeV.  We take all available photons (\texttt{FRONT + BACK}), excluding those arriving with zenith angles greater than 90$^\circ$ in order to avoid the contamination from the Earth limb, and apply the \texttt{$(\text{DATA}\_\text{QUAL}>0)\&\&(\text{LAT}\_\text{CONFIG}==1)$} filter to ensure data quality. Moreover, we choose a region of interest (ROI) of 15$^\circ$ x 15$^\circ$ around the source of interest, with an angular bin size of 0.01$^\circ$  and 8 evenly spaced logarithmic energy bins. We utilize the \texttt{P8R3\_SOURCE\_V3} instrumental response functions (IRFs), along with the Galactic diffuse emission model \texttt{gll\_iem\_v07.fits}, the isotropic diffuse model \texttt{iso\_P8R3\_SOURCE\_V3\_v1.txt}, and the most recent point-source catalog, the \textit{Fermi} Large Area Telescope Fourth Source Catalog Data Release 4 (\texttt{4FGL-DR4}) \cite{ballet2023fermi}. A summary of the binned analysis setup is shown in Table \ref{table:table_setup}.

\begin{table}[h!]
\begin{center}
\begin{tabular}{ | c | c | }
\hline Time domain (Gregorian) & 2008-08-04 to 2023-04-01 \\ \hline
Time domain (MET) & 239557417 to 702032312 \\ \hline Energy range & 500 MeV - 500 GeV \\ \hline IRF & P8R3\_SOURCE\_V3 \\
\hline Event type & FRONT + BACK\\
\hline Point-source catalog & 4FGL-DR4\\
\hline ROI size & 15$^\circ$ x 15$^\circ$\\
\hline Angular bin size & 0.01$^\circ$ \\
\hline Bins per energy decade & 8 \\
\hline Galactic diffuse model & gll\_iem\_v07.fits\\
\hline Isotropic diffuse model & iso\_P8R3\_SOURCE\_V3\_v1.txt\\
\hline

\end{tabular}
\end{center}
\caption{Summary of the \textit{Fermi}-LAT analysis technical setup.}
    \label{table:table_setup}
\end{table}

\subsection{General methodology}
\label{subsec:metho}
We perform a point-source analysis of the ROI centered in the core of each stream (see Table \ref{table:table_sample2} for the location considered in each case), both for the \textit{golden} and \textit{silver} samples, with the aforementioned technical setup. 

The analysis can be briefly described as follows. First of all, we run \texttt{GTAnalysis} and define the full analysis setup with \texttt{GTAnalysis.setup}. Then, we run the \texttt{GTAnalysis.optimize} routine in order to perform an automatic optimization of the ROI by fitting all sources. This way, we ensure that all parameters are close to their global likelihood maxima. Next, we free the normalization and spectral shape of 4FGL-DR4 catalog sources within 3 degrees of the ROI center using the \texttt{GTAnalysis.free\_sources} module. We also free the normalization and spectral index of the Galactic diffuse component, as well as the normalization of the isotropic diffuse template. After freeing all these parameters, we need to optimize the model again by calling \texttt{GTAnalysis.fit}.

Then, after freeing the sources and optimizing the model, we look for new sources above $\sim 5 \sigma$ (being $\sigma$ the significance expressed in standard deviations) -- since this is the usual threshold above which sources are included in \textit{Fermi}-LAT catalogs -- with the \allowbreak{\texttt{GTAnalysis.find\_sources}} module. To compute the detection significance, we define the Test Statistic (TS) as:
\begin{equation}
    TS = 2 \, \log \left(\frac{\mathcal{L} (H_1)}{\mathcal{L} (H_0)} \right),
    \label{eq:ts}
\end{equation}
where $\mathcal{L} (H_0)$ represents the likelihood under the null (no  source) hypothesis and $\mathcal{L} (H_1)$ represents
the likelihood of the alternative (existing source).  
In an ideal scenario where all the model components were perfectly known, the TS values for a gamma-ray analysis should follow a Poissonian distribution. In such idealized situation, the Chernoff theorem \cite{chernoff} holds and the TS distribution is well described by a $\chi^2$ distribution for \textit{n} degrees of freedom divided by two. This, in turn, allows for the computation of the detection significance simply as $\sqrt{TS}$ (i.e., $\sqrt{TS} \sim \sigma$).
  
Should we find a new unidentified (unID) source at a distance of less than 3 degrees from the center of our ROI, we perform a dedicated detailed analysis to it, starting by recalculating its sky position with the \texttt{GTAnalysis.localize} module, and following with both a spectral and spatial analysis. The reason a specific analysis of potential new sources in the vicinity of the ROI center is required is that one of the dominant uncertainties is associated with the location of the streams' core. However, as we are using almost the same amount of data as the one used for the 4FGL-DR4 catalogue, we do not expect a significant number of new sources. After the search, in most cases we find at least a new source in the ROI, but none of them is located in the innermost 3 degrees, so we will not give them relevance in this study.\footnote{The exception is Sagittarius, for which we find a new source at 2.5 degrees from the center of the ROI, with a low TS value $\sim33$. The analysis of this source did not provide any relevant results in the context of our DM search.}

Once we have built the best-fit model for the ROI, and if no source is found at the stream core (i.e., the center of the ROI), we obtain 95\% confidence level (C.L.) upper limits to the flux, for which we use the \texttt{GTAnalysis.sed} module. We do so by placing a putative DM source in the ROI center, whose spectrum is modelled with the \texttt{DMFitFunction} \cite{Jeltema_2008} routine.\footnote{The inclusion of more refined and updated DM spectra such as those given by CosmiXs \cite{arina2023cosmixscosmicmessengerspectra} would not be relevant in this context, as no source is really detected.}

\subsection{Analysis results}
\label{sec:results}
No gamma-ray emission is detected from any of the stream cores.  
Therefore, in the next Section \ref{sec:constraints} we proceed and use our analysis results, namely the 95\% C.L. upper limits to the flux, to set constraints on the DM annihilation cross-section. Here, we show as an illustrative example of our analysis the main results obtained for LMS-1, one of the streams in the \textit{golden} sample. We present in Appendix \ref{sec:appendix_fermipy} the main results obtained for the rest of the streams considered in this work.

\begin{figure}[h!]
    \centering
    \includegraphics[width=1.0\textwidth]{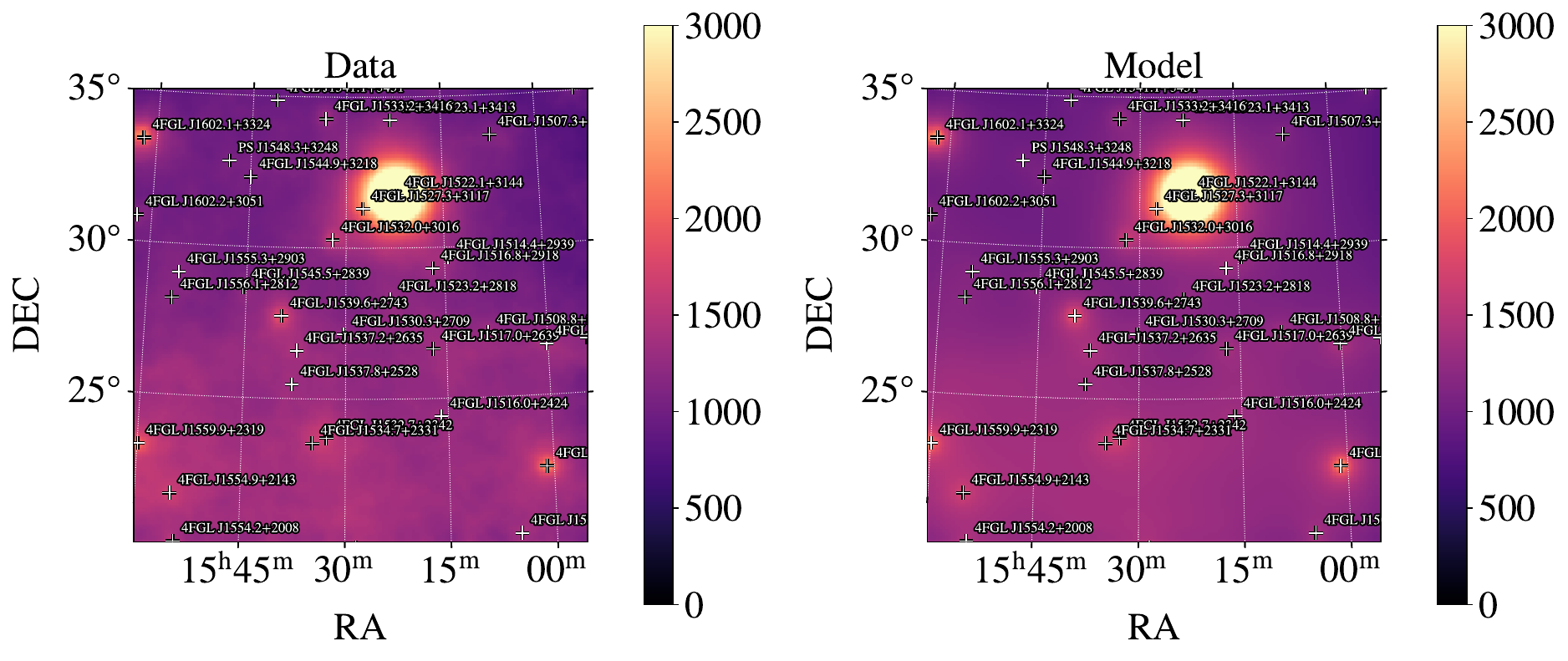}
    \caption{Smoothed data map (left) and smoothed model map (right) for a 15$^\circ$ x 15$^\circ$ ROI centered at the core of LMS-1 stellar stream. See text for details.}
    \label{fig:datamodel}
\end{figure}

\begin{figure}[h!]
    \centering
    \includegraphics[width=0.5\textwidth]{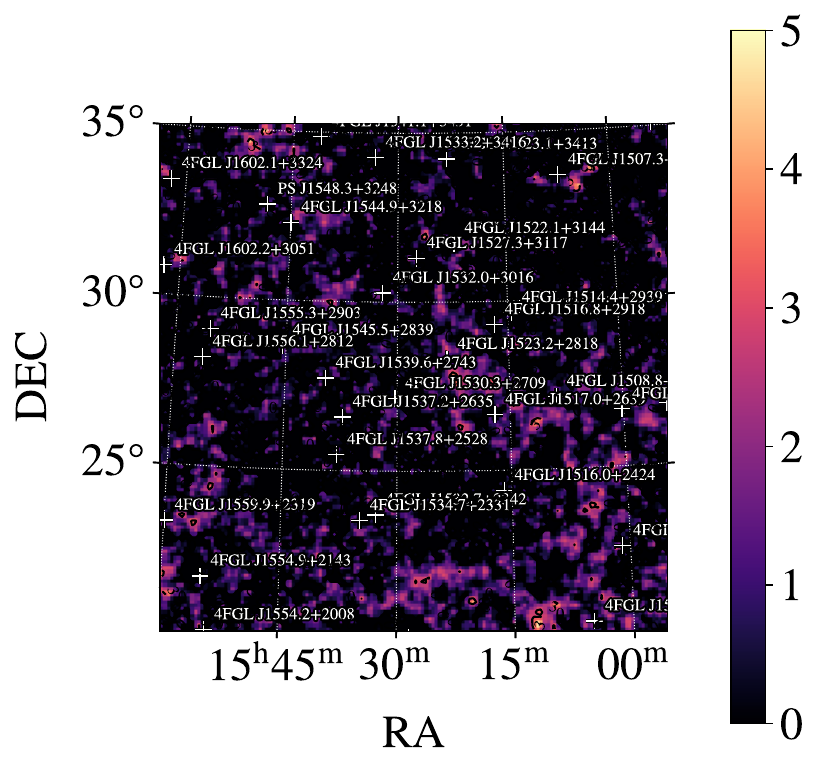}
    \caption{Significance map (i.e., $\sqrt{TS}$) for the 15$^\circ$ x 15$^\circ$ ROI centered at the core of LMS-1 stellar stream.}
    \label{fig:tsmap}
\end{figure}

\begin{figure}[h!]
    \centering
    \includegraphics[width=0.75\textwidth]{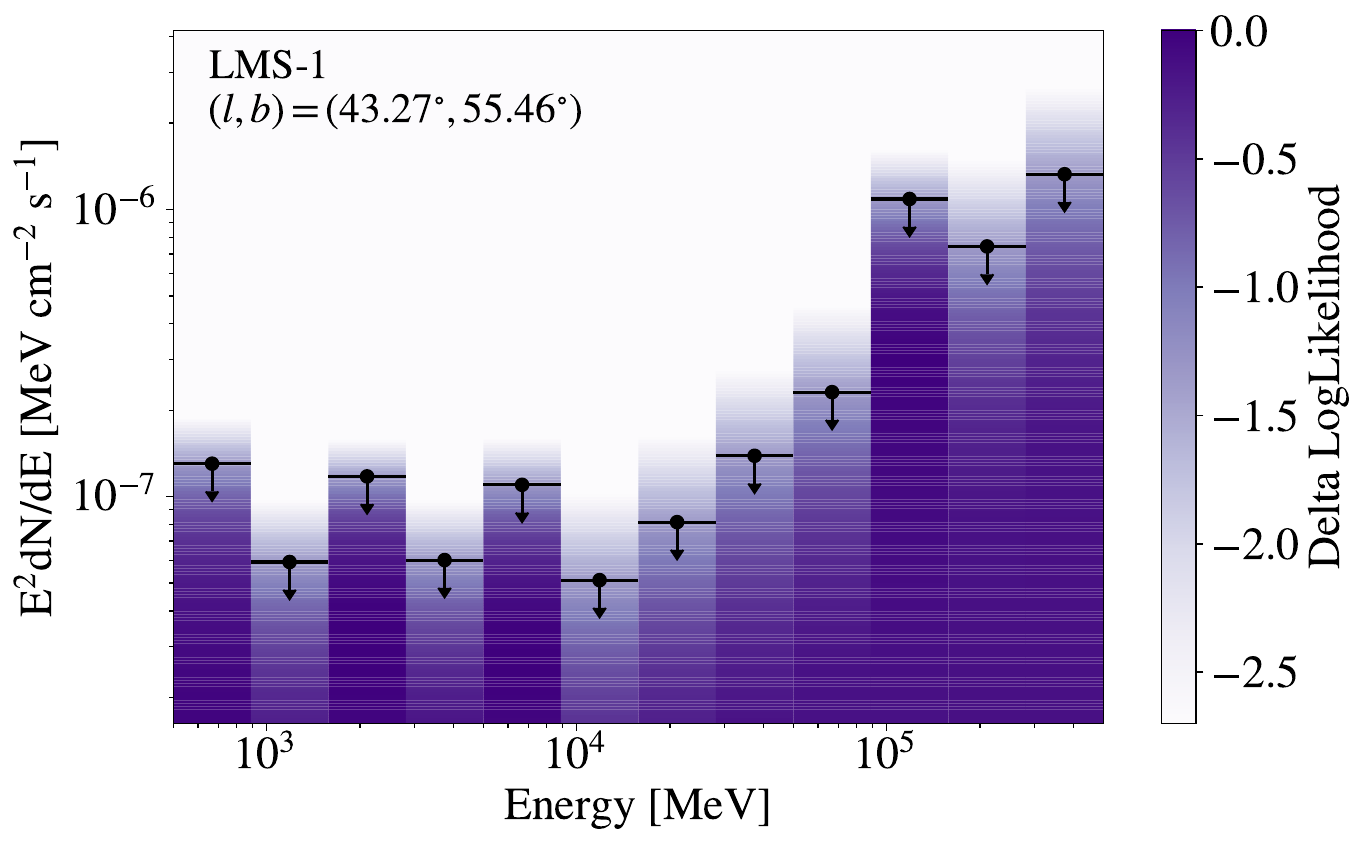}
    \caption{Spectral analysis of the core of the LMS-1 stellar stream, showing 95\% C.L. upper limits to the flux and delta likelihood values.}
    \label{fig:sed}
\end{figure}

Figure \ref{fig:datamodel} shows the skymaps of the 15$^\circ$ x 15$^\circ$ region around the core of LMS-1, obtained using the \texttt{GTAnalysis.residmap} tool, after having performed the fitting procedure outlined above, and after having looked for new sources. 
Figure \ref{fig:tsmap} depicts the TS map of this same region, obtained by means of the \texttt{GTAnalysis.tsmap} routine. We derive very similar results for all the streams in our sample. In all cases, we do not find significant excesses across the ROI. In particular, no source at the position of the stream's core is found. In the absence of a signal we compute 95\% C.L. upper limits on the flux using the \texttt{GTAnalysis.sed} module, by placing in the ROI center a putative DM source\footnote{In particular, we introduce a putative DM source at the ROI center annihilating to $b\bar{b}$ and with a mass of 100 GeV. Note that choosing another value for the mass will not alter the result.} modelled with the function \texttt{DMFitFunction}. The resulting SED with 4 bins per decade showing these upper limits appears in Figure \ref{fig:sed}.


Notwithstanding the above, Sagittarius constitutes the exception. Consistent with what has been reported in the literature (e.g. \cite{Viana2012, Crocker_2022}), we found emission just $\sim$0.2 degrees from the Sgr core coming from the source 4FGL J1855.1-3025, which is associated with the globular cluster M54 in the 4FGL-DR4 catalog \cite{ballet2023fermi}. There is currently discussion about whether the emission from this source may be due to the presence of millisecond pulsars inside M54 or to DM annihilation at the core of Sgr (see for instance Ref. \cite{Evans_2023}). However, the recent study presented in Ref. \cite{venville2023prospective} seems to disfavor the scenario in which the emission is due to DM annihilation, since this would imply a cross section incompatible with existing constraints. We leave for the future a detailed analysis of this region in which to investigate whether this emission could be due to DM annihilation associated to the Sgr core in some way. Instead, we proceed as in the rest of the streams and introduce a putative DM source at the exact position of the Sgr core, i.e., (l, b) = (6.01$^{\circ}$, -14.89$^{\circ}$). As in the rest of cases, we find no emission and obtain the corresponding 95$\%$ C.L. upper limits on the flux.

\section{Dark matter constraints}
\label{sec:constraints}

\subsection{Method to set the limits}
\label{subsec:method_limits}
In the absence of a gamma-ray signal from the direction of the stream cores, we place constraints on the DM particle properties. More precisely, we set constraints on the DM annihilation cross-section, adopting the $b\bar{b}$ and $\tau^{+}\tau^{-}$ annihilation channels, since they are representative of the hadronic and leptonic channels, respectively, see e.g.~\cite{Charles_2016}. 
We adopt a grid of 25 DM mass values between 10 GeV up and $10^4$ GeV. We compute the individual log-likelihood profile for each of the streams as a function of DM mass and annihilation cross-section (a couple of examples of such log-likelihood profiles are provided in Appendix~\ref{sec:appendix:lhl}). The likelihood for a given DM mass and cross-section is computed by introducing the expected DM-induced gamma-ray flux given by Eq. \ref{eq:flux} in the likelihood profile in flux-energy space, i.e., $\mathcal{L}(\mathcal{F}, E)$, and summing over all energy bins as: 
\begin{equation}
    \mathcal{L}(\langle \sigma v \rangle, m_{\chi})=\sum_{E_{j}}\mathcal{L}\left(\mathcal{F}(\langle \sigma v \rangle, m_{\chi}, E_{j}), E_{j}  \right).
\end{equation}

Then, the 95$\%$ C.L. upper limit on the cross-section (for each DM mass and annihilation channel) is set as the value at which the profile crosses $\Delta\mathcal{L}=2.71/2$ (see Appendix~\ref{sec:appendix:lhl} for details).

Once we have obtained individual upper limits on the cross-section for each stream, we combine the results by summing up together their individual log-likelihood profiles independently for each energy bin, to obtain a global likelihood:
\begin{equation}
    \log(\mathcal{L}_{j}(\mu , \theta_{j} | \mathcal{D}_{j})) = \sum_{i} \log(\mathcal{L}_{i,j}(\mu , \theta_{i,j} | \mathcal{D}_{i,j})),
    \label{eq:comb_lhl}
\end{equation}
being $\log(\mathcal{L}_{j}(\mu , \theta_{j} | \mathcal{D}_{j}))$ the combined likelihood for a particular DM annihilation channel as a function of the DM mass and annihilation cross-section for all targets. The index $i$ runs over the targets list, whereas $j$ is the index of each energy bin of the \textit{Fermi}-LAT data ($\mathcal{D}$). $\mu$ represents the DM parameters ($\langle \sigma v \rangle$ and $m_{\chi}$), and $\theta$ corresponds to the parameters in the background model, i.e., the nuisance parameters. 

It is common to include the systematic uncertainty on the J-factor (see, e.g., Ref.~\cite{Ackermann_2015}). However, when compared to the overall systematic uncertainty arising from the use of different M/L scenarios for each stream (i.e., Low, Benchmark, High; see Section~\ref{sec:modeling}), typically with a difference of $1 - 2$ orders of magnitude, the uncertainty on individual J-factor values is expected to be subdominant. Also, as we will focus the discussion on the Benchmark case, we are confident that the corresponding results will always be contained within the M/L uncertainty band, 
the uncertainty on individual J-factors not significantly altering our conclusions. 

In the following, we show combined results for DM masses in the range from 10 to $10^4$ GeV, and for two annihilation channels: $b\bar{b}$ and $\tau^{+}\tau^{-}$, both for the \textit{golden} and \textit{silver} samples.

\subsection{Individual and combined DM limits}

\begin{figure}[h!]
\begin{center}
        \captionsetup[subfigure]{labelformat=empty}
		\begin{subfigure}{0.495\textwidth}
			\includegraphics[width=\textwidth]{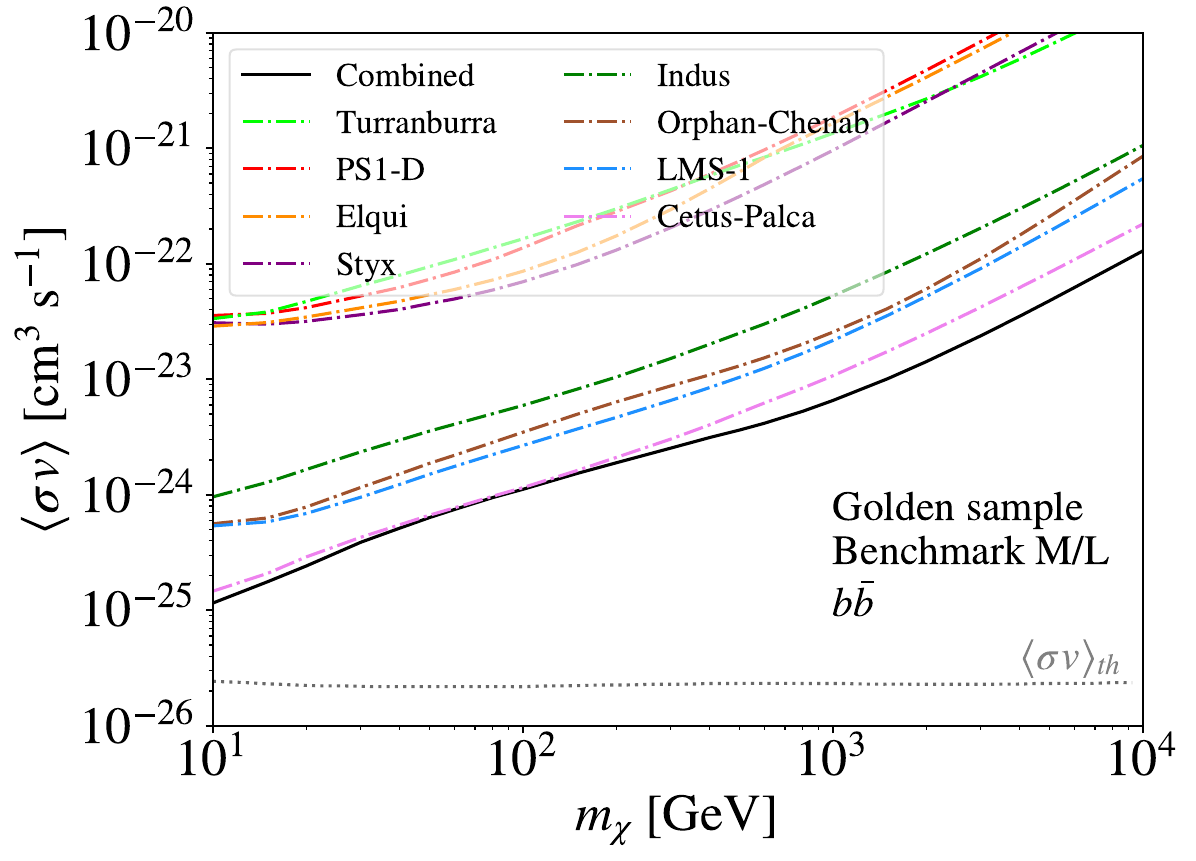}
			\caption{}
			\label{fig:combined_indviduales_int_golden_bb_new}
		\end{subfigure}
		  \begin{subfigure}{0.495\textwidth}
		      	\includegraphics[width=\textwidth]{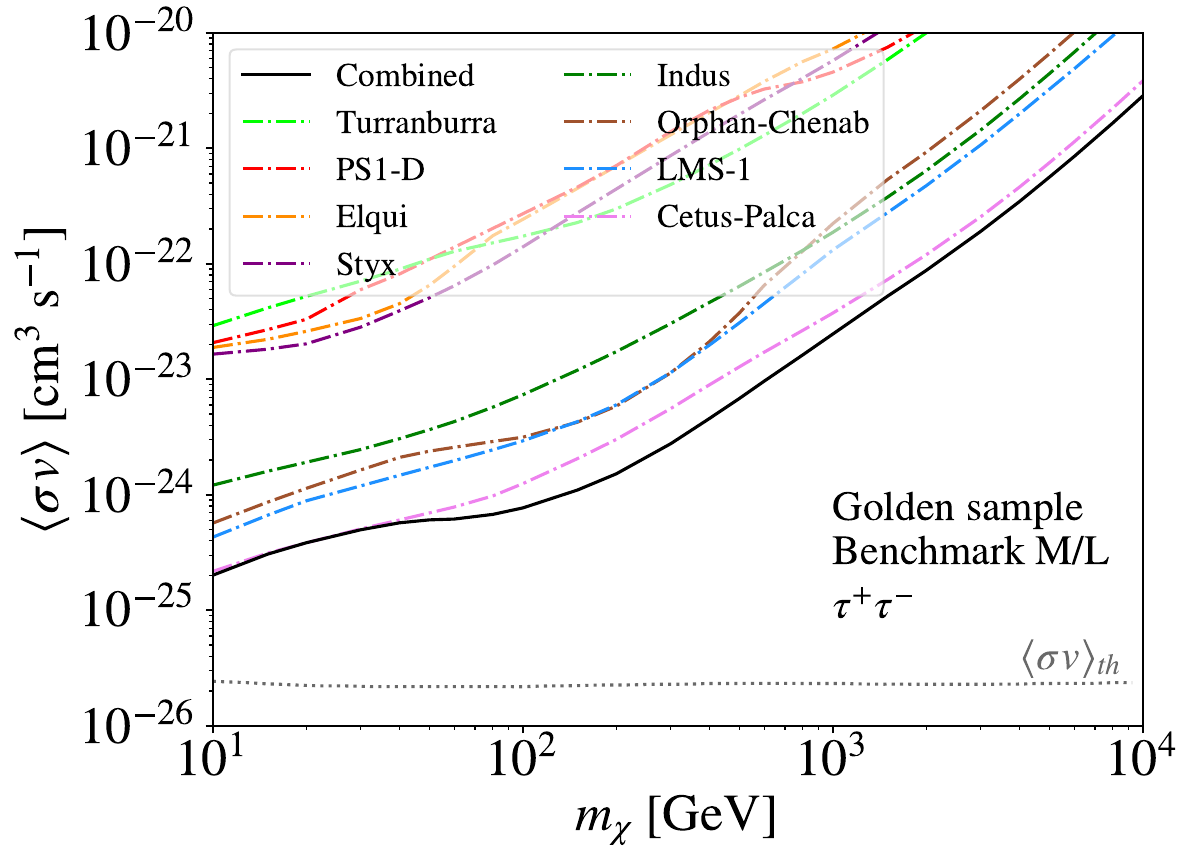}
		        \caption{}
	            \label{fig:combined_indviduales_int_golden_taus}
		  \end{subfigure}
        \caption{Constraints on the DM annihilation cross section for each individual stream in the \textit{golden} sample (Table \ref{table:table_sample2}) and the Benchmark M/L scenario described in Section \ref{sec:modeling}. The combined limit is also depicted as a thick solid line. Left and right panels refer to the $b\bar{b}$ and $\tau^{+}\tau^{-}$ annihilation channels, respectively. The gray dotted line represents the thermal relic cross-section \cite{Steigman_2012}.}
		\label{fig:combined_indviduales_int_golden}
\end{center}
\end{figure}

Figure \ref{fig:combined_indviduales_int_golden} shows both individual and combined DM limits for our \textit{golden} sample of streams, in the case of the Benchmark M/L scenario, for both the $b\bar{b}$ and $\tau^{+}\tau^{-}$ annihilation channels. 
The best individual limits are those from the Indus, Orphan-Chenab, LMS-1 and Cetus-Palca streams. The latter stream dominates the combined result, which is at most about a factor 2 more constraining than Cetus-Palca individual limits, for the $b\bar{b}$ ($\tau^{+}\tau^{-}$) channel at the highest (intermediate) considered DM masses. The combined constraints do not reach the thermal relic cross-section value, being $\mathcal{O}(10)$ above it for the lowest considered WIMP masses and both channels. 

\begin{figure}[h!]
\begin{center}
		\begin{subfigure}{0.495\textwidth}
			\includegraphics[width=\textwidth]{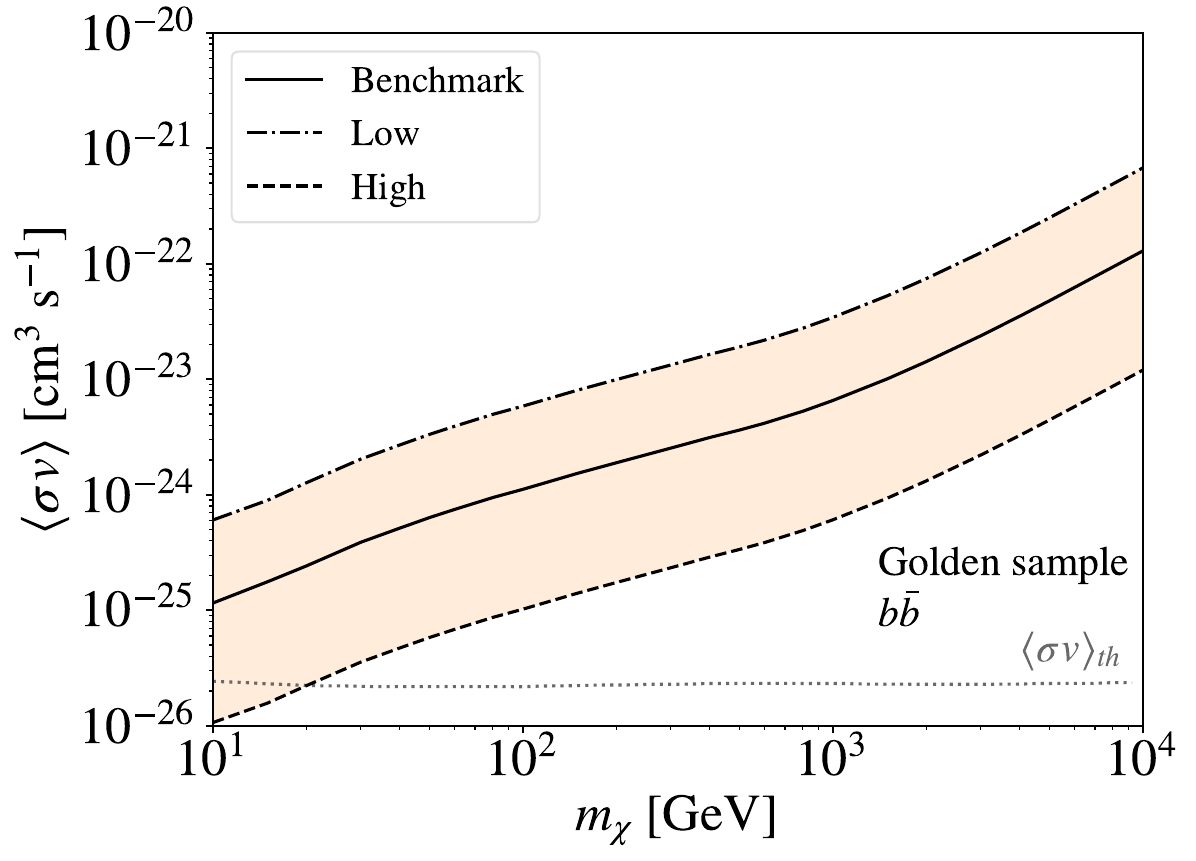}
		\end{subfigure}
		  \begin{subfigure}{0.495\textwidth}
		      	\includegraphics[width=\textwidth]{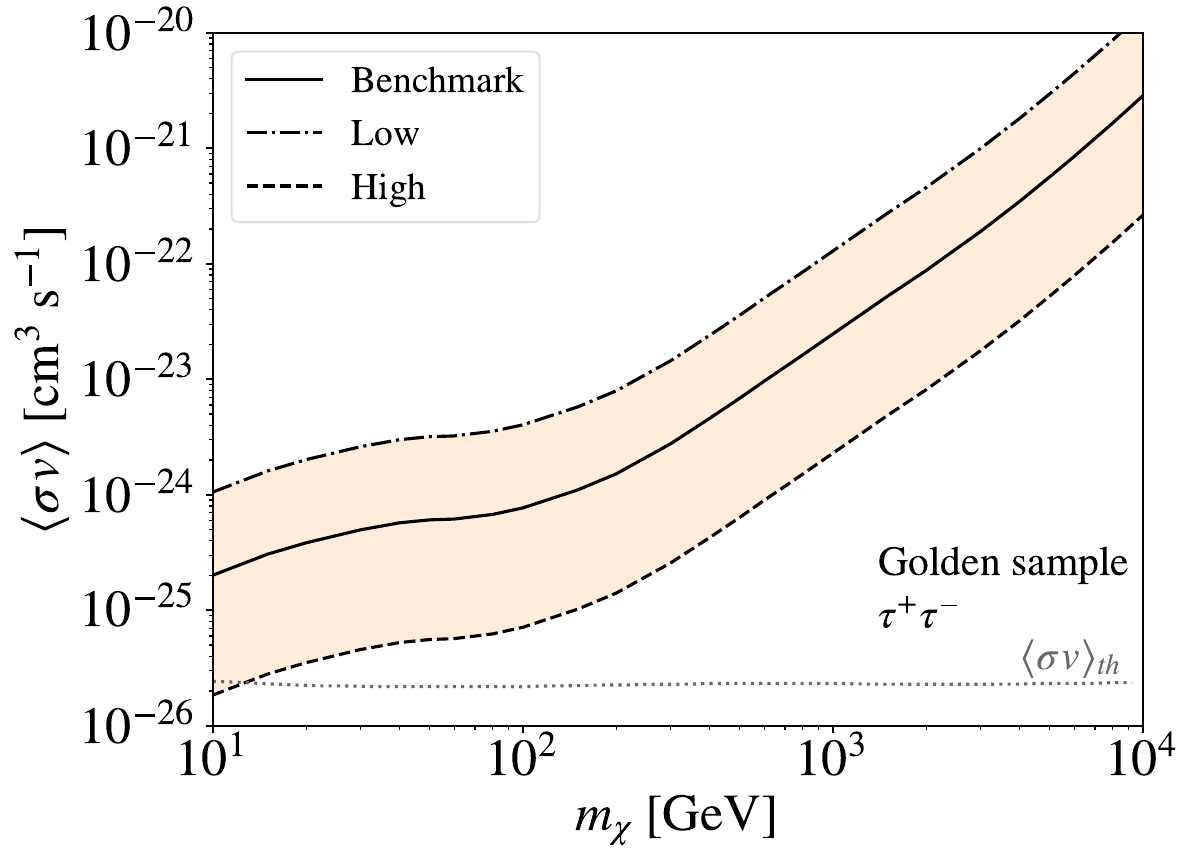}
		  \end{subfigure}
        \caption{Limits on the DM annihilation cross-section obtained in the combined analysis of our \textit{golden} sample of streams (Table \ref{table:table_sample2}), for the three M/L scenarios considered in this work (see Section \ref{sec:modeling}). Left and right panels refer to the $b\bar{b}$ and $\tau^{+}\tau^{-}$ annihilation channels, respectively.}
		\label{fig:combined_Lw_int_Up_golden}
		
\end{center}
\end{figure}

Figure \ref{fig:combined_Lw_int_Up_golden} shows the combined constraints obtained for the \textit{golden} sample and the three M/L scenarios considered in this work, again for the $b\bar{b}$ and $\tau^{+}\tau^{-}$ annihilation channels. Note that the upper limits for $\langle \sigma v \rangle$ reach the thermal relic cross-section at lower masses when considering the High scenario, ruling out WIMPs with masses below 20 GeV (10 GeV) for the $b\bar{b}$ ($\tau^{+}\tau^{-}$) channel. Overall, this systematic uncertainty due to the lack of knowledge on the precise M/L stream values translates into an uncertainty of $\mathcal{O}(100)$ in the DM limits. We note that this uncertainty band may not yet fully encapsulate the actual uncertainty, as it just refers to the range of M/L (from 2 to 50) considered in this work. 
Individual and combined \textit{golden} sample DM limits for both the Low and High scenarios are provided in Appendix~\ref{sec:appendix_golden}.

\begin{figure}[h!]
\begin{center}
        \captionsetup[subfigure]{labelformat=empty}
		\begin{subfigure}{0.495\textwidth}
			\includegraphics[width=\textwidth]{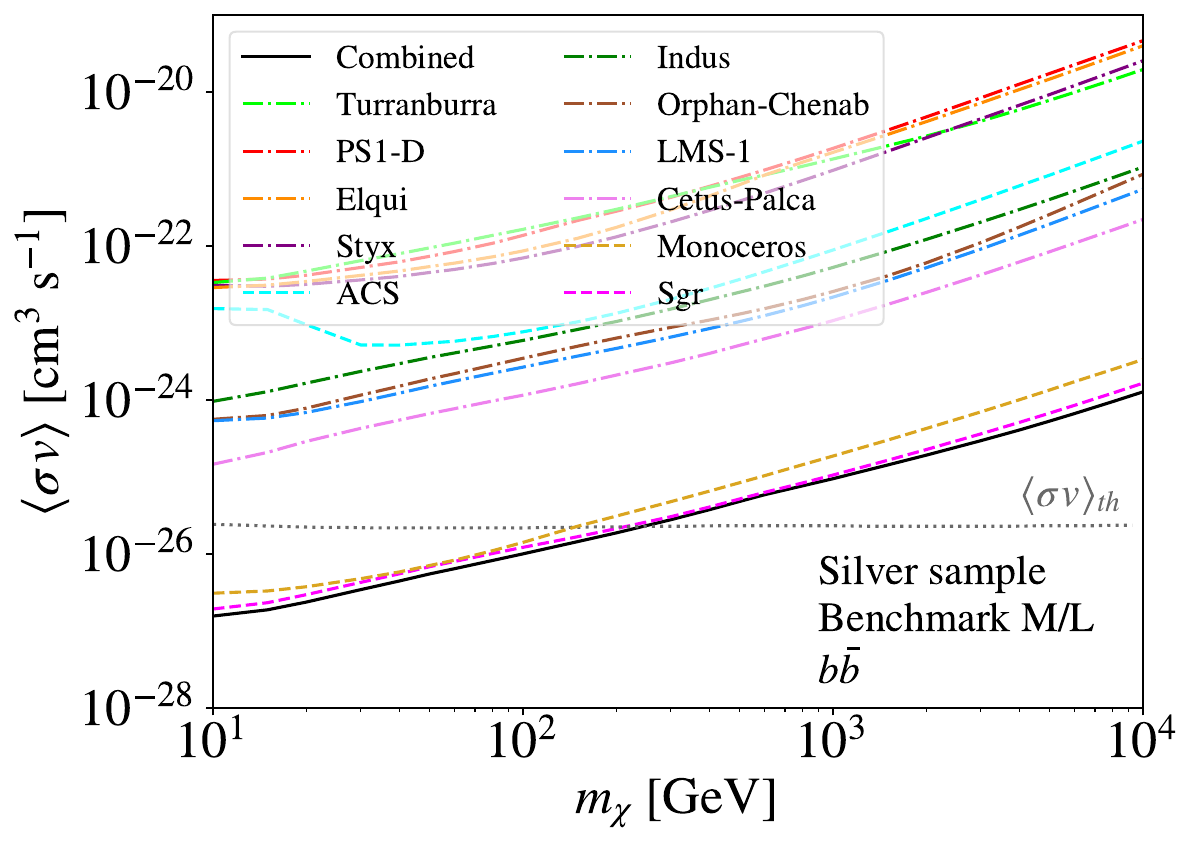}
			\caption{}
			\label{fig:combined_indviduales_int_silver_bb}
		\end{subfigure}
    \begin{subfigure}{0.495\textwidth}
		      	\includegraphics[width=\textwidth]{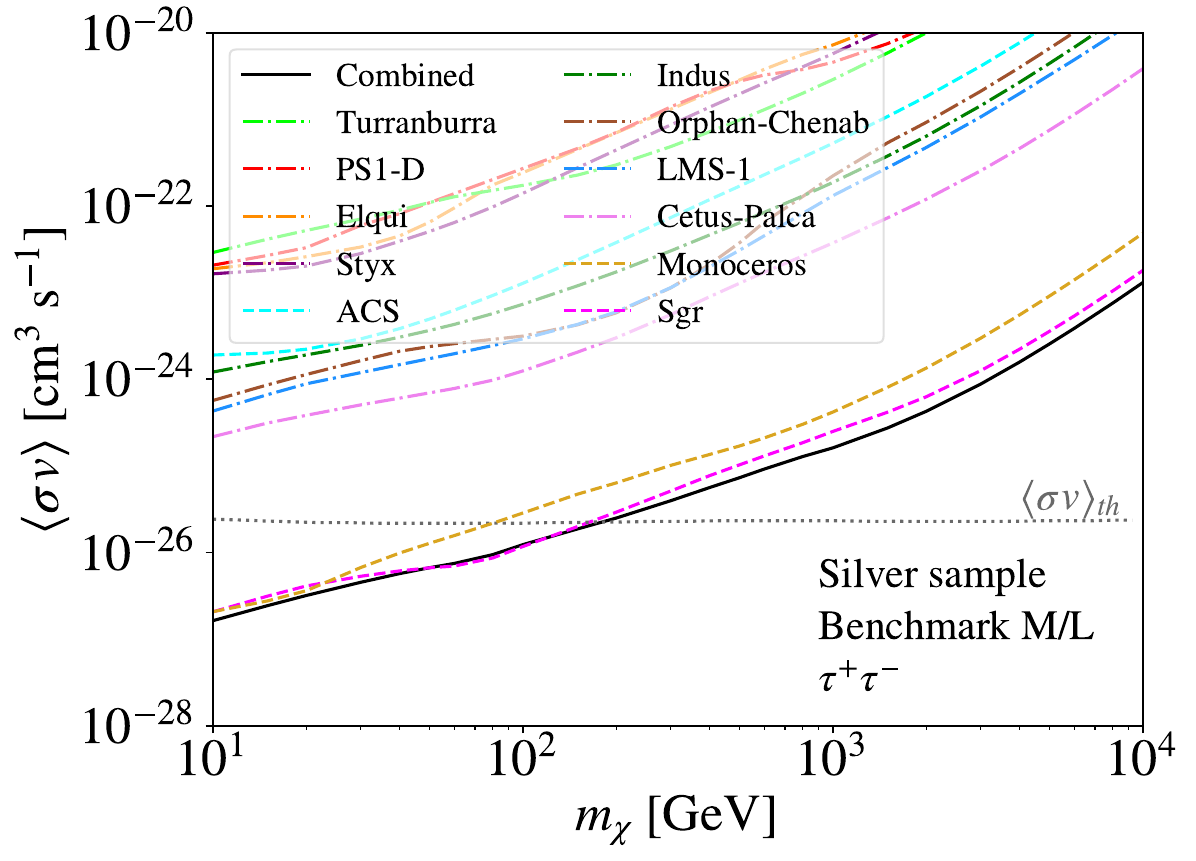}
		        \caption{}
	            \label{fig:combined_individuales_int_silver_taus}
		  \end{subfigure}
        \caption{Constraints on the DM annihilation cross section for each individual stream in the \textit{silver} sample (Table \ref{table:table_sample2}) and the Benchmark M/L scenario described in Section \ref{sec:modeling}. The combined limit is also depicted as a thick solid line. Left and right panels refer to the $b\bar{b}$ and $\tau^{+}\tau^{-}$ annihilation channels, respectively. The gray dotted line represents the thermal relic cross-section \cite{Steigman_2012}.}
		\label{fig:combined_individuales_int_silver}
\end{center}
\end{figure}

We also compute DM limits for the streams in our \textit{silver} sample. Figure \ref{fig:combined_individuales_int_silver} displays both the individual and combined DM limits in this case, as obtained for the Benchmark M/L scenario and the $b\bar{b}$ and $\tau^{+}\tau^{-}$ annihilation channels. 
When considering the \textit{silver} sample, the Monoceros and Sagittarius streams (both included in the \textit{silver} sample, but not in the \textit{golden} one) turn out to be the most constraining ones and, thus, the ones dominating the combined DM limits (which are, indeed, very similar to the individual result obtained for Sagittarius). 
The DM limits reach the thermal relic cross-section value at low WIMP masses, for both the $b\bar{b}$ and $\tau^{+}\tau^{-}$ channels. More precisely, this sample allows to rule out thermal WIMPs up to $\sim 200$ GeV for both channels. We recall that we consider the \textit{silver} sample to be not as reliable as the \textit{golden} sample in terms of their DM modelling (see Section~\ref{sec:modeling}), thus the corresponding DM limits are expected to be subject to even larger uncertainties in comparison to those obtained for the streams in the \textit{golden} sample. 

Appendix~\ref{sec:appendix_silver} includes individual and combined limits for the \textit{silver} sample for the other two considered M/L scenarios (High and Low).

\section{Discussion and conclusions}
\label{sec:concl}
In this work, we have proposed and scrutinized a new astrophysical target to shed further light on the nature of DM, assuming that it is composed of WIMPs. We searched for a gamma-ray signal coming from WIMPs annihilation within a sample of stellar streams of the Milky Way. First, we built our \textit{golden} and \textit{silver} samples of streams, which comprise the streams most suitable for this type of search, according to a set of DM-motivated criteria and to our level of confidence in them. More precisely, we focus only on those streams whose progenitors are thought to be dwarf galaxies (thus DM-dominated systems), lie at small ($< 100$ kpc) distances, and whose stellar mass is known (then we can make a correspondence between the stellar and the DM mass). Stellar streams belonging to both the \textit{golden} and \textit{silver} samples fulfill all our criteria, but the difference between them is based on our confidence in them, since streams in the \textit{silver} sample have some caveats that we have explained in Section \ref{sec:sample}.

Lacking a precise knowledge of the amount and distribution of DM within streams at the moment, it was necessary to make some important assumptions in our work. For the DM distribution in every stream, we adopted an NFW DM density profile entirely truncated beyond the scale radius of the host DM subhalo (i.e., the progenitor), this way reflecting at first order the extreme tidal stripping these objects have been subject to since accretion time to present. As for the DM amount associated to every stream, we considered three scenarios -- Low, Benchmark, High -- based on three different assumptions of the M/L ratio, namely 2 (same DM than baryonic mass), 5, and 50. Current observations only provide information about the luminous matter in the streams (see for instance Ref.~\cite{Shipp_2018}), but there are no estimates on the M/L ratios of these objects.

We have analyzed our selected streams (a total of 11) using almost 15 years of \textit{Fermi}-LAT data in the energy range between 500 MeV and 500 GeV. No emission is detected from the core of any of the streams, in both the \textit{golden} and the \textit{silver} samples.
In the absence of a signal, we use the 95$\%$ C.L. flux upper limits from our LAT analysis to place constraints on the DM particle properties. In particular, we set constraints on the WIMP annihilation cross-section vs. mass parameter space and show the main results for the \textit{golden} sample in Figures \ref{fig:combined_indviduales_int_golden} and \ref{fig:combined_Lw_int_Up_golden}, while Figure \ref{fig:combined_individuales_int_silver} is devoted to the \textit{silver} sample. Additional plots showing DM limits for both samples are displayed in Appendix \ref{sec:appendix_golden} and \ref{sec:appendix_silver}, for other possible combinations of the involved parameters (annihilation channels, M/L ratios, etc.). Our key findings in terms of DM limits can be summarized as follows:
\begin{itemize}
    \item The most reliable constraints on the DM annihilation cross-section are those obtained for the \textit{golden} sample when considering the Benchmark scenario. In this particular case, the combined constraints are $\mathcal{O}(10)$ above the thermal relic cross-section for low WIMP masses, for both the $b\bar{b}$ and $\tau^{+}\tau^{-}$ annihilation channels.
    
    \item A critical improvement is achieved when considering the \textit{silver} sample of streams, which allows to rule out WIMPs up to masses of $\sim 200$ GeV for both the $b\bar{b}$ and $\tau^{+}\tau^{-}$ channels in the Benchmark M/L scenario. We recall though that the results obtained for the \textit{silver} sample are not as reliable as those for the \textit{golden} sample, since they have associated larger uncertainties. See also the discussion further below.

    \item Considering the three M/L scenarios for the \textit{golden} sample translates in an uncertainty of $\mathcal{O}(100)$ in the DM limits. As mentioned above, constraints for the Benchmark case are $\mathcal{O}(10)$ above the thermal cross-section for low WIMP masses, while the constraints derived for the High M/L scenario are able to rule out thermal WIMPs with masses below 20 GeV (10 GeV) for the $b\bar{b}$ ($\tau^{+}\tau^{-}$) channel. The constraints for the Low M/L scenario remain $\mathcal{O}(100)$ above the thermal cross-section at lower masses. Same $\mathcal{O}(100)$ uncertainty in the DM limits applies to the \textit{silver} sample when considering the three M/L scenarios. In this case, the limits are able to rule out thermal WIMPs with masses up to $\sim 20 - 30$ GeV, $\sim 200$ GeV, and $\sim 1000 - 2000$ GeV in the Low, Benchmark and High scenarios, respectively, for both the $b\bar{b}$ and $\tau^{+}\tau^{-}$ channels.

\end{itemize}

\begin{figure}[h!]
\begin{center}
        \captionsetup[subfigure]{labelformat=empty}
		\begin{subfigure}{0.495\textwidth}
			\includegraphics[width=\textwidth]{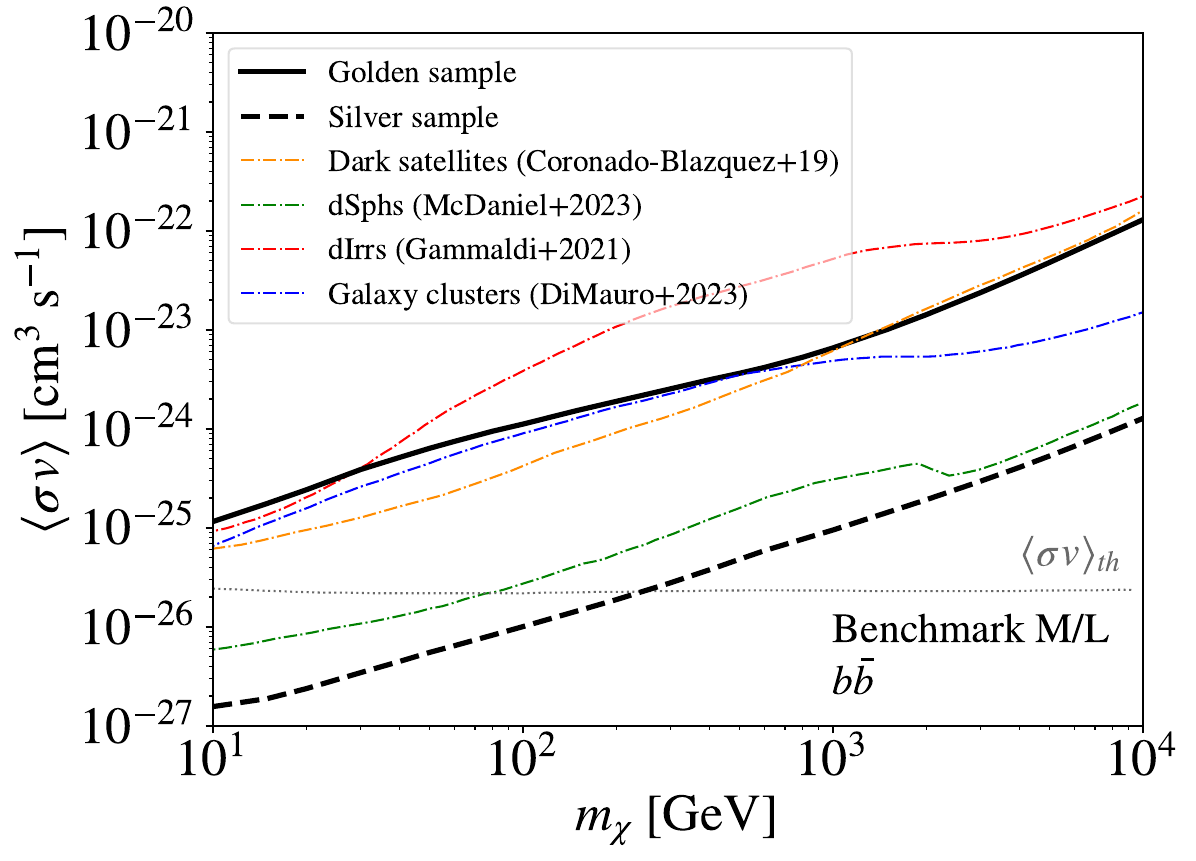}
			\caption{}
			\label{fig:combined_targets_bb}
		\end{subfigure}
		  \begin{subfigure}{0.495\textwidth}
		      	\includegraphics[width=\textwidth]{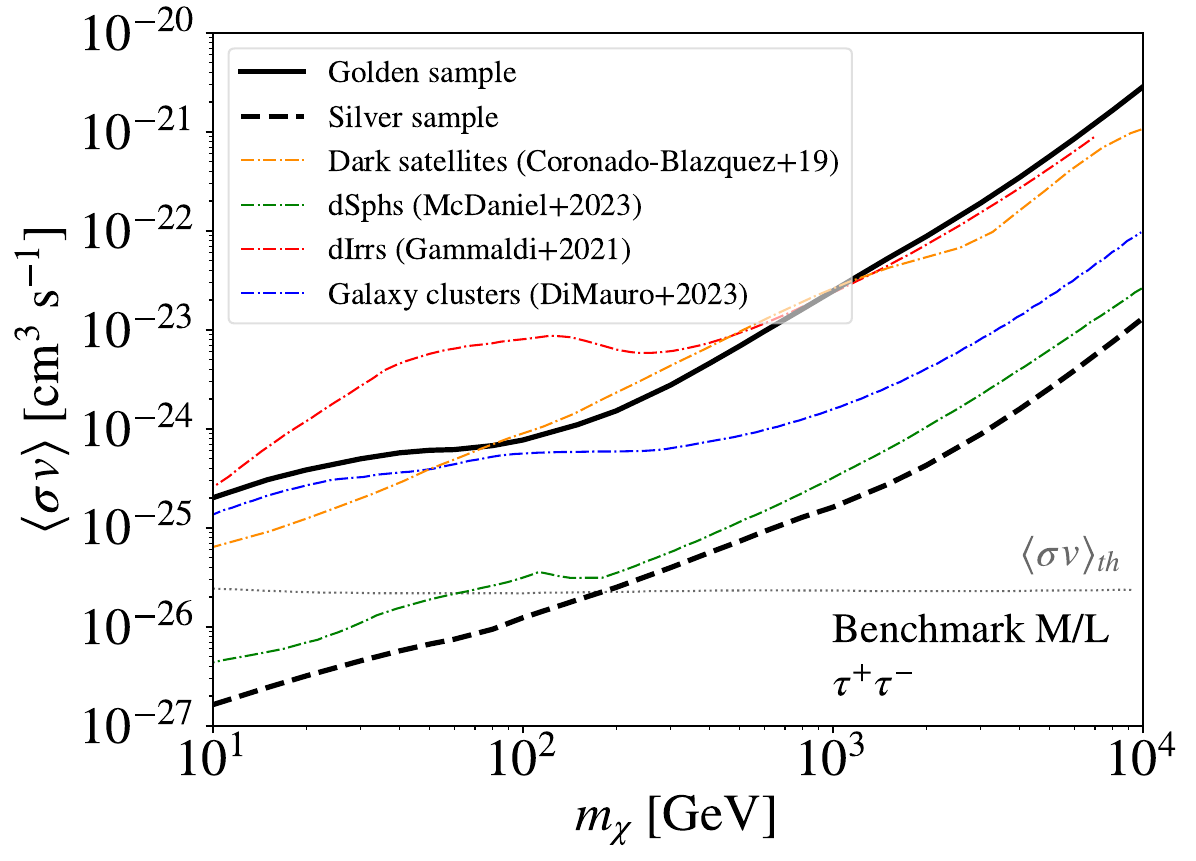}
		        \caption{}
	            \label{fig:combined_targets_taus}
		  \end{subfigure}
        \caption{DM limits obtained in this work using \textit{Fermi}-LAT data from stellar streams, in comparison with others obtained with LAT data from different targets, for both the $b\bar{b}$ (left pannel) and $\tau^{+}\tau^{-}$ (right pannel) annihilation channels. The thick solid black line represents the constraints for the \textit{golden} sample, whereas the thick dashed black line is for the \textit{silver} sample, both in our Benchmark M/L scenario. These are compared to those obtained with \textit{Fermi} LAT from dwarf galaxies~\cite{mcdaniel2023legacy} (green dot-dashed line), dark satellites~\cite{Coronado_Bl_zquez_2022} (orange dot-dashed line), dIrr galaxies~\cite{Gammaldi_2021} (red dot-dashed line), and galaxy clusters~\cite{Di_Mauro_2023} (blue dot-dashed line).}
		\label{fig:combined_targets}
\end{center}
\end{figure}

The DM limits derived in this work are placed into a more general context in Figure \ref{fig:combined_targets}, where we compare them to those obtained from different targets that have been widely used for DM searches involving gamma rays. We display the constraints obtained for the \textit{golden} and the \textit{silver} samples of streams in the Benchmark scenario, along with the most recent DM limits for dwarf spheroidal satellite galaxies \cite{mcdaniel2023legacy}, dark satellites \cite{Coronado_Bl_zquez_2022}, dwarf irregular galaxies \cite{Gammaldi_2021} and galaxy clusters \cite{Di_Mauro_2023}, all of them obtained using \textit{Fermi}-LAT data. 
As we expect a DM subhalo to host the streams' core, a direct comparison between the constraints obtained in this work with those of dSphs and dark satellites seems particularly appropriate. The best DM limits presented in Ref. \cite{Coronado_Bl_zquez_2022} for dark satellites (the ones when considering a single subhalo among the poll of LAT unIDs) are similar to the ones obtained for our \textit{golden} sample of streams; whereas the limits coming from dSphs provide the most stringent constraints (indeed, the latter being considered the best and most reliable ones in the field so far). 
Nevertheless, when considering the \textit{silver} sample of streams, the DM limits turn out to be as competitive as those for dSphs.
We can also compare our results with those of dIrr galaxies and local-volume galaxy clusters.
The results obtained for the streams in our \textit{golden} Benchmark scenario improve the limits achieved with dIrr galaxies, while they are similar to those for clusters of galaxies. 

Our results, although very promising and highly competitive, are subject to significant uncertainties, further work being necessary in several fronts to decrease them. Yet, to our knowledge this study represents the first attempt to use stellar streams as an additional target to constrain the properties of the DM particle with gamma rays. Indeed, from our work we conclude that stellar streams should be perceived as a complementary and competitive tool for indirect DM searches, thanks to both their proximity to us and their expected reservoirs of DM. 
However, as said, there are still some important caveats to address in the future:
\begin{itemize}
    \item The location of the streams' core is uncertain in most cases.  In fact, in those cases where the location of the progenitor is unknown, motivated by other studies like Ref.~\cite{Malhan_2021}, we consider the mid-point of the stream as the location of its most likely progenitor (see Section \ref{sec:sample} for details). Yet in our analysis we scrutinize a 15$^\circ$ x 15$^\circ$ region around each presumable core, and look for new sources within the innermost 3 degrees (no new sources are found, but in one single case). This should help mitigate the problem, however we recall that some of the streams subtend angles much larger than 15$^\circ$ in the sky (see Table~\ref{table:table_sample2}), thus, although improbable, we could have missed the ``actual'' core of the progenitor in our data analyses. Future high-resolution spectroscopic observations should help decreasing the uncertainty in the location of the cores of streams in our sample.

    \item As explained in Section \ref{sec:modeling}, both the stellar and DM mass of the streams are currently not well known. In turn, this lack of knowledge can lead to overestimating or underestimating the J-factor significantly, which results into large uncertainties in the DM limits. 
    In this work, we intentionally adopted a broad, conservative range of 2-50 M/L values so that we could safely encapsulate this source of systematic uncertainty (which, we recall, translates into an $\mathcal{O}(100)$ uncertainty in the DM limits). Again, further spectroscopic measurements, such as those ongoing with DESI, are expected to be key in addressing this issue.

    \item Throughout this work we adopted an NFW profile truncated beyond its scale radius, introduced to account for extreme material removal in these objects due to tidal stripping (see Section~\ref{sec:modeling}). Though a reasonable first-order approximation, the precise properties of the underlying DM density profile in these objects remains unclear. High-resolution hydrodynamic simulations arise as the perfect tool to address this and related questions -- e.g., what is the exact DM amount that remains today given a particular orbital configuration and infall time of the progenitor; how this is distributed; what is the dynamical state of the core. This more accurate modelling of the DM density profile in streams is already ongoing using state-of-the-art hydrodynamical simulations and will be presented elsewhere.  
       
\end{itemize}
 
In conclusion, our study demonstrates the viability of using stellar streams for indirect DM searches, evidencing their usefulness to shed further light on the properties of the DM particle. Indeed, despite  the important existing caveats, the obtained results are promising and nicely complementary to others obtained by other methods and targets. Further work -- both with observations and simulations -- is mandatory though should we want to make further progress and to produce robust DM limits from streams. Work is already ongoing in this direction. 
Additionally, complementary WIMP searches in stellar streams could also be performed in the TeV mass range by current or future ground-based facilities. Imaging Atmospheric Cherenkov Telescopes like MAGIC~\cite{Mirzoian:2004bk}, HESS~\cite{Hinton_2004}, VERITAS~\cite{Weekes_2002} or, in the near future, CTAO~\cite{2018cta}, would necessarily focus on the most promising and best characterized streams in our sample, given both the narrow field of view of such instruments and low duty cycles. Others such as LHAASO~\cite{Ma_2022}, HAWC~\cite{Abeysekara_2017}, or the future SWGO~\cite{abreu2019southernwidefieldgammarayobservatory} could also explore and search for DM signals in a larger area of each stream, thanks to their wide field of view, while waiting for a more accurate localization of the streams' core.

\section*{Acknowledgements}

The authors would like to thank Alejandra Aguirre-Santaella and all the DAMASCO group \footnote{https://projects.ift.uam-csic.es/damasco/} for fruitful discussions, as well as N\'estor Mirabal and Regina Caputo for their valuable comments and insightful feedback on the manuscript. The work of CFS was supported by Programa Investigo 2022 Comunidad de Madrid with ref. A113, funded by European Union - NextGenerationEU, and by the ``La Caixa'' Foundation (ID 100010434) and the European Union's Horizon~2020 research and innovation program under the Marie Skłodowska-Curie grant agreement No~847648, fellowship code LCF/BQ/PI21/11830030. The work of CFS and MASC was supported by the grants PID2021-125331NB-I00 and CEX2020-001007-S, both funded by \allowbreak{MCIN/AEI/10.13039/501100011033} and by ``ERDF A way of making Europe''. CFS and MASC also acknowledge the MultiDark Network, ref. RED2022-134411-T. 

The \textit{Fermi} LAT Collaboration acknowledges generous ongoing support
from a number of agencies and institutes that have supported both the
development and the operation of the LAT as well as scientific data analysis.
These include the National Aeronautics and Space Administration and the
Department of Energy in the United States, the Commissariat \`a l'Energie Atomique
and the Centre National de la Recherche Scientifique / Institut National de Physique
Nucl\'eaire et de Physique des Particules in France, the Agenzia Spaziale Italiana
and the Istituto Nazionale di Fisica Nucleare in Italy, the Ministry of Education,
Culture, Sports, Science and Technology (MEXT), High Energy Accelerator Research
Organization (KEK) and Japan Aerospace Exploration Agency (JAXA) in Japan, and
the K.~A.~Wallenberg Foundation, the Swedish Research Council and the
Swedish National Space Board in Sweden.
Additional support for science analysis during the operations phase is gratefully
acknowledged from the Istituto Nazionale di Astrofisica in Italy and the Centre
National d'\'Etudes Spatiales in France. This work performed in part under DOE
Contract DE-AC02-76SF00515.

\appendix

\section{Additional data analysis results}
\label{sec:appendix_fermipy}

This appendix is devoted to show the main results obtained in the data analysis for all the streams considered in this work, except those of the LMS-1 stream, whose results are shown in Section \ref{sec:analysis}.

\begin{figure}
 \begin{center}
		\begin{subfigure}{0.4\textwidth}
			
            \includegraphics[width=\textwidth]{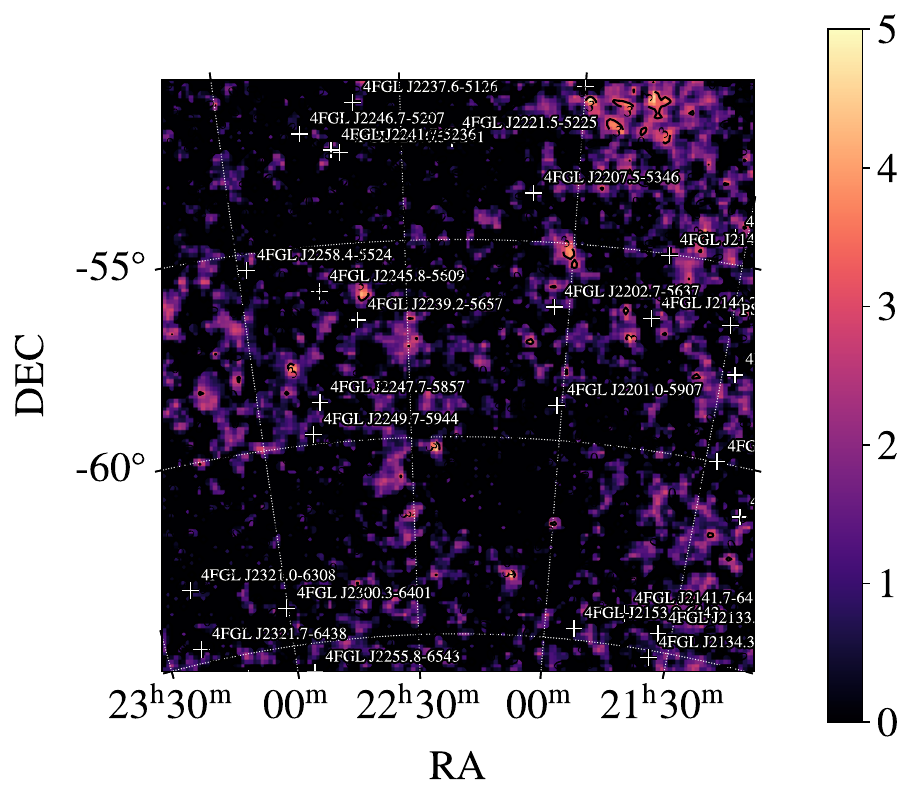}
		\end{subfigure}
        \hspace{0.1cm}
		  \begin{subfigure}{0.4\textwidth}
		      
              \includegraphics[width=\textwidth]{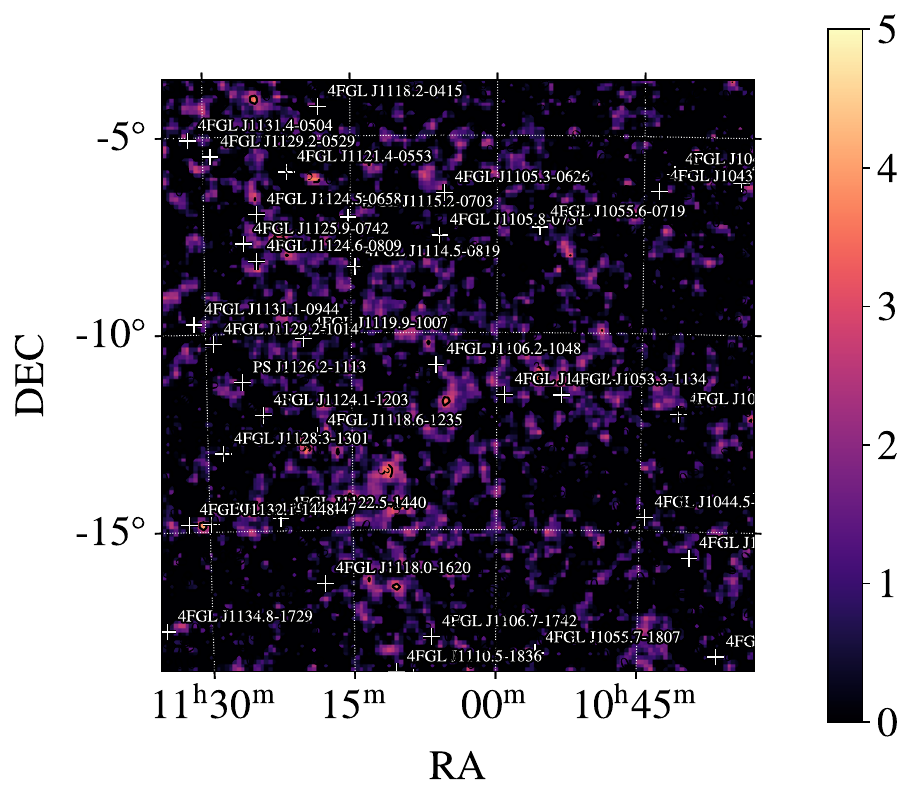}
		  \end{subfigure}

        
        \begin{subfigure}{0.4\textwidth}
			
            \includegraphics[width=\textwidth]{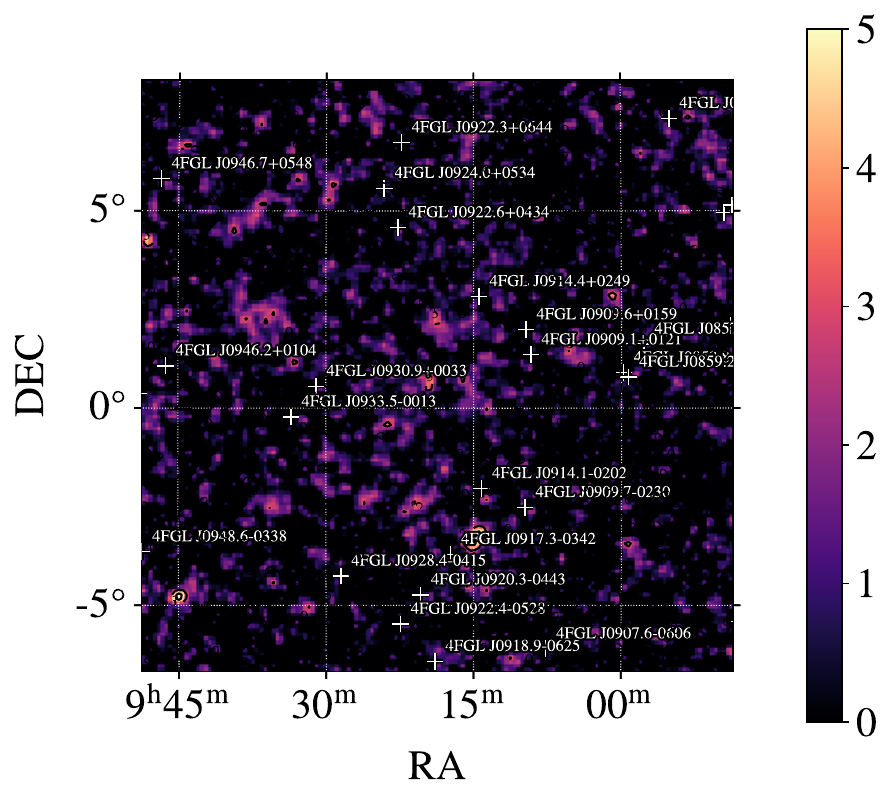}
		\end{subfigure}
        \hspace{0.1cm}
		  \begin{subfigure}{0.4\textwidth}
		      	\includegraphics[width=\textwidth]{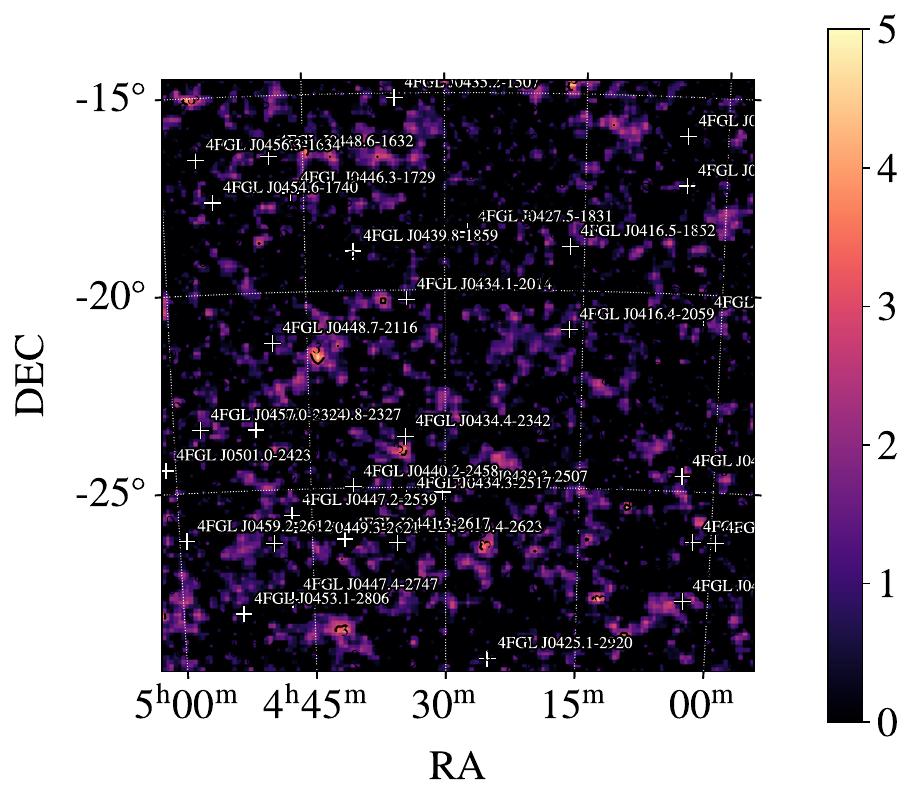}
		  \end{subfigure}
        
        \begin{subfigure}{0.4\textwidth}
			
            \includegraphics[width=\textwidth]{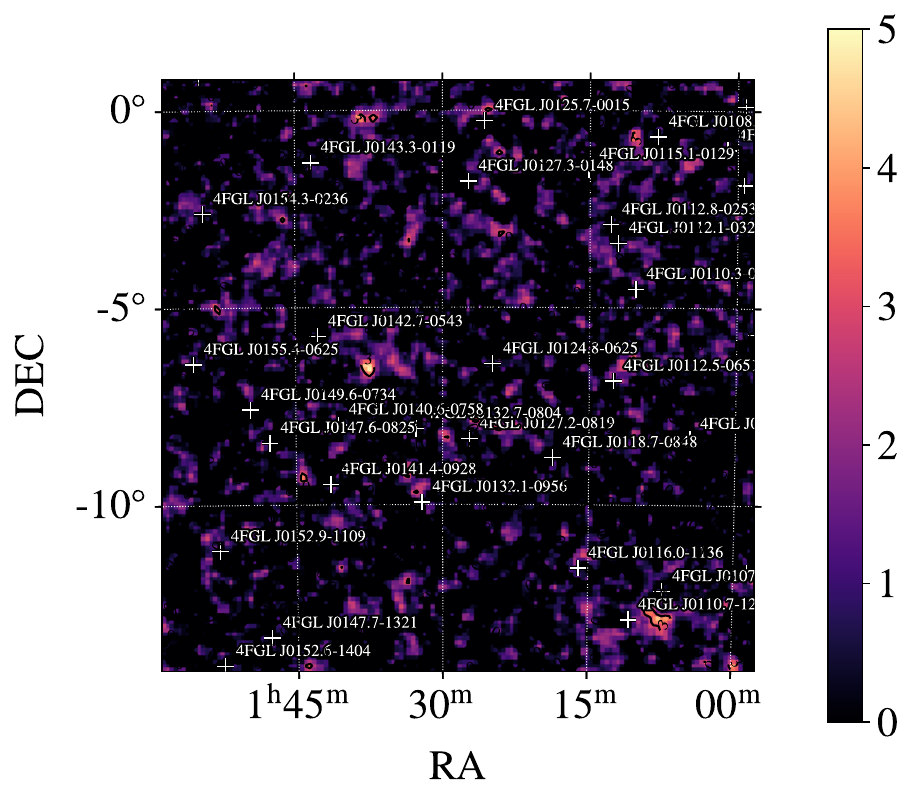}
		\end{subfigure}
        \hspace{0.1cm}
		  \begin{subfigure}{0.4\textwidth}
		      	\includegraphics[width=\textwidth]{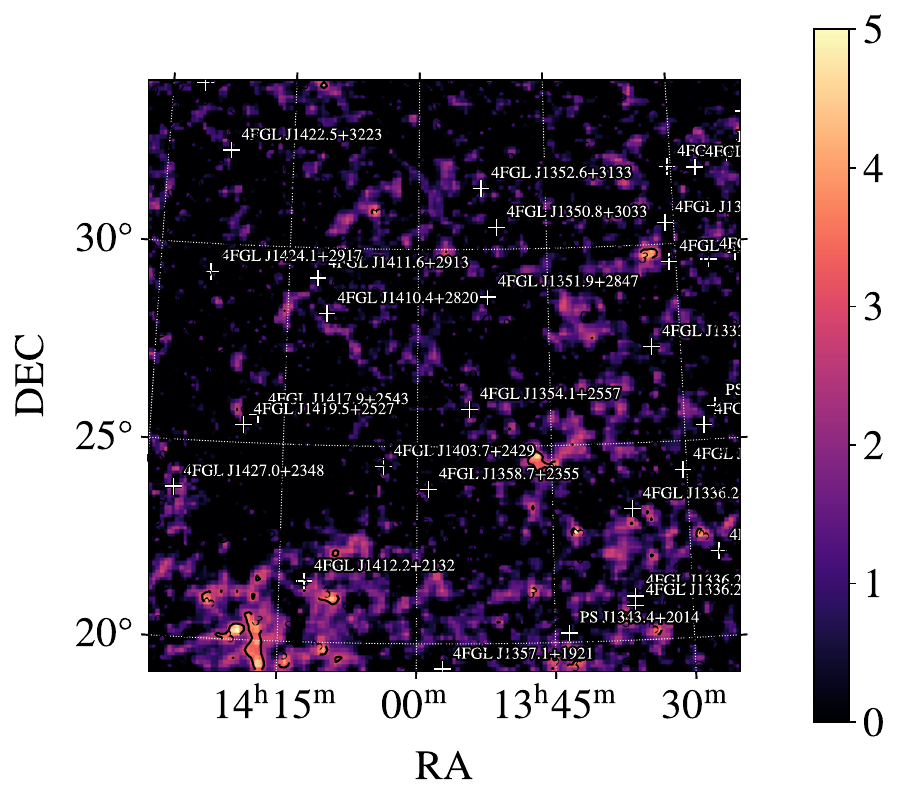}
		  \end{subfigure}


        \begin{subfigure}{0.4\textwidth}
			
            \includegraphics[width=\textwidth]{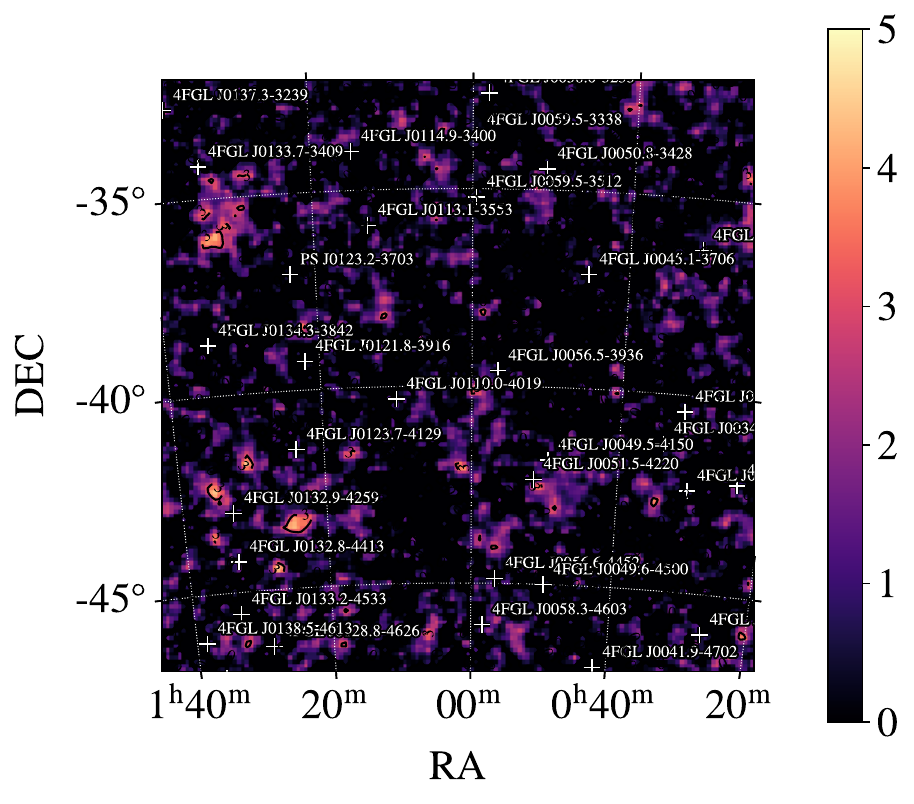}
		\end{subfigure}
       \caption{Significance (i.e., $\sim\sqrt{TS}$) map for the streams in the \textit{golden} sample. Panels show, from left to right and top to bottom, Indus, Orphan-Chenab, PS1-D, Turranburra, Cetus-Palca, Styx, and Elqui.
       }
		\label{fig:tsmap_golden}
  \end{center}
\end{figure}

\begin{figure}
 \begin{center}    
		  \begin{subfigure}{0.4\textwidth}
		      	\includegraphics[width=\textwidth]{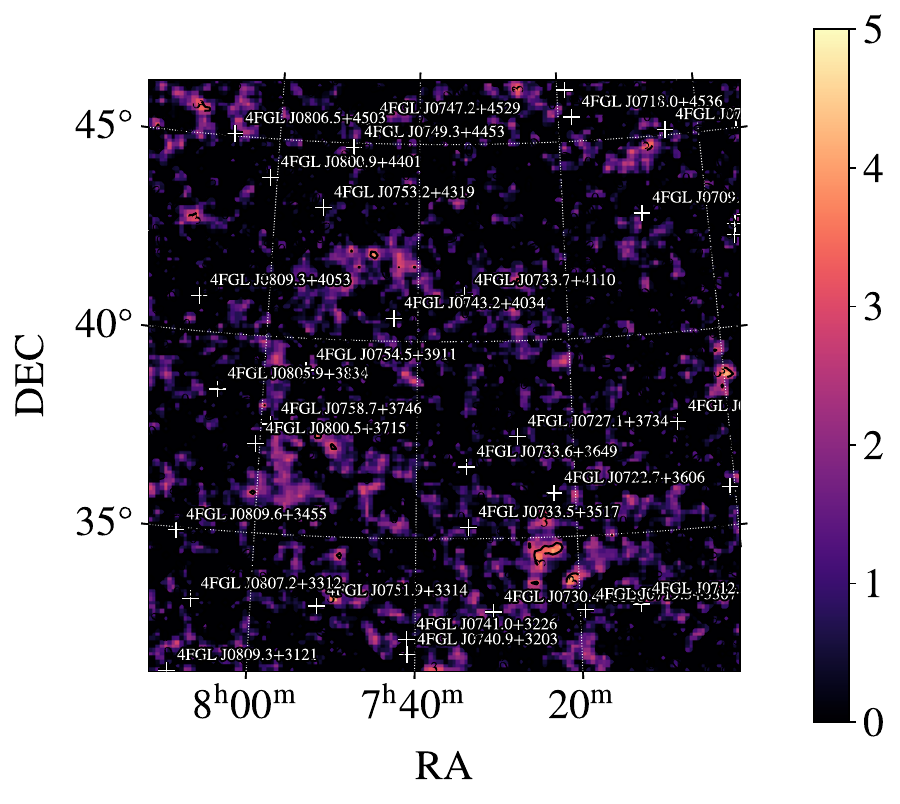}
		  \end{subfigure}
       \hspace{0.1cm}
        \begin{subfigure}{0.4\textwidth}
			
            \includegraphics[width=\textwidth]{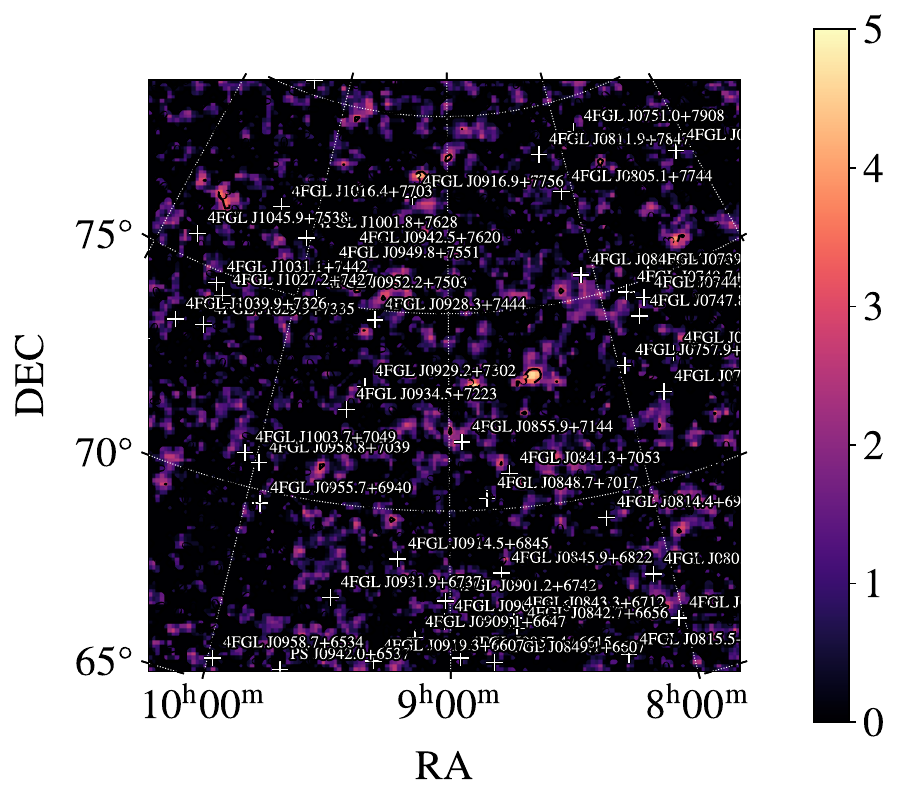}
		\end{subfigure}
  
   
		  \begin{subfigure}{0.4\textwidth}
		      	
               \includegraphics[width=\textwidth]{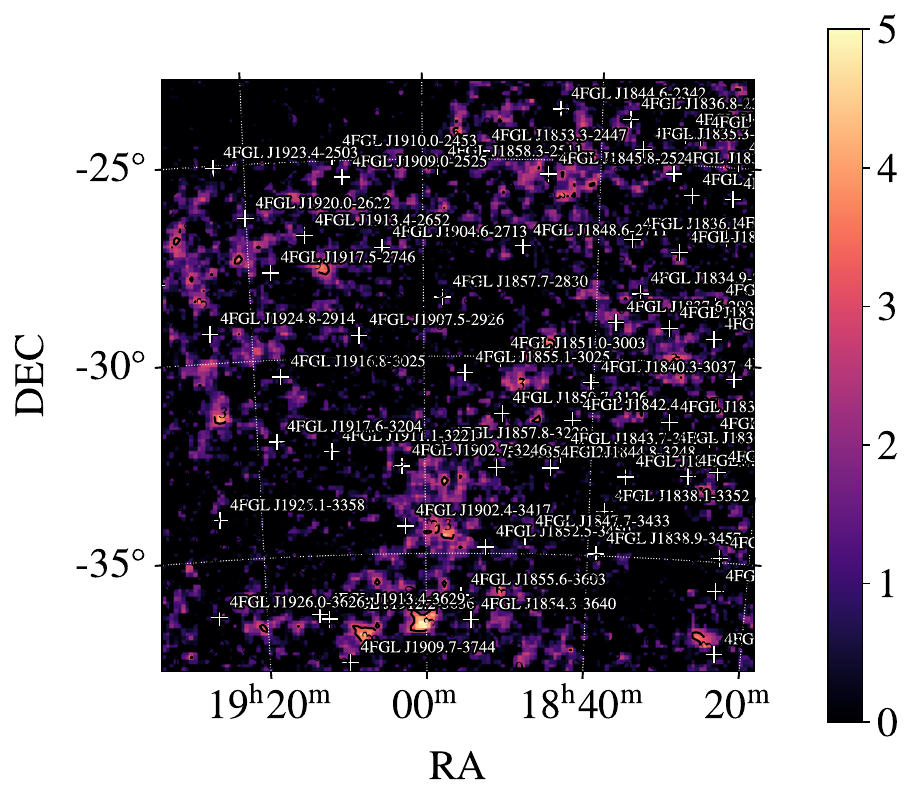}
		  \end{subfigure}
        \caption{Significance (i.e., $\sqrt{TS}$) map for the three additional streams in the \textit{silver} sample. Panels show, from left to right and top to bottom, Monoceros, AntiCenter, and Sagittarius.}
		\label{fig:tsmap_silver}
		
\end{center}
\end{figure}

\begin{figure}
 \begin{center}
		\begin{subfigure}{0.4\textwidth}
			
            \includegraphics[width=\textwidth]{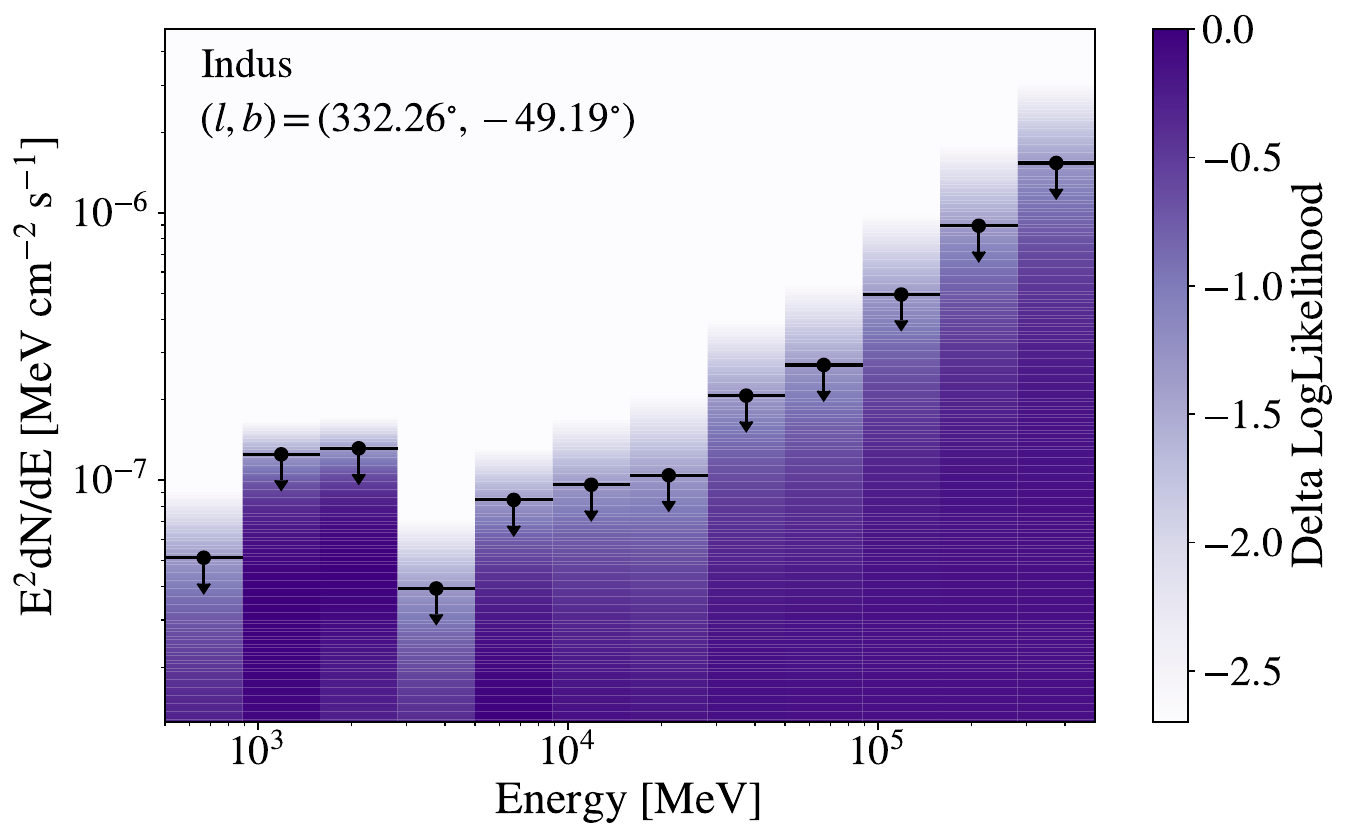}
		\end{subfigure}
        \hspace{0.1cm}
		  \begin{subfigure}{0.4\textwidth}
		      
              \includegraphics[width=\textwidth]{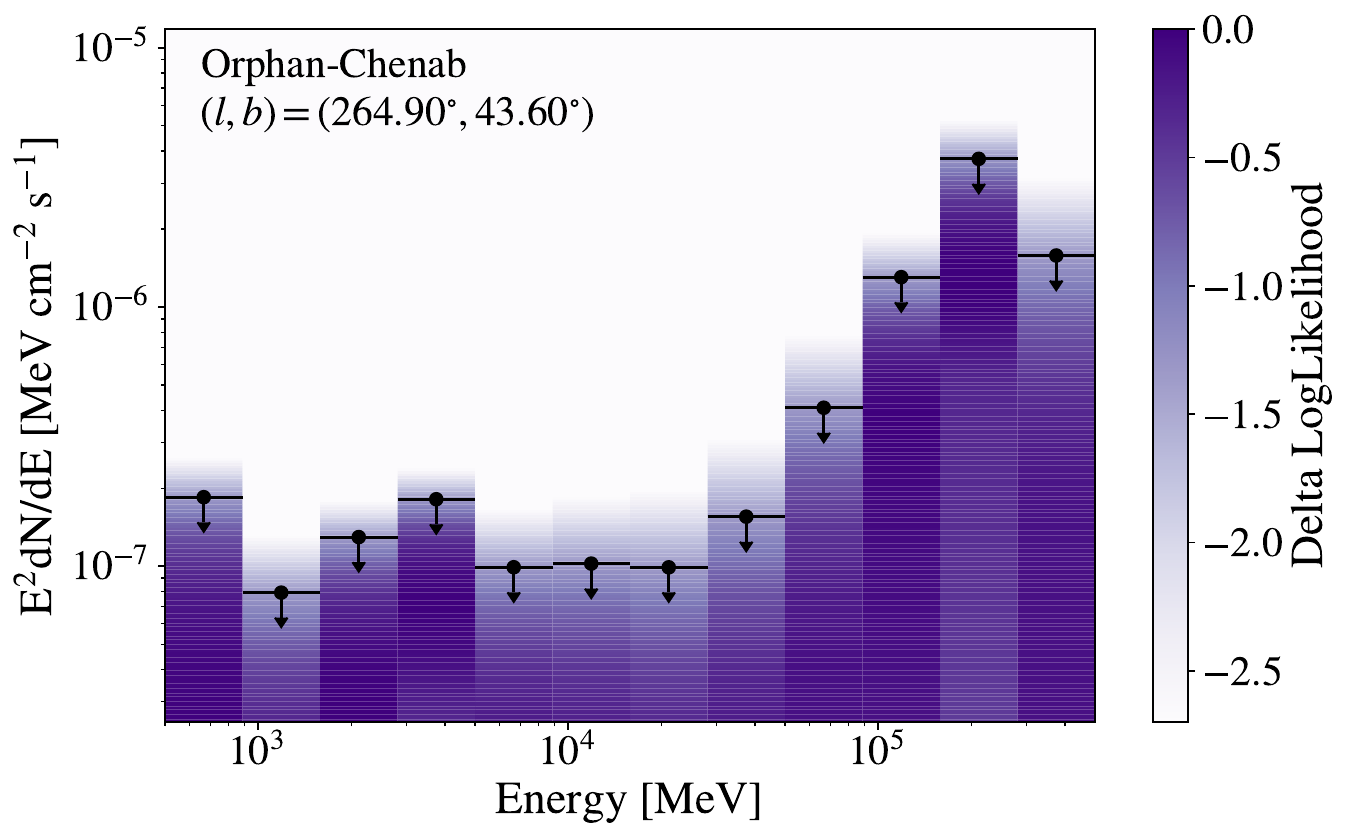}
		  \end{subfigure}

        
        \begin{subfigure}{0.4\textwidth}
			
            \includegraphics[width=\textwidth]{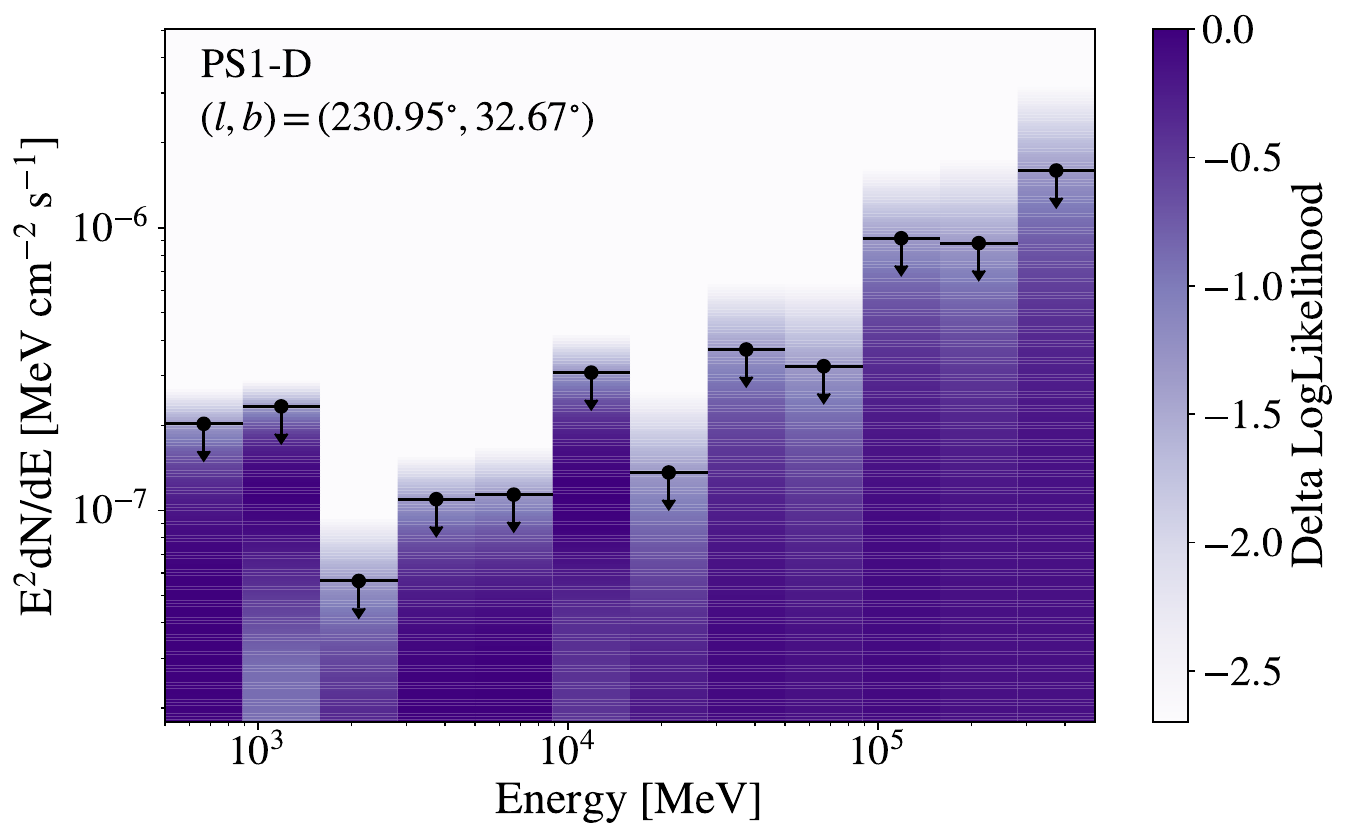}
		\end{subfigure}
        \hspace{0.1cm}
		  \begin{subfigure}{0.4\textwidth}
		      	\includegraphics[width=\textwidth]{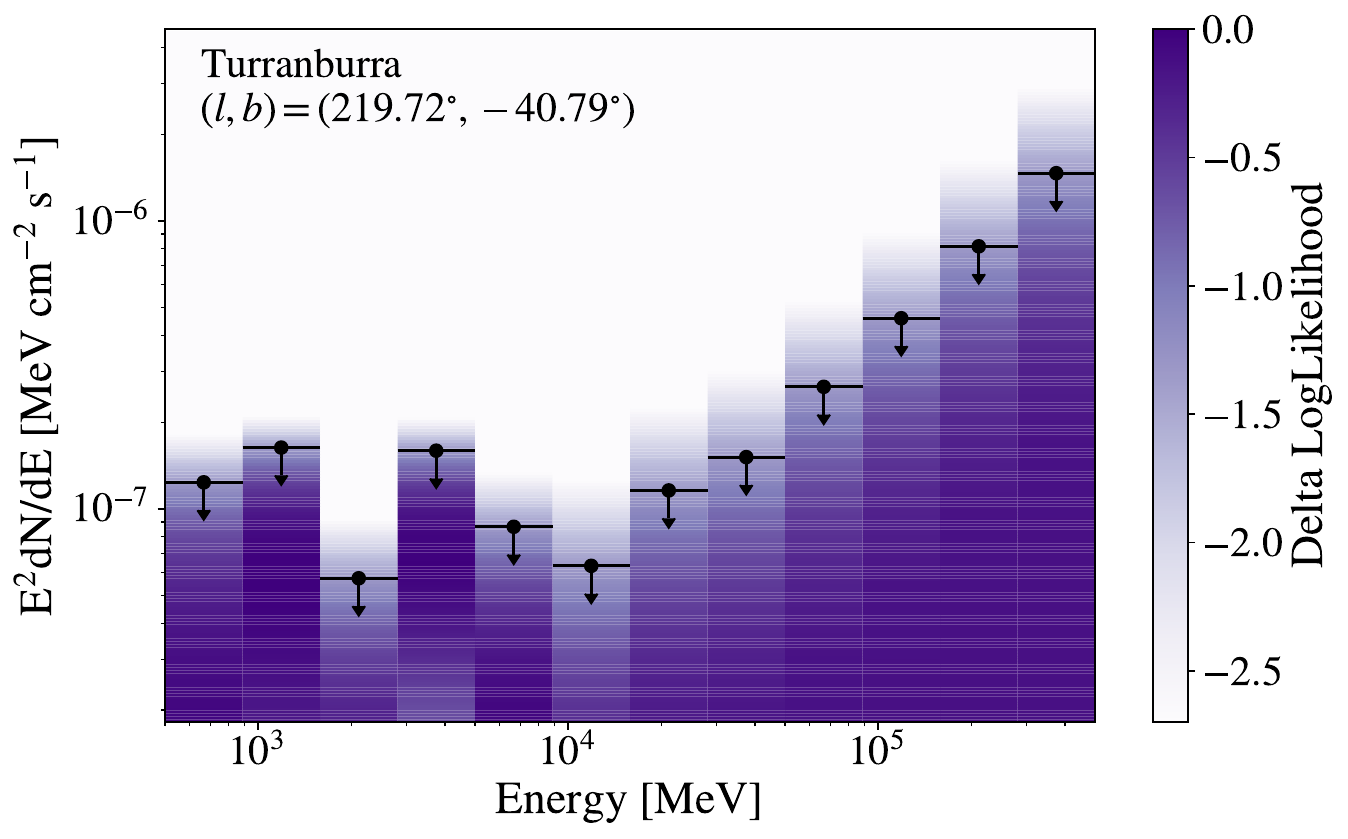}
		  \end{subfigure}
        
        \begin{subfigure}{0.4\textwidth}
			
            \includegraphics[width=\textwidth]{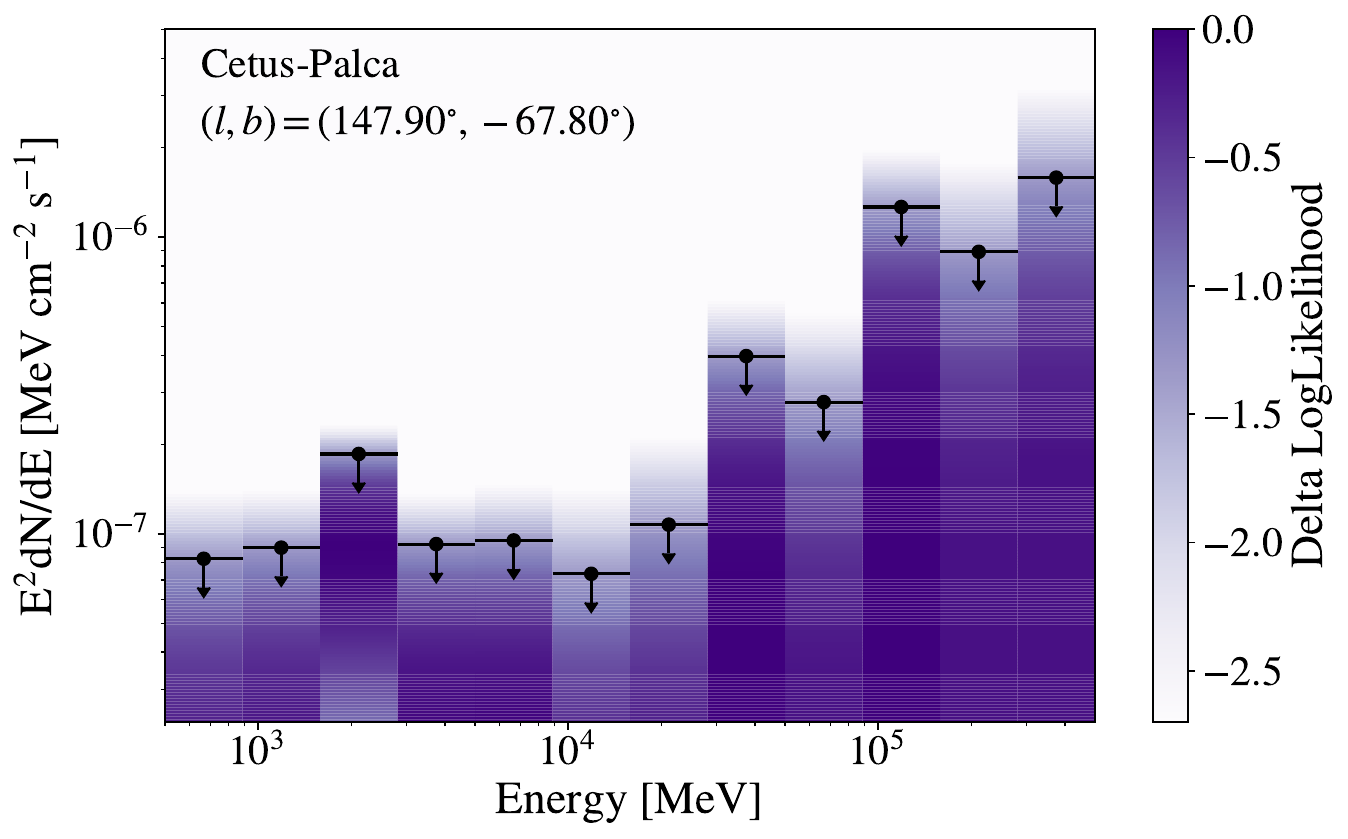}
		\end{subfigure}
        \hspace{0.1cm}
		  \begin{subfigure}{0.4\textwidth}
		      	
            \includegraphics[width=\textwidth]{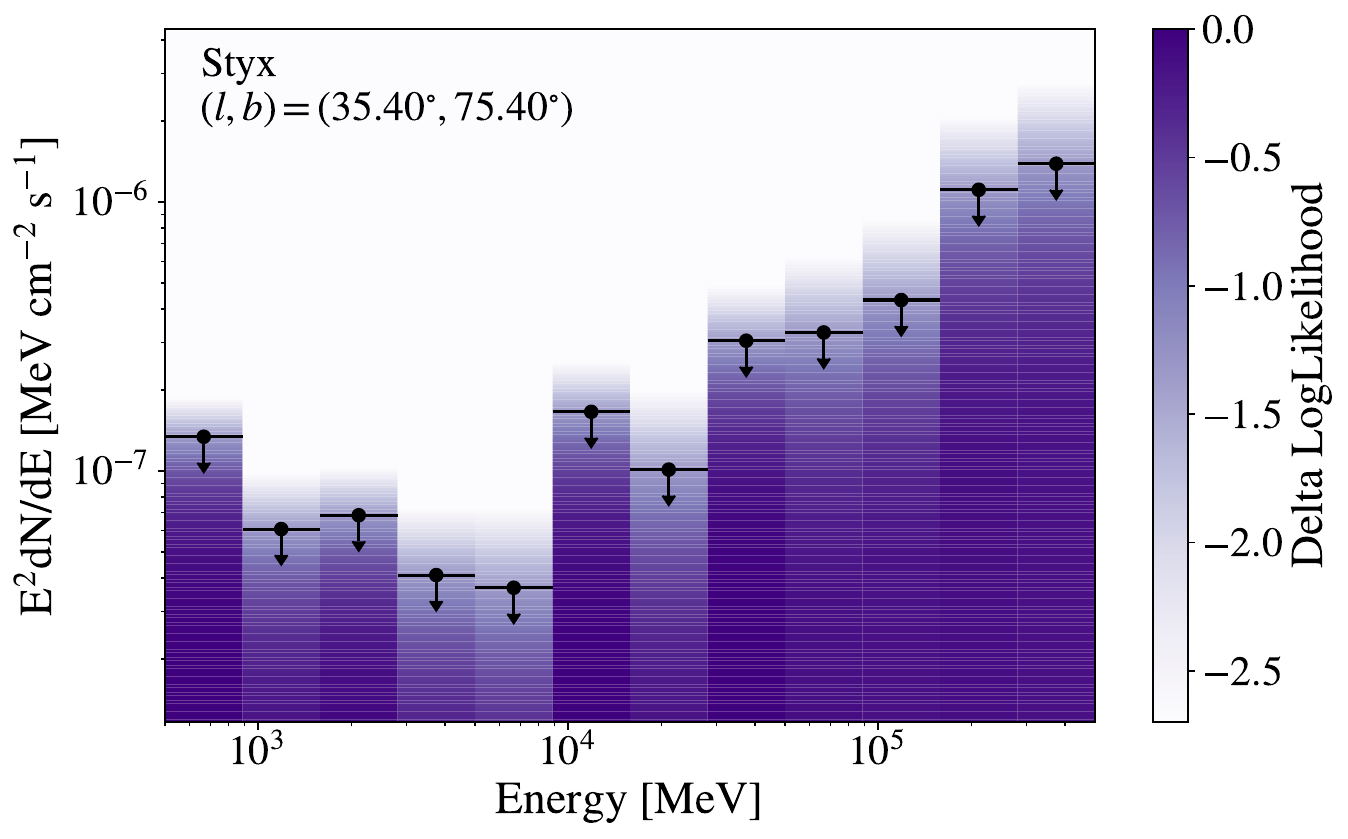}
		  \end{subfigure}


        \begin{subfigure}{0.4\textwidth}
			
            \includegraphics[width=\textwidth]{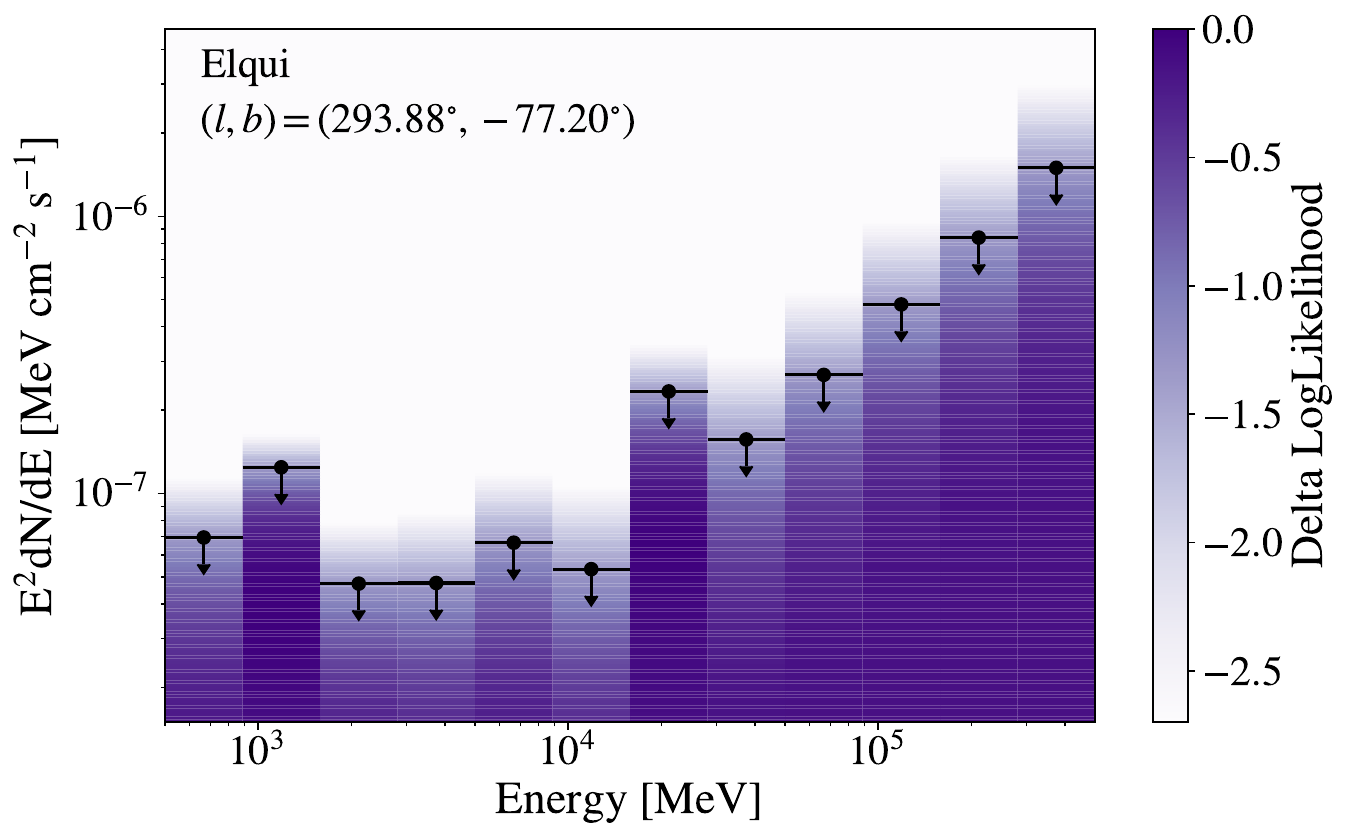}
		\end{subfigure}
       \caption{95$\%$ C.L. upper limits to the flux and likelihood values for the streams in the \textit{golden} sample. Panels show, from left to right and top to bottom, Indus, Orphan-Chenab, PS1-D, Turranburra, Cetus-Palca, Styx, and Elqui.}
		\label{fig:SED_golden}
  \end{center}
\end{figure}

\begin{figure}
 \begin{center}    
		  \begin{subfigure}{0.4\textwidth}
		      	\includegraphics[width=\textwidth]{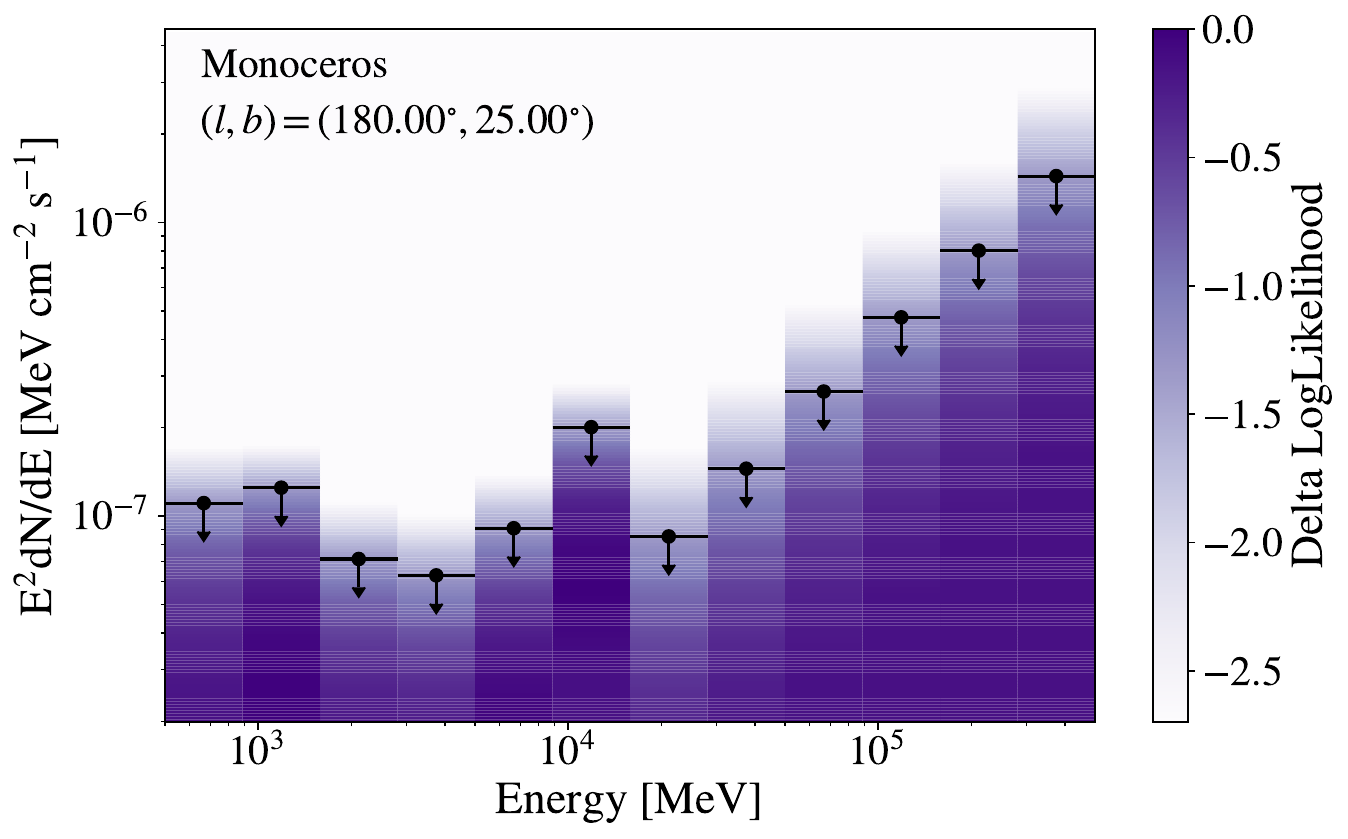}
		  \end{subfigure}
       \hspace{0.1cm}
        \begin{subfigure}{0.4\textwidth}
			
            \includegraphics[width=\textwidth]{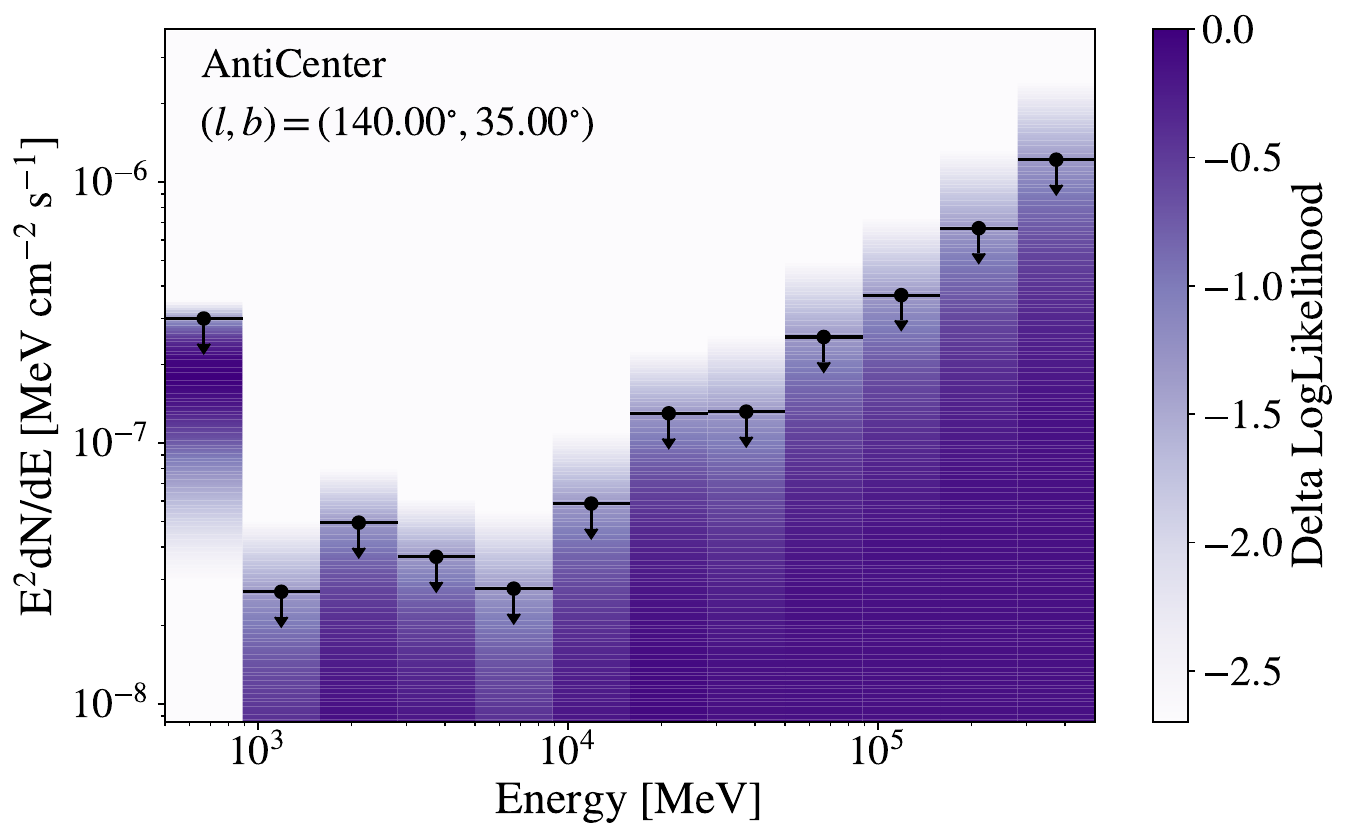}
		\end{subfigure}
  
   
		  \begin{subfigure}{0.4\textwidth}
		      	
            \includegraphics[width=\textwidth]{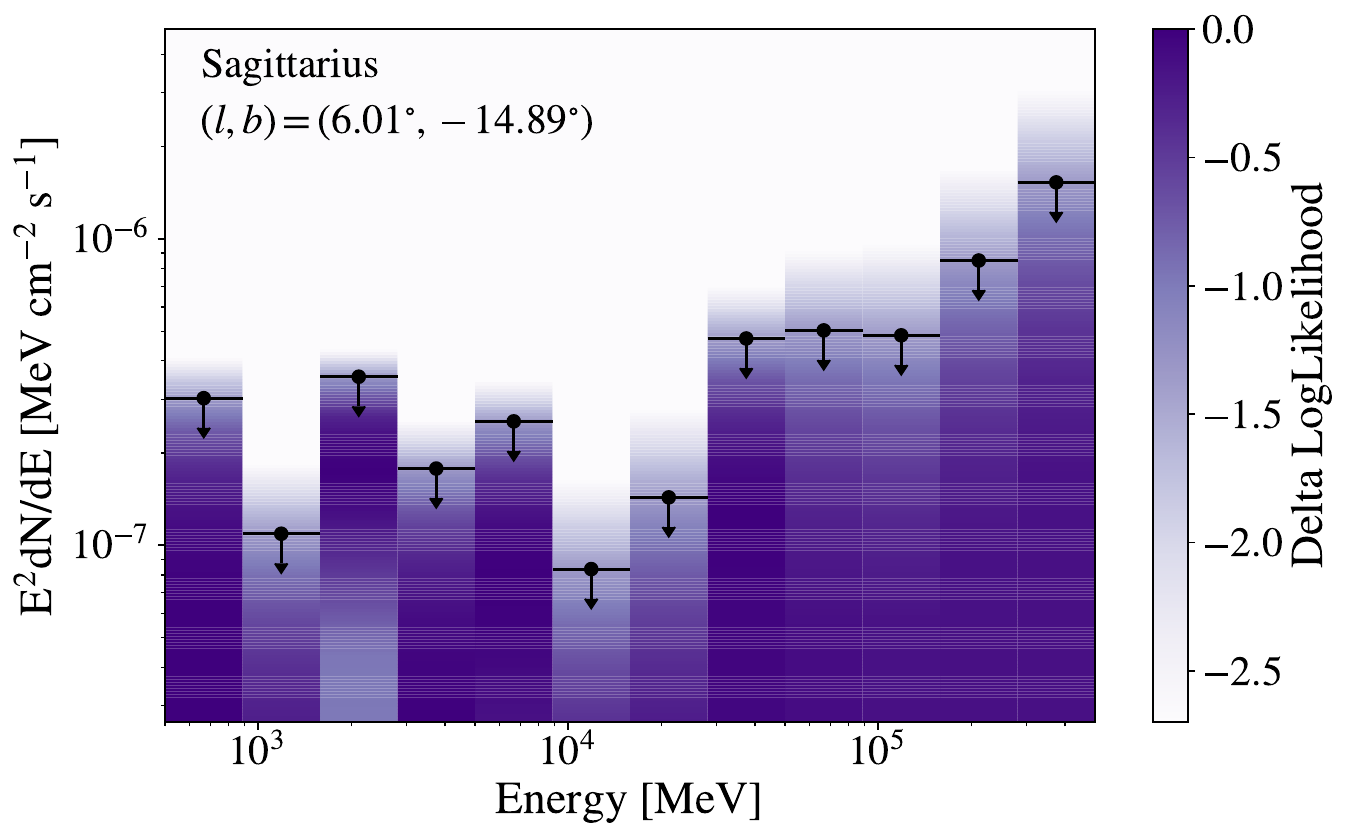}
		  \end{subfigure}
        \caption{95$\%$ C.L. upper limits to the flux and likelihood values for the three additional streams in the \textit{silver} sample. Panels show, from left to right and top to bottom, Monoceros, AntiCenter, and Sagittarius.}
		\label{fig:SED_silver}
		
\end{center}
\end{figure}

Figure \ref{fig:tsmap_golden} displays the TS map of the 15$^\circ$ x 15$^\circ$ region around the core of each stream in the \textit{golden} sample, whereas Figure \ref{fig:tsmap_silver} shows the TS map for the additional streams in the \textit{silver} sample. No significant excesses are observed across the ROI in all cases.  
The 95$\%$ C.L. flux upper limits found for each stream's core in the analysis are shown in Figure \ref{fig:SED_golden} for the \textit{golden} sample. Figure \ref{fig:SED_silver} depicts the same, but only for the three additional streams in the \textit{silver} sample.

\newpage
\section{Individual and combined log-likelihood profiles}
\label{sec:appendix:lhl}
\begin{figure}[h!]
\begin{center}
        \captionsetup[subfigure]{labelformat=empty}
		\begin{subfigure}{0.49\textwidth}
			\includegraphics[width=\textwidth]{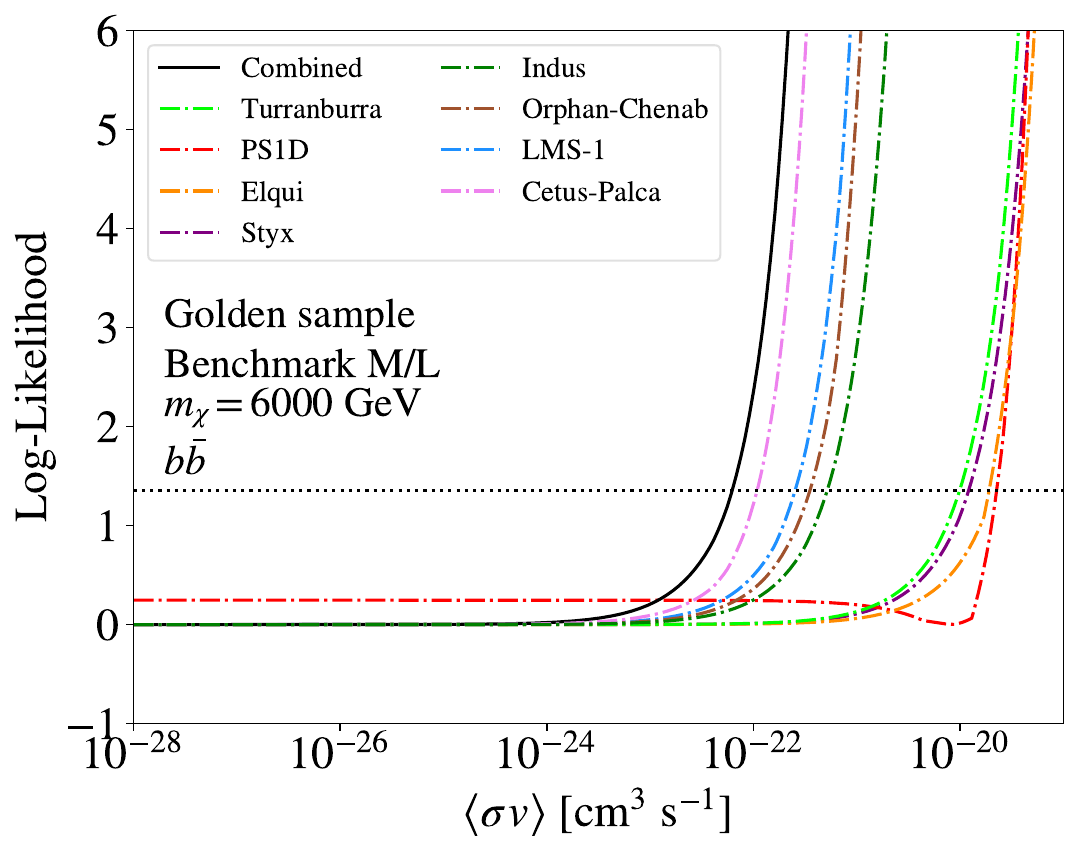}
			\caption{}
			\label{fig:lhl_bb}
		\end{subfigure}
		  \begin{subfigure}{0.5\textwidth}
		      	\includegraphics[width=\textwidth]{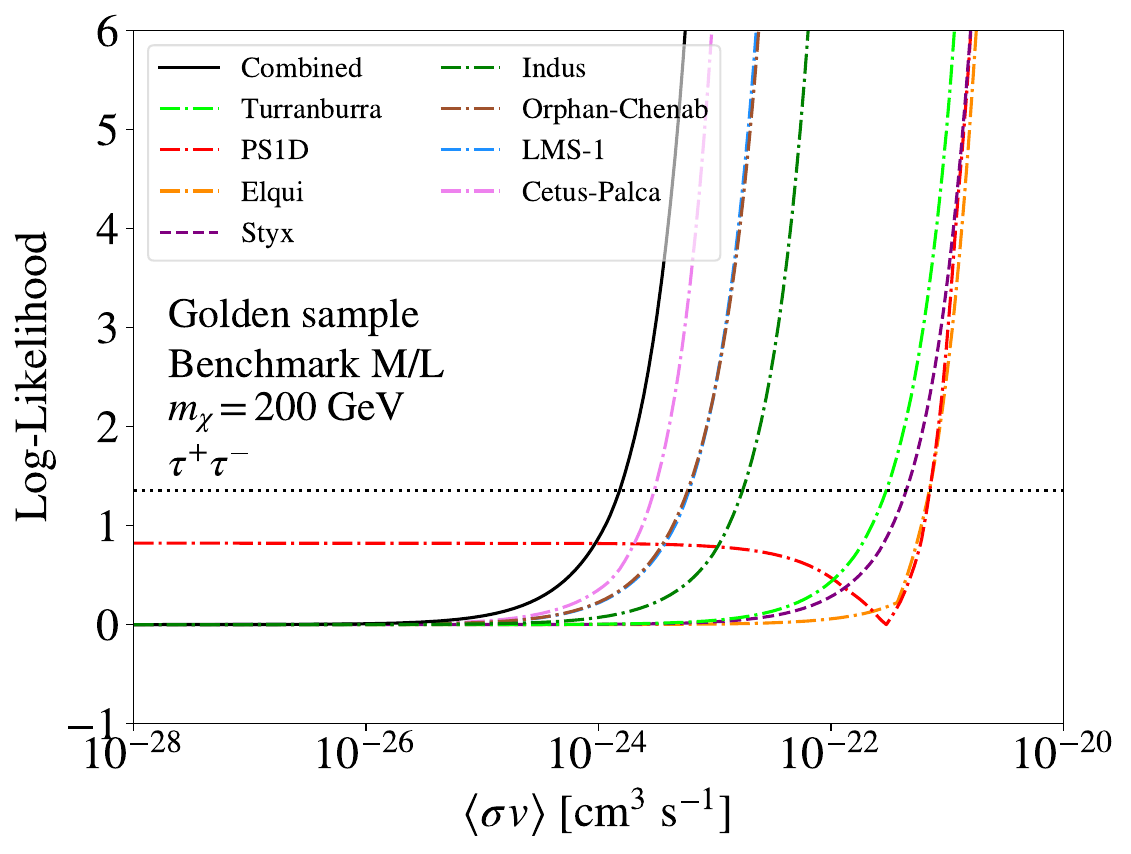}
		        \caption{}
	            \label{fig:lhl_taus}
		  \end{subfigure}
        \caption{Log-likelihood profile for a DM mass of 6000 GeV and $b\bar{b}$ annihilation channel (right panel), and 200 GeV and the $\tau^{+}\tau^{-}$ channel (left panel), for each individual stream in the \textit{golden} sample and the Benchmark M/L scenario. The combined profile is depicted as a thick solid line (to which a shift has been applied to normalize it to the value of zero). The gray dotted line represents a $\Delta\mathcal{L}=2.71/2$, 
        i.e., the value used to set the 95$\%$ C.L. upper limit on the DM annihilation cross-section (which is the value that crosses that line).}
		\label{fig:lhl}
\end{center}
\end{figure}

In this Appendix, we provide examples of the annihilation cross-section log-likelihood profiles described in Subsec. \ref{subsec:method_limits}. More precisely, Figure \ref{fig:lhl} shows individual and combined log-likelihood profiles for the streams in the \textit{golden} sample and the Benchmark scenario as a function of the annihilation cross-section, for a DM mass of 6000 GeV in the $b\bar{b}$ annihilation channel (right panel) and for 200 GeV in the $\tau^{+}\tau^{-}$ channel (left panel). These were chosen as interesting cases according to Fig.~\ref{fig:combined_indviduales_int_golden}. Log-likelihood profiles like these ones are the ones used to obtain 95$\%$ C.L. upper limit on the cross-section for each considered DM mass and annihilation channel.

\newpage
\section{Additional DM modeling parameters and DM constraints for the \textit{golden} sample}
\label{sec:appendix_golden}

Table~\ref{table:table_modeling} of Section \ref{sec:modeling} provides NFWt density profile parameters for the streams in the \textit{golden} sample and the Benchmark scenario. In this Appendix, we provide NFWt parameters for the other two M/L cases, i.e. Low and High. Results are provided in Table \ref{table:table_modeling_golden_lw_up_2}.

\begin{table}[h!]
\begin{adjustwidth}{}{}
\begin{center}
\begin{tabular}{ c  c  c  c  c  c c c }
\hline\hline
Stream & {$\frac{M_{200}}{10^4} \, (M_{\odot})$} & {$R_{200}$ (kpc)} & {$c_{200}$}  & {$\frac{\rho_{0}}{10^8} \, (\frac{M_{\odot}}{kpc^3})$} & {$\frac{M_{s}}{10^4} \, (M_{\odot})$} & {$r_{s}$ (pc)}\\ \hline\hline

\multicolumn{7}{c}{\rule{7cm}{0.4pt} \textit{Low} \rule{7cm}{0.4pt}}\\
Indus & 3.40 & 0.67 & 57.4 & 5.55 & 0.23 & 12.0 \\
LMS-1 & 10.0  & 0.96 & 57.2 & 5.56 & 0.63 & 16.7 \\
Orphan-Chenab & 16.00 & 1.12 & 57.3 & 5.58 & 1.00 & 19.5 \\
PS1-D & 0.75 & 0.40 & 53.3 & 4.58 & 0.05 & 7.6 \\
Turranburra & 0.76 & 0.41 & 50.1 & 3.86 & 0.05 & 8.0 \\
Cetus-Palca & 150.00 & 2.36 & 52.5 & 4.41 & 9.69 & 44.9\\
Styx & 1.80 & 0.54 & 47.2 & 3.21 & 0.12 & 11.4 \\
Elqui & 1.04 & 0.45 & 47.2 & 3.30 & 0.07 & 9.5 \\

\multicolumn{7}{c}{\rule{7cm}{0.4pt} \textit{High} \rule{7cm}{0.4pt}}\\
Indus & 170.00 & 2.46 & 59.8 & 6.26 & 10.47 & 41.0 \\
LMS-1 & 500.00  & 3.52 & 60.1 & 6.33 & 30.92 & 58.6 \\
Orphan-Chenab & 800.00 & 4.12 & 60.3 & 6.40 & 49.28 & 68.2 \\
PS1-D & 37.50 & 1.48 & 55.1 & 5.01 & 2.39 & 27.0 \\
Turranburra & 38.00 & 1.49 & 51.8 & 4.25 & 2.46 & 28.8 \\
Cetus-Palca & 7500.00 & 8.68 & 55.9 & 5.22 & 472.72 & 155.1 \\
Styx & 90.00 & 1.99 & 49.0 & 3.67 & 5.92 & 40.5 \\
Elqui & 52.00 & 1.65 & 48.9 & 3.64 & 3.41 & 33.8 \\
 
\end{tabular}
\end{center}

\caption{DM modelling parameters for our \textit{golden} sample of stellar streams (Tab.~\ref{table:table_sample2}) in the Low and High M/L scenarios, computed following the procedure described in Section~\ref{sec:modeling} and Eqs. (\ref{eq:rs}-\ref{eq:fc200}). The values of $M_s$ are also given.}
\label{table:table_modeling_golden_lw_up_2}
\end{adjustwidth}
\end{table}

\begin{figure}[h!]
\begin{center}
		\begin{subfigure}{0.495\textwidth}
			\includegraphics[width=\textwidth]{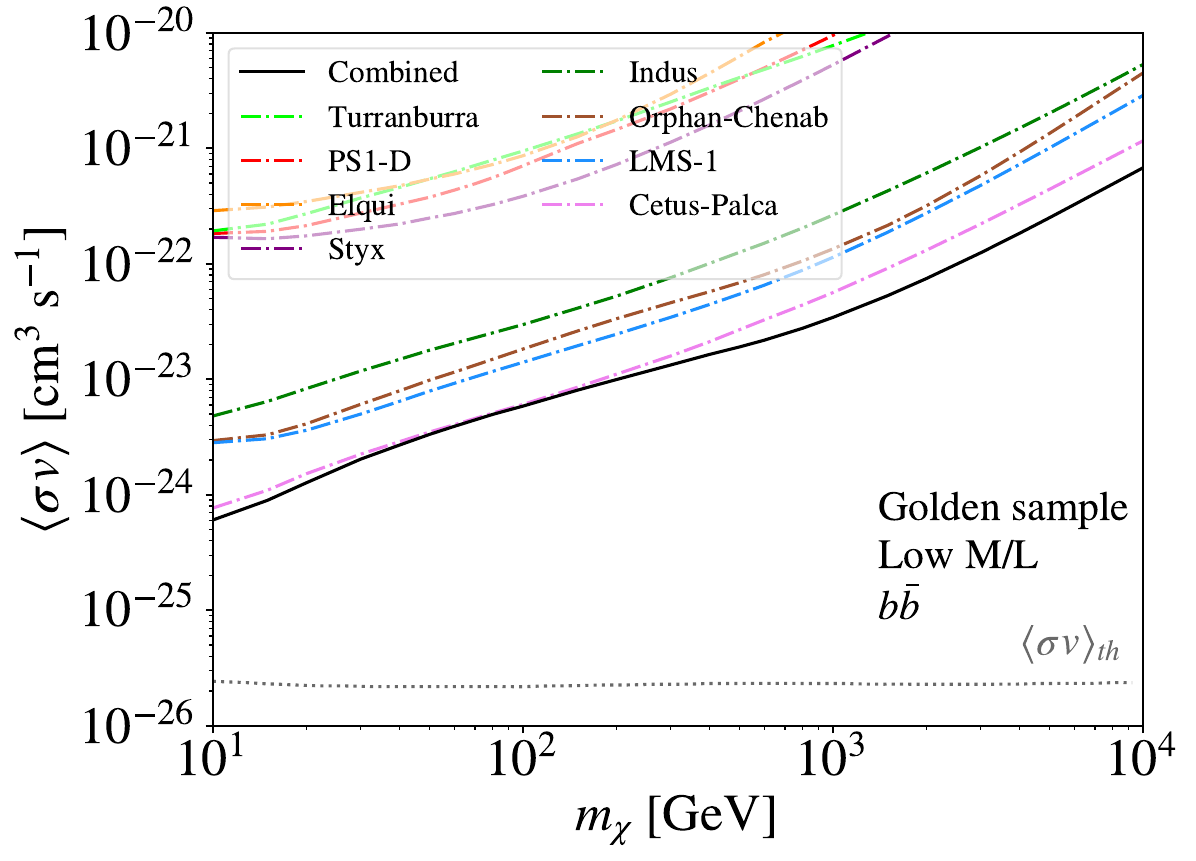}
		\end{subfigure}
		  \begin{subfigure}{0.495\textwidth}
		      	\includegraphics[width=\textwidth]{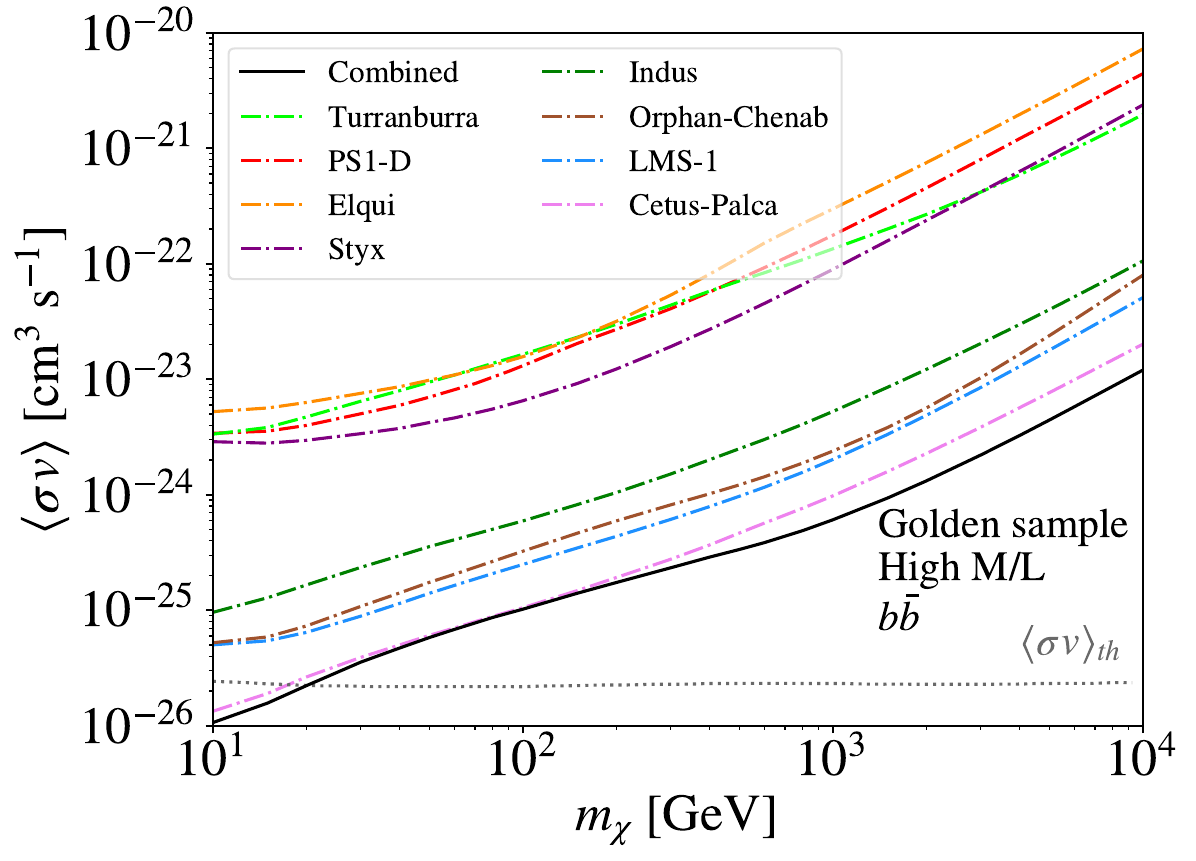}
		  \end{subfigure}
    \caption{Constraints on the DM annihilation cross section for each individual stream in the \textit{golden} sample (Table \ref{table:table_sample2}) and the Low (left) and High (right) M/L scenario described in Section \ref{sec:modeling}. The combined limit is also depicted as a thick solid line. Both panels are for the $b\bar{b}$ annihilation channel. The gray dotted line represents the thermal relic cross-section \cite{Steigman_2012}.}
		\label{fig:combined_Lw_Up_golden_bb}
		
\end{center}
\end{figure}

\begin{figure}[h!]
\begin{center}
		\begin{subfigure}{0.495\textwidth}
			\includegraphics[width=\textwidth]{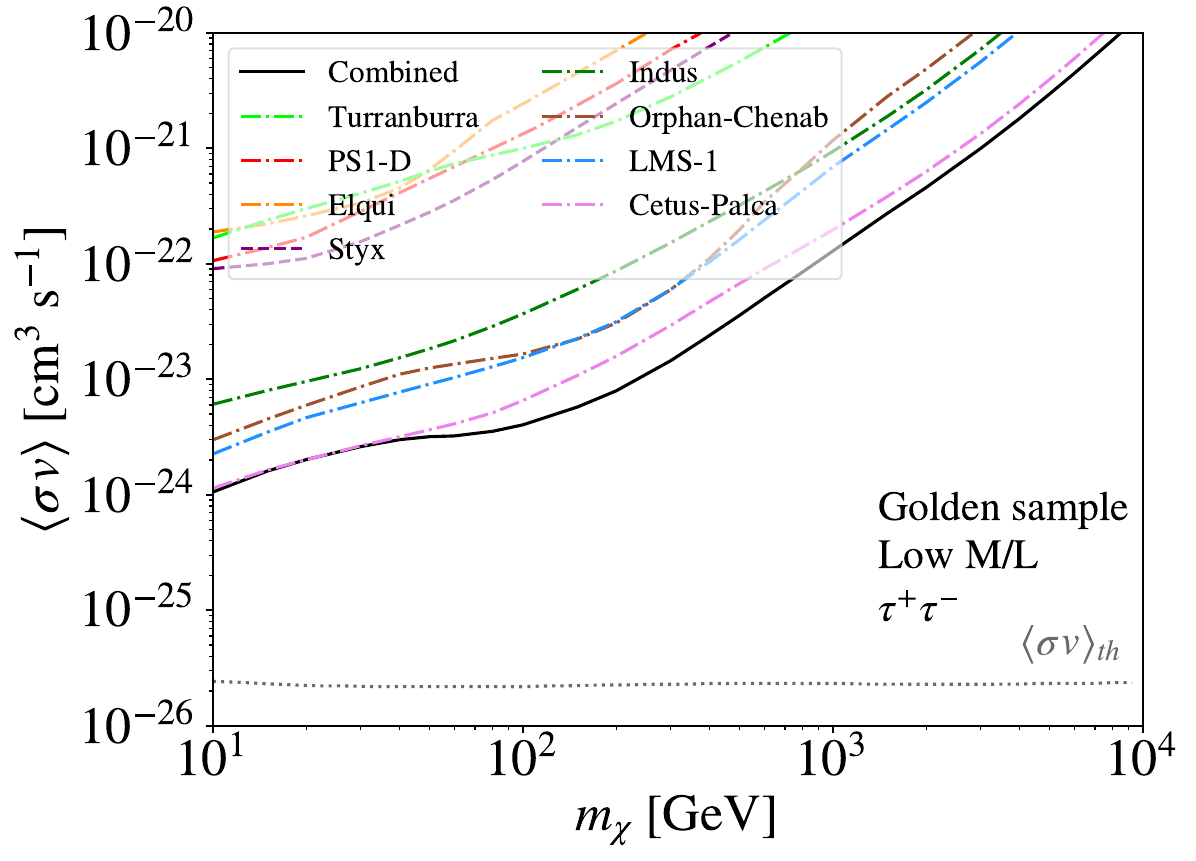}
		\end{subfigure}
		  \begin{subfigure}{0.495\textwidth}
		      	\includegraphics[width=\textwidth]{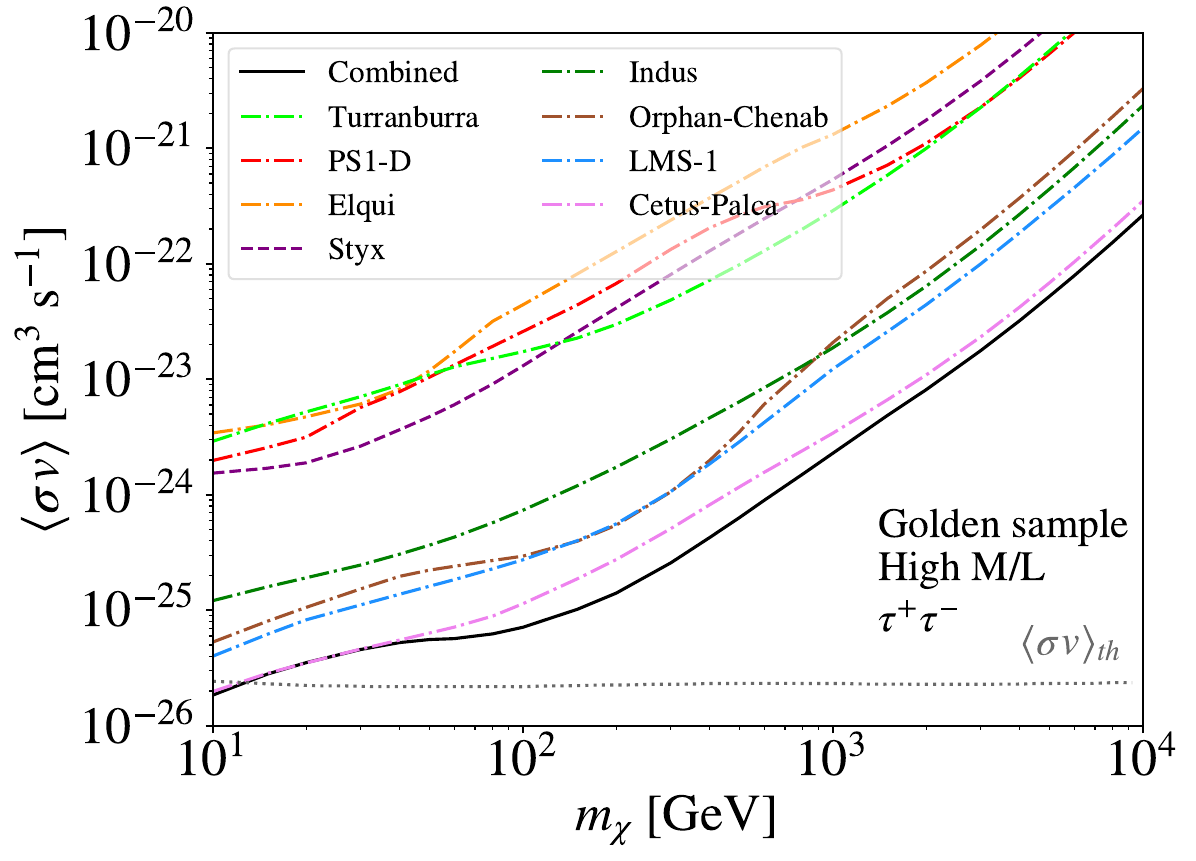}
		  \end{subfigure}
    \caption{Constraints on the DM annihilation cross section for each individual stream in the \textit{golden} sample (Table \ref{table:table_sample2}) and the Low (left) and High (right) M/L scenario described in Section \ref{sec:modeling}. The combined limit is also depicted as a thick solid line. Both panels are for the $\tau^{+}\tau^{-}$ annihilation channel. The gray dotted line represents the thermal relic cross-section \cite{Steigman_2012}.}
		\label{fig:combined_Lw_Up_golden_taus}
		
\end{center}
\end{figure}

Moreover, we include in this same appendix the corresponding DM limits for the Low and High scenarios, both for the \textit{golden} sample. Figures \ref{fig:combined_Lw_Up_golden_bb} and~\ref{fig:combined_Lw_Up_golden_taus} display, respectively, such DM constraints for each individual stream and for the combined sample, for the $b\bar{b}$ and $\tau^{+}\tau^{-}$ annihilation channels. In Low M/L case, independently of the channel, the combined constraints are $\mathcal{O}(10^2)$ above the thermal relic cross-section for low WIMP masses. When considering the High scenario, the combined limit crosses the thermal relic cross-section at low masses, ruling out thermal WIMPs up to 20 GeV (10 GeV) for the $b\bar{b}$ ($\tau^{+}\tau^{-}$) channel.

\newpage
\section{Additional DM modeling parameters and DM constraints for the \textit{silver} sample}
\label{sec:appendix_silver}

This appendix provides additional information about the \textit{silver} sample that has not been included in the main text of the paper. Specifically, in Table \ref{table:table_modeling_silver_lw_bench_up}, we show the obtained results for the three considered M/L scenarios (Low, Benchmark and High), of the DM modelling of the three additional streams that are included in the \textit{silver} sample. Finally, Table \ref{table:table_jfactor_silver} provides angular sizes and J-factor values computed for these streams under the three considered M/L scenarios.

\begin{table}[h!]
\begin{center}
\begin{tabular}{ c  c  c  c  c  c c c }
\hline\hline
Stream & {$\frac{M_{200}}{10^4} \, (M_{\odot})$} & {$R_{200}$ (kpc)} & {$c_{200}$}  & {$\frac{\rho_{0}}{10^8} \, (\frac{M_{\odot}}{kpc^3})$} & {$\frac{M_{s}}{10^4} \, (M_{\odot})$} & {$r_{s}$ (pc)} \\ \hline\hline
\multicolumn{7}{c}{\rule{5.5cm}{0.4pt} \textit{Benchmark} \rule{5.5cm}{0.4pt}}\\
Monoceros & 3000.00 & 6.39 & 62.0 & 6.89 & 182.74 & 103.0  \\
AntiCenter & 4.65 & 0.74 & 59.1 & 6.01 & 0.15 & 10.0  \\
Sagittarius & 65000.00 & 17.82 & 59.8 & 6.25 & 4014.46 & 298.0  \\

\multicolumn{7}{c}{\rule{6cm}{0.4pt} \textit{Low} \rule{6cm}{0.4pt}}\\
Monoceros & 600.00 & 3.74 & 60.4 & 6.41 & 37.08 & 62.0 \\
AntiCenter & 0.93 & 0.43 & 58.4 & 5.84 & 0.58 & 7.4 \\
Sagittarius & 13000.00 & 10.42 & 57.9 & 5.73 & 811.09 & 180.0 \\

\multicolumn{7}{c}{\rule{6cm}{0.4pt} \textit{High} \rule{6cm}{0.4pt}}\\
Monoceros & 30000.00 & 13.78 & 64.7 & 7.74 & 1810.31 & 212.8 \\
AntiCenter & 46.50 & 1.59 & 60.5 & 6.44 & 2.88 & 26.4 \\
Sagittarius & 650000.00 & 38.40 & 62.9 & 7.17 & 39500.89 & 610.0 \\
 
\end{tabular}
\end{center}

\caption{DM modelling parameters for the three additional stellar streams that we include in the \textit{silver} sample (Tab.~\ref{table:table_sample2}) in the three considered M/L scenario (Low, Benchmark and High), computed following the procedure described in Section~\ref{sec:modeling} and Eqs. (\ref{eq:rs}-\ref{eq:fc200}). The values of $M_s$ are also given. 
}
\label{table:table_modeling_silver_lw_bench_up}
\end{table}

\begin{table}[h!]
    \begin{center}
        
    \begin{tabular}{| c | c | c | c | c | c | c |}
        \hline
        Stream & \multicolumn{3}{c|}{$\theta_{s} \, (^{\circ})$} & 
        
        \multicolumn{3}{c|}{$\log_{10}J_s \, (GeV^{2} cm^{-5})$} 
        \\
        \cline{2-7}
        & \textit{Low} & \textit{Bench.} & \textit{High} & 
    
        \textit{Low} & \textit{Bench.} &\textit{High}
        \\
        \hline
        Monoceros & 0.34 & 0.56 & 1.15  & 18.91 & 19.63 & 20.68\\
        AntiCenter & 0.04 & 0.05 & 0.13 & 15.97 & 16.39 & 17.72\\
        Sagittarius & 0.41 & 0.68 & 1.38 & 19.45 & 20.18 & 21.23\\
        
        \hline
        
    \end{tabular} 
    \end{center} 
    \caption{Values of the angular extension of the streams' core ($\theta_{s}$), and the astrophysical J-factors integrated up to $r_{s}$ ($J_s$) for the three additional stellar streams that we include in the \textit{silver} sample, in the three considered M/L scenarios (Low, Benchmark, High).}
    \label{table:table_jfactor_silver}
\end{table}

\begin{figure}[h!]
\begin{center}
		\begin{subfigure}{0.495\textwidth}
			\includegraphics[width=\textwidth]{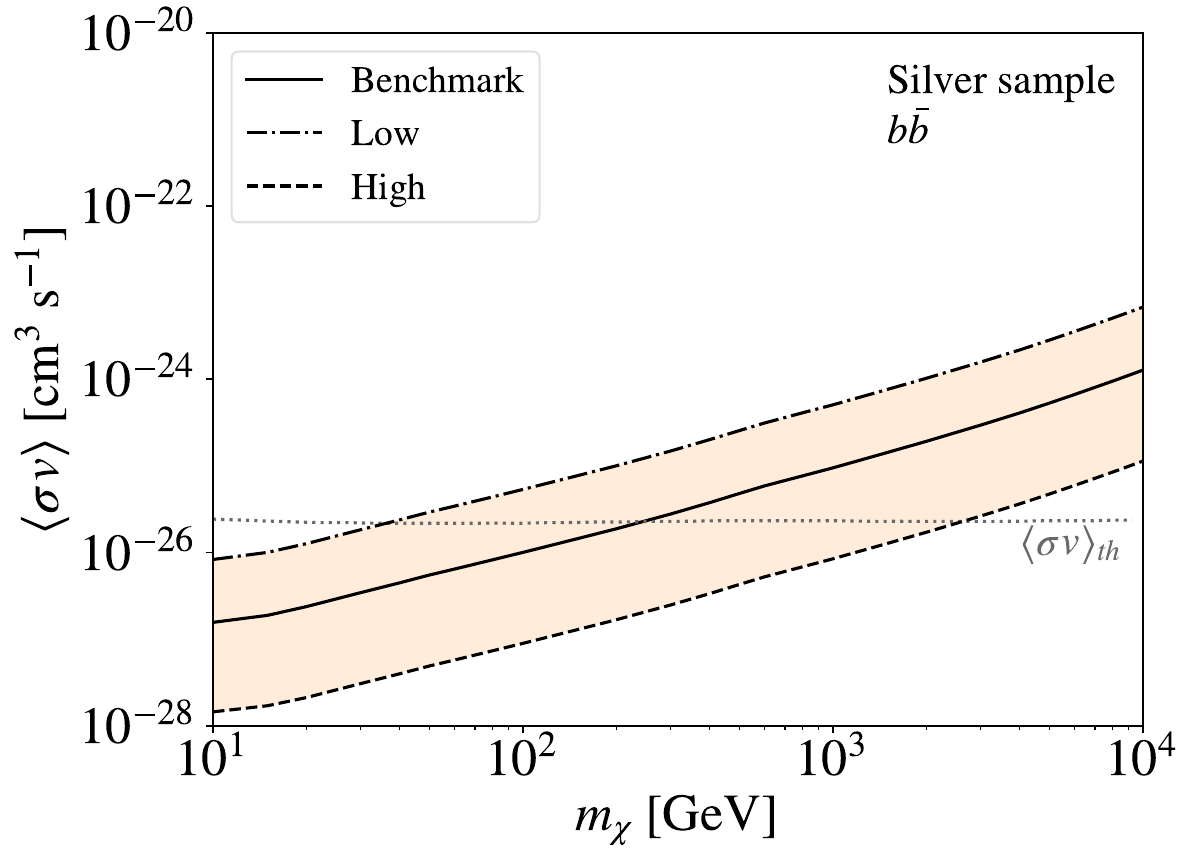}
		\end{subfigure}
		  \begin{subfigure}{0.495\textwidth}
		      	\includegraphics[width=\textwidth]{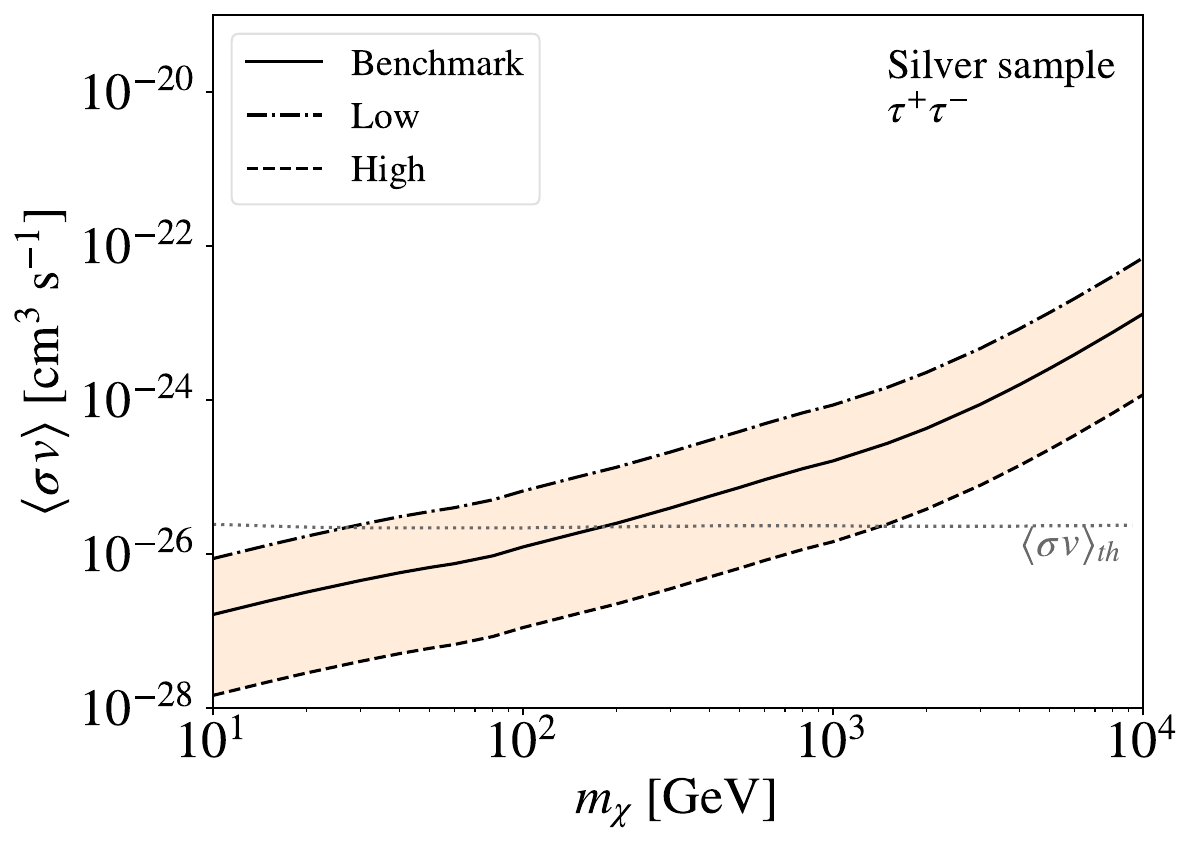}
		  \end{subfigure}
        \caption{Limits on the DM annihilation cross-section obtained in the combined analysis of our \textit{silver} sample of streams (Table \ref{table:table_sample2}), for the three M/L scenarios considered in this work (see Section \ref{sec:modeling}). Left and right panels refer to the $b\bar{b}$ and $\tau^{+}\tau^{-}$ annihilation channels, respectively.}
        
		\label{fig:combined_Lw_int_Up_silver}
		
\end{center}
\end{figure}

We also provide in this appendix further information regarding the DM limits obtained for the \textit{silver} sample. Figure \ref{fig:combined_Lw_int_Up_silver} shows the combined limits obtained in each of the three M/L scenarios considered in this work, both for $b\bar{b}$ and $\tau^{+}\tau^{-}$ annihilation channels. Focusing on the Benchmark scenario, the DM limits allow to rule out thermal WIMPs up to $\sim 200$ GeV for both channels. 

\begin{figure}[h!]
\begin{center}
		\begin{subfigure}{0.495\textwidth}
			\includegraphics[width=\textwidth]{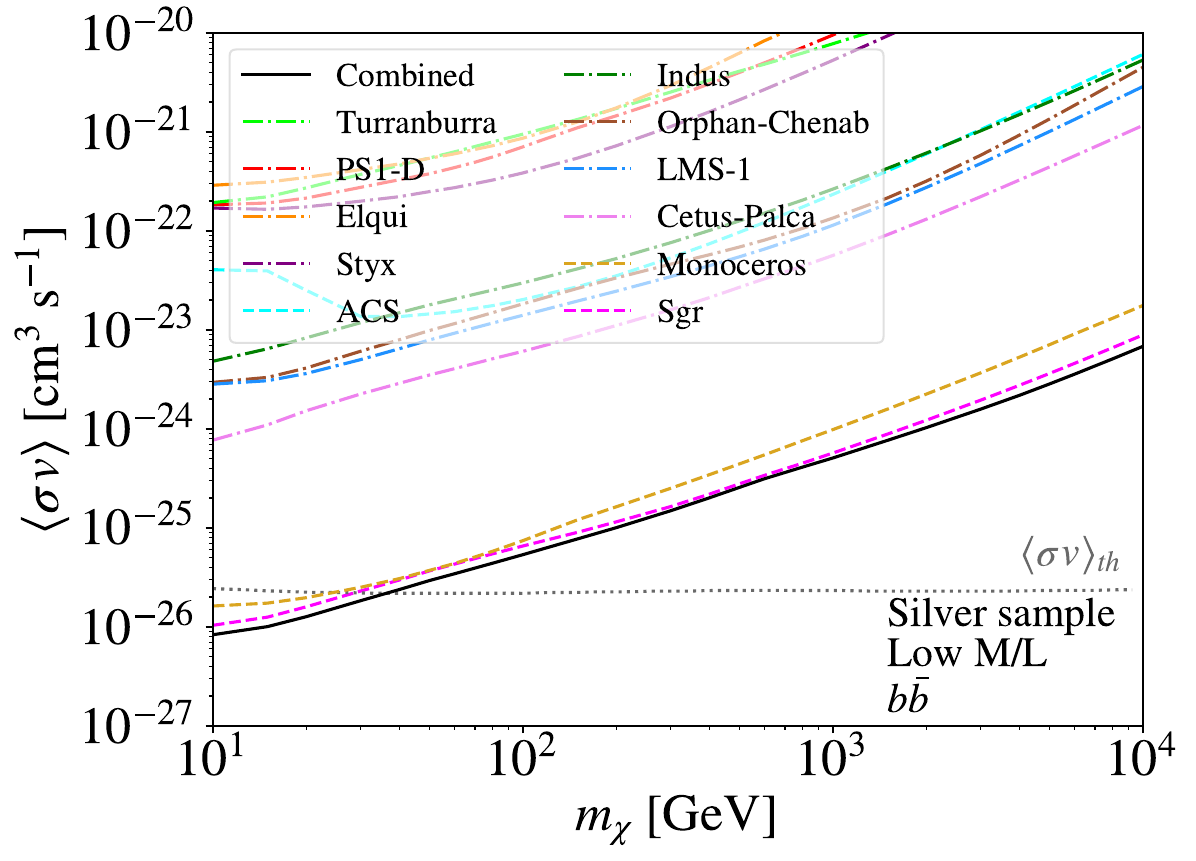}
		\end{subfigure}
		  \begin{subfigure}{0.495\textwidth}
		      	\includegraphics[width=\textwidth]{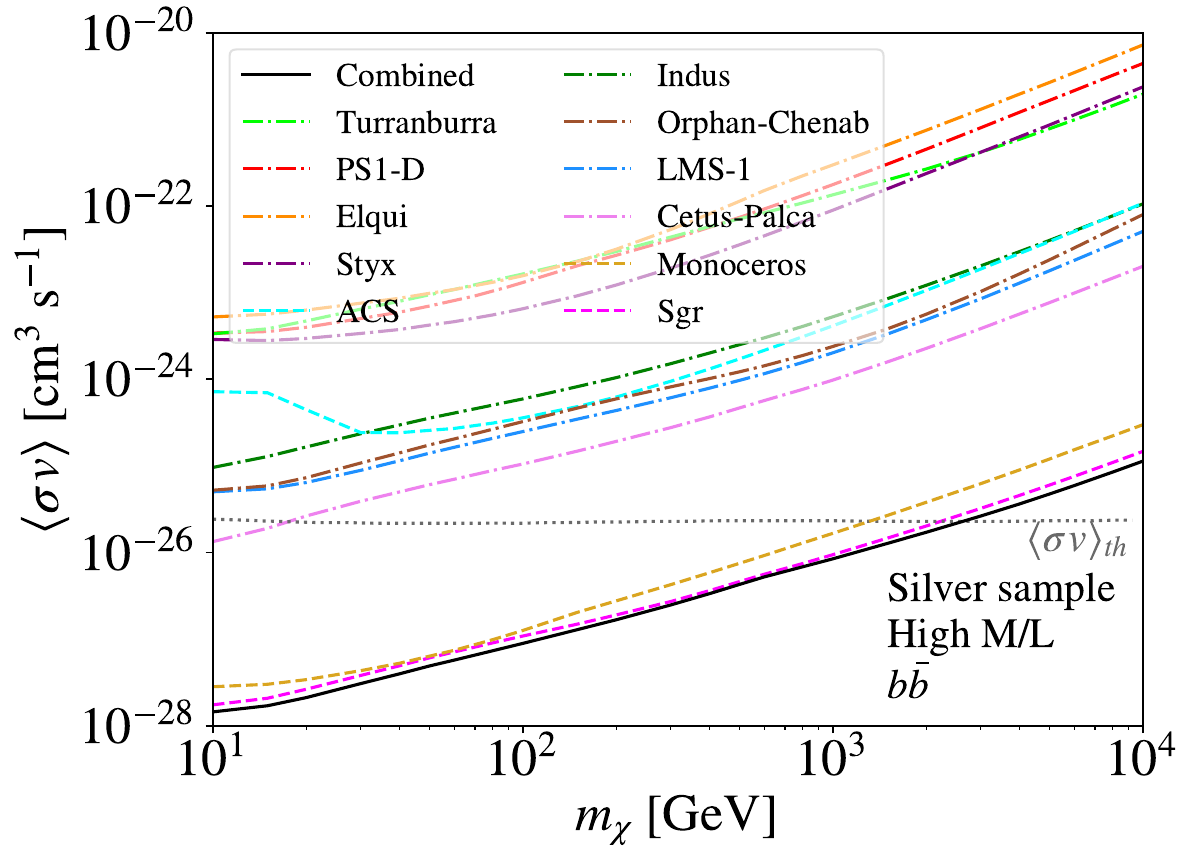}
		  \end{subfigure}
    \caption{Constraints on the DM annihilation cross section for each individual stream in the \textit{silver} sample (Table \ref{table:table_sample2}) and the Low (left) and High (right) M/L scenario described in Section \ref{sec:modeling}. The combined limit is also depicted as a thick solid line. Both panels are for the $b\bar{b}$ annihilation channel. The gray dotted line represents the thermal relic cross-section \cite{Steigman_2012}.}

		\label{fig:combined_Lw_Up_silver_bb}
		
\end{center}
\end{figure}

\begin{figure}[h!]
\begin{center}
		\begin{subfigure}{0.495\textwidth}
			\includegraphics[width=\textwidth]{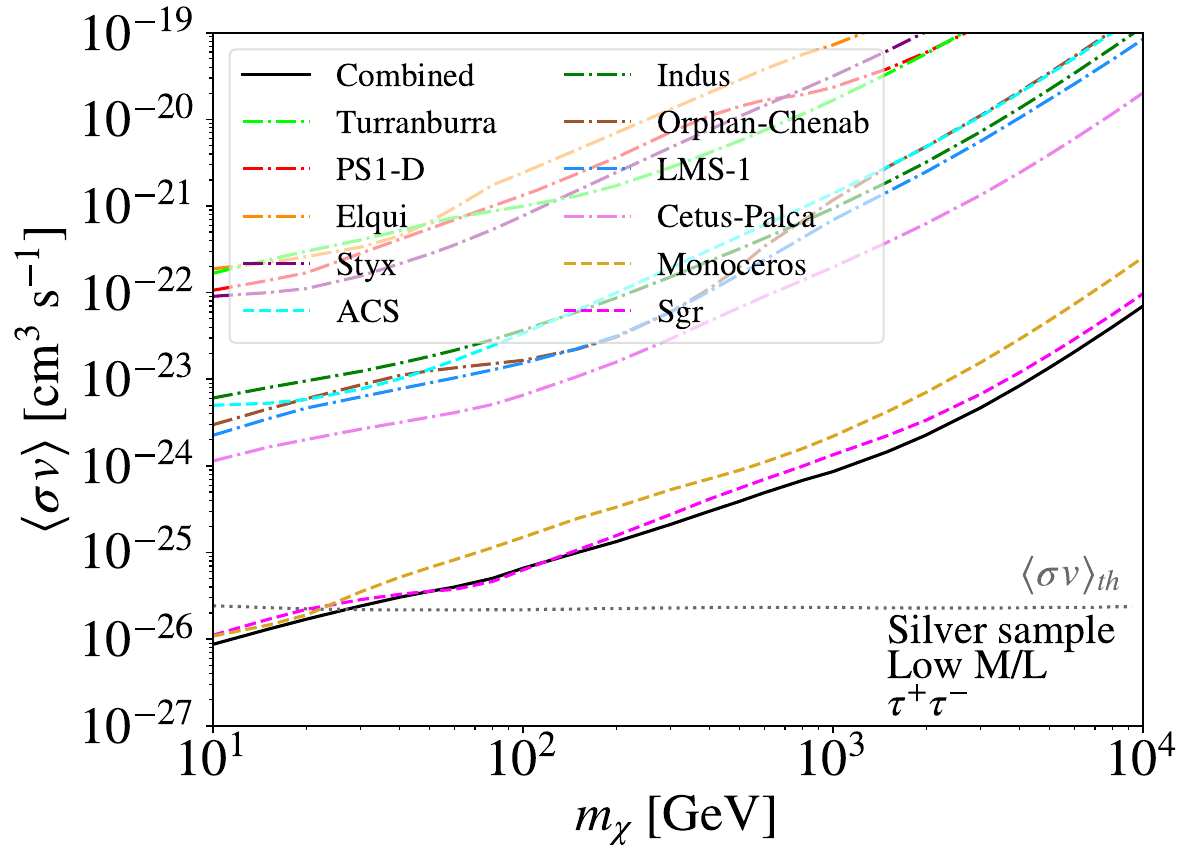}
		\end{subfigure}
		  \begin{subfigure}{0.495\textwidth}
		      	\includegraphics[width=\textwidth]{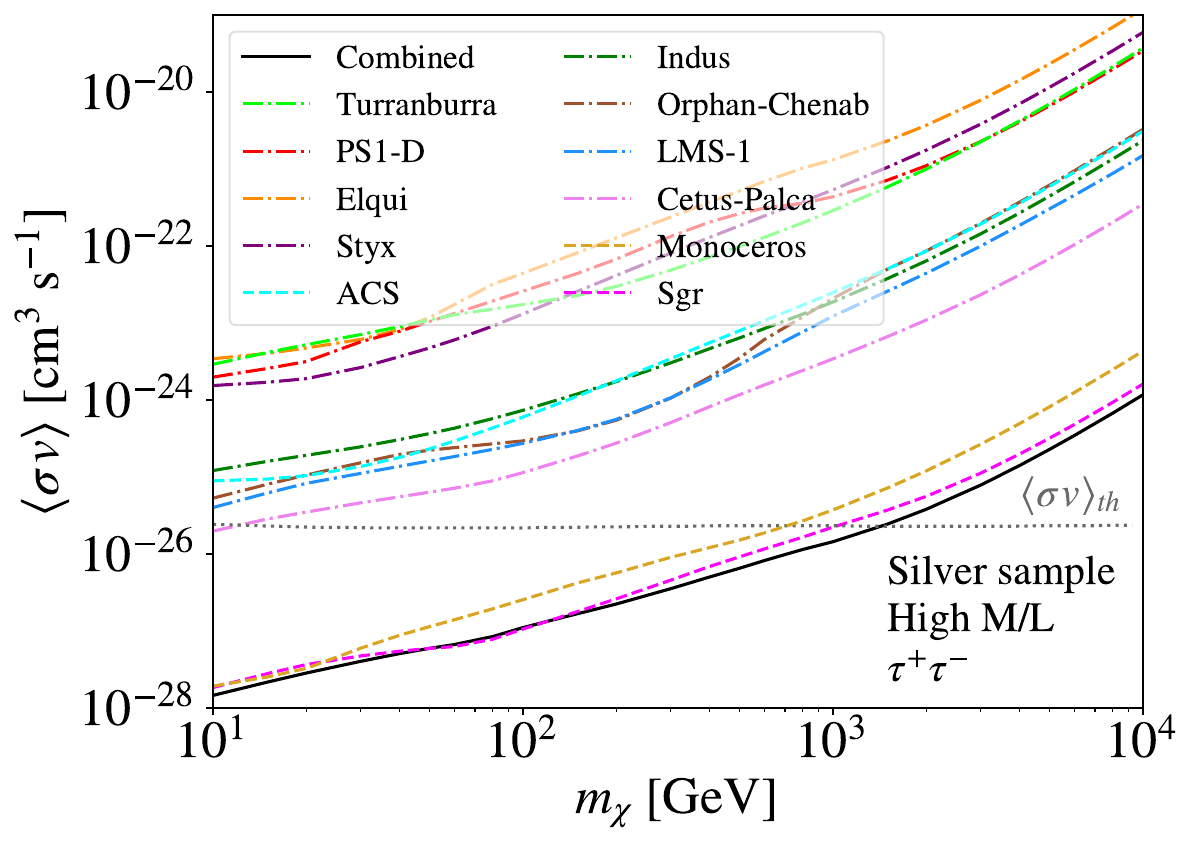}
		  \end{subfigure}
    \caption{Constraints on the DM annihilation cross section for each individual stream in the \textit{silver} sample (Table \ref{table:table_sample2}) and the Low (left) and High (right) M/L scenario described in Section \ref{sec:modeling}. The combined limit is also depicted as a thick solid line. Both panels are for the $\tau^{+}\tau^{-}$ annihilation channel. The gray dotted line represents the thermal relic cross-section \cite{Steigman_2012}.}

		\label{fig:combined_Lw_Up_silver_taus}
		
\end{center}
\end{figure}

Figures \ref{fig:combined_Lw_Up_silver_bb} and \ref{fig:combined_Lw_Up_silver_taus} show the constraints on the DM annihilation cross-section for each individual stream and the combined analysis for the $b\bar{b}$ and the $\tau^{+}\tau^{-}$ channels, respectively, and in the Low and High M/L scenarios. The combined limit obtained in the case of the Low scenario rules out thermal WIMPs up to $\sim 20 - 30$ GeV in both channels. The High scenario represents the most extreme (thus probably the less robust) J-factor case in this whole study, allowing to discard thermal WIMPs with masses up to $\sim 1000 - 2000$ GeV for both channels. 

\newpage


\bibliographystyle{JHEP}
\bibliography{biblio.bib}



\end{document}